\documentclass[10pt]{amsart}
\usepackage{amsmath,amssymb,eucal}
\usepackage[dvips]{epsfig}
\newtheorem{theorem}{Theorem}[section]
\newtheorem{lemma}[theorem]{Lemma}
\newtheorem{corollary}[theorem]{Corollary}
\newtheorem{proposition}[theorem]{Proposition}
\newtheorem{definition}{Definition}[section]
\newtheorem{example}[theorem]{Example}
\newtheorem{remark}[theorem]{Remark}
\allowdisplaybreaks
\numberwithin{equation}{section}
\newcommand{\rnc}{\renewcommand}
\newcommand{\nc}{\newcommand}
\newcommand{\be}{\boldsymbol{\epsilon}}
\newcommand{\bd}{\boldsymbol{\delta}}
\nc{\pq}{\left[^{\boldsymbol{\delta}}_{\boldsymbol{\epsilon}}\right]}
\nc{\mpq}{\left[^{-\boldsymbol{\delta}}_{-\boldsymbol{\epsilon}}\right]}
\nc{\pqt}{\left[^{\tilde{\boldsymbol{\delta}}}_{\tilde{\boldsymbol{\epsilon}}}
\right]}
\nc{\de}{\left[^{\delta}_{\epsilon}\right]}
\newcommand{\res[1]}{\operatornamewithlimits{Res}_{#1}}
\nc{\no}{\nonumber} \nc{\IM}{\hbox{\mathrm{Im}}\,}
\nc{\A}{\mathcal{A}}
\nc{\ID}{{\mathbf 1}} \nc{\RE}{\mathrm{Re}\,}
\nc{\Id}{\mathrm{Id}\,}
\nc{\C}{{\mathbb C}}
\nc{\T}{{T}}
\nc{\Cp}{{\mathbbC\rm I\!P}}
\nc{\Z}{\mathbb{Z}}
\nc{\bv}{\boldsymbol{v}}
\newcommand{\Rn}{{\rm I\!R}}
\nc{\e}{\mbox{e}}
\rnc{\i}{\imath}
\nc{\Uc}{\boldsymbol{\mathcal{U}}}
\nc{\Ac}{\mathcal{A}}
\nc{\Bc}{\mathcal{B}}
\nc{\Cc}{\mathcal{C}}
\nc{\tAc}{\tilde{\mathcal{A}}}
\nc{\tBc}{\tilde{\mathcal{B}}}
\nc{\tCc}{\tilde{\mathcal{C}}}
\nc{\p}{\bf p}
\nc{\q}{\bf q}
\rnc{\d}{\mathrm{d}}
\nc{\lb}{\lambda}
\rnc{\a}{\alpha}
\rnc{\b}{\beta}
\nc{\hS}{\hat{S}}
\nc{\g}{\it{g}}
\nc{\Si}{\Sigma}
        \newcommand{\li}{\mathcal{L}}

\newcommand{\ra}{\rightarrow}
\begin{document}

\title[Singular $Z_N$ curves, Riemann-Hilbert problem, Schlesinger equation]
{Singular $Z_N$ curves, Riemann-Hilbert problem and  Schlesinger equations}

\author[Enolski]{V.Z. Enolski}
\address{Dipartimento di Fisica ``E.R.Caianiello", Universit\`a di Salerno\\
Via S.Allende -84081 Baronissi (SA), ITALY}
\email{enolskii@sa.infn.it,vze@ma.hw.ac.uk}

\author[Grava]{T.Grava}
\address{SISSA, via Beirut 2-4 340104 Trieste, Italy\\
E-mail: grava@sissa.it}

\thanks  {
VZE is  grateful to Abdus Salam Center for Theoretical Physics and
SISSA for support in 2003 year}

\begin{abstract}
We are solving the classical Riemann-Hilbert  problem of rank $N>1$ on the extended  complex plane punctured in  $2m+2$  points, for
 $N\times N$ quasi-permutation monodromy  matrices.  
Our approach  is based on the finite gap
integration method    applied to
study the Riemann-Hilbert by Kitaev and Korotkin 
\cite{kk98}, Deift, Its, Kapaev and Zhou \cite{dikz99} and
Korotkin, \cite{kor01}.
This permits us to solve the  Riemann-Hilbert problem 
in terms of the Szeg\"o kernel of  certain Riemann surfaces 
branched over the given $2m+2$ points.
The monodromy group of  these Riemann surfaces 
is determined  from the quasi-permutation monodromy  matrices  of the Riemann-Hilbert problem by setting all their non-zero entries equal to one. 
In our case, the monodromy group of the Riemann surfaces 
turns out to be the cyclic subgroup  $Z_N$  of the symmetric group $S_N$  
and for this reason these Riemann surfaces of genus $N(m-1)$ have $Z_N$ symmetry.
This fact enables us to write  the matrix entries of the solution of
the $N\times N$   Riemann-Hilbert problem as a product of an algebraic function
and $\theta$-function quotients. The algebraic function is
 related to the  Szeg\"o kernel with zero characteristics.

From the solution of the Riemann-Hilbert problem we
automatically obtain a particular solution of the Schlesinger system.
The $\tau$-function of the Schlesinger system is computed
explicitly in terms of $\theta$-functions and the holomorphic 
projective connection of the Riemann surface. 
In the course of the computation we also derive Thomae-type formulae for
a class of non-singular $1/N$-periods.

Finally we study in detail the solution of  the rank $3$  problem
with four singular points $(\lb_1,\lb_2,\lb_3,\infty)$.
The corresponding Riemann surface $\mathcal {C}_{3,1}$ is of genus two
branched at the above four  points  and admits the dihedral
group $D_3$ of automorphisms. This implies that $\mathcal{C}_{3,1}$
is a $2$-sheeted cover of two elliptic curves
which  are 3-isogenous. As a result, the corresponding solution of
the  Riemann-Hilbert problem and the Schlesinger system is given in
terms of Jacobi's $\vartheta$-function with modulus
$T=T(t)$, $t=\frac{\lb_2-\lb_1}{\lb_3-\lb_1}$ and $\mathrm{Im}\,T>0$.
 The inverse function $t=t(T)$ is automorphic under the action of
  the subgroup $\Gamma_0(3)$ of the modular group
  and generates a  solution of a general
Halphen system. The analytic counterpart of this
picture is given by Goursat's higher identities for hypergeometric
functions.
\end{abstract}

\maketitle
\tableofcontents

\section{Introduction}
The Riemann-Hilbert problem  (R-H problem) in its
classical formulation consists of
 deriving a linear differential equation of Fuchsian type with a given
set $D$ of singular points and  a given monodromy representation
\begin{equation}
\label{rep}
\mathcal{M}\;:\;
\pi_1(\mathbb{CP}^1\backslash D,\lambda_0)\rightarrow
GL(N,\C),\quad N\geq 2,
\end{equation}
of the fundamental group $\pi_1(\mathbb{CP}^1\backslash
D,\lambda_0)$. An element $\gamma$ of  the group
$\pi_1(\mathbb{CP}^1\backslash D,\lambda_0)$ is a loop contained in
$\mathbb{CP}^1\backslash D$ with initial and end
point $\lambda_0\in \mathbb{CP}^1$, $\lambda_0 \notin  D$.
Not all the representation (\ref{rep}) can be realized as the
monodromy representation of a Fuchsian system.
For $N=3,4$ representations (\ref{rep})  for which the R-H problem cannot be
solved are given in \cite{an90} and \cite{gla00} respectively. In
dimension $N=2$ the R-H problem is always solvable \cite{dek79} for
an arbitrary number of singular points. For $N\geq 3$,  every irreducible
representation (\ref{rep}) can be realized as the monodromy
representation of some Fuchsian system \cite{an90},\cite{anbo94},\cite{ko92}.
In general, among the solvable cases, the solution of the matrix R-H
problem cannot be computed analytically in terms of
known special functions \cite{ok87},\cite{um90}. Nevertheless,
there are special cases when the R-H problem can be solved
explicitly in terms of $\theta$-functions
\cite{kk98},\cite{dikz99},\cite{kor01}. We
discuss one of these cases.

\begin{figure}[htb]
\centering
\mbox{\epsfig{figure=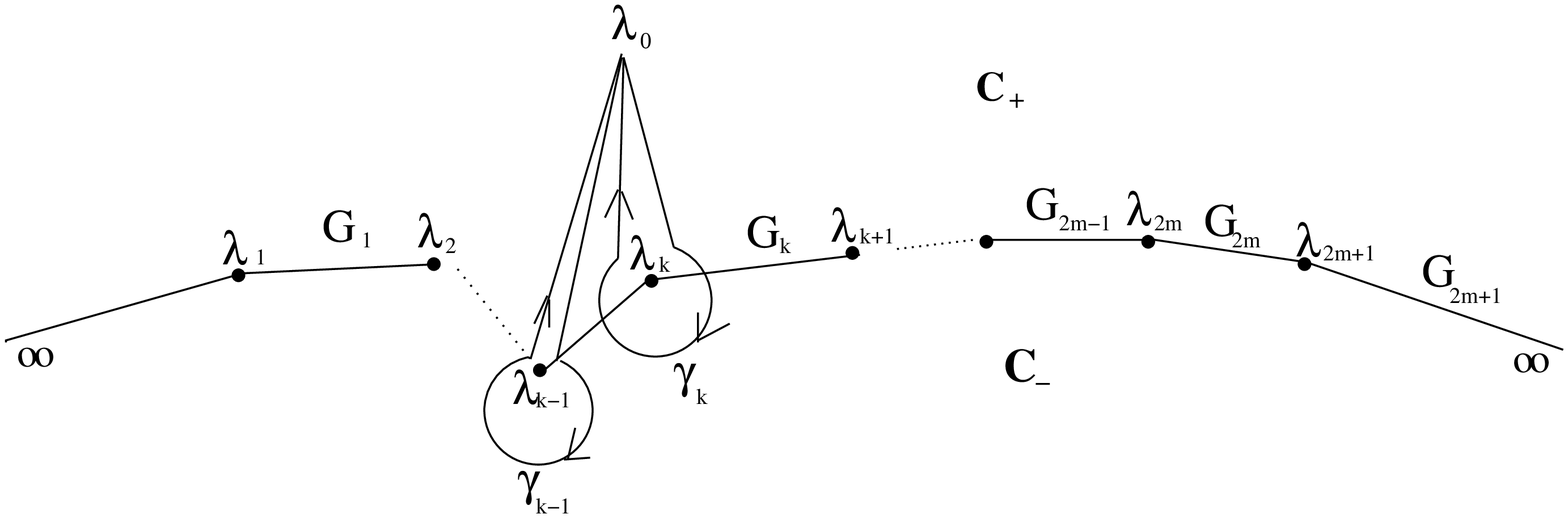,width=1.0\textwidth}}
     \caption{The contour $\li$\label{fig1}}
\end{figure}

The method of solution proposed by Plemelj \cite{pl64} consists
of reducing  the R-H problem to a homogeneous boundary value
problem in the complex plane for a $N\times N$   matrix  function
$Y(\lb)$. The boundary can be chosen in the form of a polygon line
$\li$, by connecting all the
singular points of the set $ D:=\{\lambda_1,\lambda_2,\dots,\lambda_{2m+1},
\lambda_{2m+2}=\infty\}$. The line $\li$ divides the
complex plane into two domains, $C_-$ and $C_+$ (see Figure.~\ref{fig1}).
Let  $\gamma_1,\gamma_2,\dots,\gamma_{2m+2}$ denote the set of generators
of the fundamental group
$\pi_1(\mathbb{CP}^1\backslash D,\lambda_0)$, i.e. the homotopy class
$\gamma_k$ corresponds to a small clock-wise loop around the
point $\lb_k$ (see Figure~\ref{fig1}).  Then the matrices $\mathcal{M}(\gamma_k)=M_k\in SL(N,\C), $ $ k=1,\dots,2m+2$, form a set of generators of the monodromy group.
Since the homotopy relation
\[
\gamma_1\circ\gamma_2\circ\dots\circ\gamma_{2m+2}\simeq \{\lb_0\},
\]
the generators $M_k$ satisfy the cyclic relation
\[
M_{\infty}M_{2m+1}\dots M_1=1_N,
\]
where $M_{2m+2}=M_{\infty}$.
Let us construct the matrices $G_k$ defined by
\begin{equation}
\begin{split}
\label{Gk}
&G_k=M_{k}M_{k-1}\dots M_1,\quad k=1,\dots,2m+2\\
&G_0=G_{2m+2}=1_N.
\end{split}
\end{equation}

The homogeneous Hilbert boundary value problem   formulated
by Plemely is the following \cite{pl64}: find the $N\times N$ matrix
 function $Y(\lb)$ which satisfies the following conditions
\vskip 0.2cm \noindent {\bf(i) } $Y(\lb)$ is analytic in
$\mathbb{CP}^1\backslash \li$;
 \vskip 0.2cm
\noindent
{\bf(ii)}
the limits $Y_{\pm}(\lb)$ as $\lb\rightarrow \li_{\pm}$
satisfy the jump conditions
\[
Y_-(\lb)=Y_+(\lb)G_k,\quad \lb\in (\lambda_k,\lambda_{k+1}),
\;\;k=0,\dots,2m+1,\;\lambda_0=\lambda_{2m+2}=\infty;
\]
\vskip 0.2cm
\noindent
{\bf (iii)} for  $0\leq \epsilon<1$,
\[
Y\left(\frac{1}{\lb}\right)\left(\frac{1}{\lb}\right)^{\epsilon}
\rightarrow 0\;\;\text{as}\;\;
 \lb\rightarrow \infty\;\;\text{ and}\;\;
Y_{\pm}(\lb)(\lb-\lambda_j)^{\epsilon}\rightarrow 0\;\;\text{ as}  \;\;\lb\rightarrow \lambda_j,
\]
over $C_+$ or $C_-$ respectively;
\vskip 0.2cm
\noindent
{\bf (iv)} $Y(\lb_0)=1_N,\quad \lb_0\in\C_+\backslash D.$
\vskip 0.2cm

 There is always a solution  of {\bf (i)-(iv)}  such
that $\det Y(\lb)\neq 0$ for $\lb\neq  D$.
The  analytic continuation of the  solution $Y(\lb)$ along a
small loop $\gamma_k$   around $\lambda_k$ is determined by the matrix
$G_kG_{k-1}^{-1}=M_k$, namely
\[
Y(\gamma_k(\lb))=Y(\lb)M_k,\quad\lb\in C_+\backslash D,\;\; k=1,\dots,2m+2.
\]
 It is possible to show
that the solution $Y(\lb)$ of the R-H problem  {\bf (i)-(iv)} satisfies
a Fuchsian equation
\begin{equation}
\label{Fuchsian0}
\dfrac{\d Y(\lb)}{\d\lb}
=\sum_{k=1}^{2m+1}\dfrac{A_k}{\lb-\lambda_k} \;Y(\lb),\quad A_k\in Mat(N,\C),
\end{equation}
if  one of the monodromy matrices is diagonalisable
\cite{aril85},\cite{tr83}. Without this condition,
Plemelj original argument does not go through.

 Kitaev and Korotkin \cite{kk98}  and  Deift, Its, Kapaev, and Zhou
\cite{dikz99}  solved the $2\times 2$ matrix R-H problem when all
the matrices  $G_{2k}$ are diagonal and
all  the matrices $G_{2k-1}$, $k=1,\dots,m+1$,  are off-diagonal.
The idea of the construction in \cite{kk98},\cite{dikz99} is to consider a
hyperelliptic covering  ${\mathcal C}$ over $\mathbb{CP}^1$ which is
ramified in $ D$ and use the natural monodromy of the  hyperelliptic
curve. The application of methods of finite-gap integration
\cite{zmnp80,bbeim94} permits to obtain a $\theta$-function
solution for the problem, depending on $2m$ parameters. 
A solution of the $2\times 2$ case appeared also
in \cite{diz97},\cite{dkmvz99} in the study of asymptotic problems 
arising in the theory of random matrix models and  in \cite{dvz97}
in the study of the small dispersion limit of the Korteweg de Vries 
equation.

The extension of the $2\times 2$ matrix R-H problem  to higher
dimensional matrices  leads naturally to non-hyperelliptic curves.
This fact was pointed out  by Zverovich
\cite{zv71}, who considered the $N\times N$ problem {\bf (i)-(iv)}
when  all the matrices $G_{2k}$ are
diagonal, and the non-zero entries  of  the matrices $G_{2k-1}$, $k=1,\dots,m+1$ are
\begin{equation}
\label{Gkm}
\begin{split}
&(G_{2k-1})_{i,i-1}\neq 0, \;\;i=2,\dots,N,\\
&(G_{2k-1})_{N,1}\neq 0,\quad k=1,\dots,m+1.
\end{split}
\end{equation}
 The solvability of the corresponding
 $N\times N $ matrix R-H problem is proved  by lifting it to a scalar problem
 on the  curve 
\begin{align}
\label{zvercurve}
&\mathcal{C}_{N,m}:=\left\{ (\lb,y),\quad y^N=q^{N-1}(\lb) p(\lb)
\right\} ,\\
\label{pq}
& q(\lb)=\prod_{j=1}^m(\lambda-\lambda_{2k}),\quad
p(\lb)=\prod_{j=0}^m(\lambda-\lambda_{2k+1}).
\end{align}
 The curve $\mathcal{C}_{N,m}$  has  singularities
at the points $(\lambda_{2k},0)$, $k=1,\dots,m$. These singularities 
can be easily resolved \cite{mi95} to give rise to a compact Riemann
surface which we still denote by $\mathcal{C}_{N,m}$.
 Such surface can be identified with $N$
copies (sheets) of the complex $\lambda$-plane cut along the segments
$\li_0=\cup_{k=1}^{m+1}  [\lambda_{2k-1},\lambda_{2k}]$ and  glued together
according to the permutation rule
$\begin{pmatrix}1&2&\dots&N-1&N\\2&3&\dots&N&1\end{pmatrix}$, that
is the first sheet is pasted to the second, the second to the
third and so on. The  pre-image  $\pi^{-1}(\lb)$,
$\lb\in\C\backslash  D$, of  the projection $\pi:
\mathcal{C}_{N,m}\rightarrow \C$, consists of $N$ points
$P^{(s)}=(\lb,\e^{2\pi\i\frac{s-1}{N}} y)$.  In this paper, we solve explicitly the $N\times N$  matrix
R-H problem considered by Zverovich.
The algebraic-geometrical approach to the R-H problem was
developed further by Ko\-rot\-kin \cite{kor01}. He showed
that for quasi-permutation monodromy matrices
(in  which each row and each column have only one non-zero
element), the R-H problem can be solved in terms of the
Szeg\"o kernel of a Riemann  surface.

The procedure to obtain the Riemann surface from the
monodromy matrices relies on the Riemann existence theorem
\cite{mi95},\cite{hur91}.
In detail the existence theorem associates a permutation representation
\[
\mathcal{S}:\pi_1(\mathbb{CP}^1\backslash D,\lambda_0)\rightarrow S_N,
\]
to a compact ramified cover $\mathcal{C}$ of degree $N$
over the  Riemann sphere with a set $ D$ of $n$
prescribed branch points
in such a way that the product
$\mathcal{S}(\gamma_1)\mathcal{S}(\gamma_2),\dots,\mathcal{S}(\gamma_n)=1_N$. The
correspondence is one-to-one between isomorphism  classes of covers and
equivalent permutation monodromy representations (this latter equivalence
relation 
simply reflects a relabeling of the points in the fiber of the covering over
the base point $\lambda_0$). The cover
$\mathcal{C}$ is connected if the only invariant subspace of the permutation
representation is the $N$-dimensional column vector $(1,1,\dots,1)^t$.
The genus $g$ of the surface $\mathcal{C}$ is obtained from the Riemann-Hurwitz relation
\begin{equation}
\label{RHu}
2(N+g-1)=\sum_{i=1}^n\text{Tran}[\mathcal{S}(\gamma_i)],
\end{equation}
where $\text{Tran}[\mathcal{S}(\gamma_i)]$ is the number of transpositions in the
permutation $\mathcal{S}(\gamma_i)$.

For a given monodromy representation  (\ref{rep}),
where the matrices $\mathcal{M}(\gamma_i)$  are
quasi-permutation, the corresponding elements  $\mathcal{S}(\gamma_i)$ of
the symmetric group $S_N$ are obtained by setting
all the non-zero entries of $\mathcal{M}(\gamma_i)$, $i=1,\dots,n$, equal to unity.
However, the Riemann existence theorem is just an {\it existence} theorem,
that is, it does not produce explicitly  algebraic equations for the coverings.
In the case under consideration,
the permutation representation induced by the monodromy
matrices $G_k G_{k-1}^{-1}$, with $G_k$  being defined in
 (\ref{Gk}), is
\begin{equation}
\label{perm0}
\begin{split}
&\mathcal{S}(\gamma_{2k-1})=
\begin{pmatrix}
1&2&\dots&N-1&N\\2&3&\dots&N&1\end{pmatrix},\quad k=1,\dots,m+1,\\
&\mathcal{S}(\gamma_{2k})=\begin{pmatrix}1&2&\dots&N-1&N\\N&1&\dots&N-2&N-1\end{pmatrix}
\quad k=1,\dots,m+1.
\end{split}
\end{equation}
 We observe that the points $P_0^{(s)}=
(\lambda_0,\e^{2\pi\i\frac{s-1}{N}} y_0)\in\mathcal{C}_{N,m}$, $s=1,\ldots, N$, $\lb\notin D$,
belonging to the pre-image
$\pi^{-1}(\lambda_0)=(P_0^{(1)},P_0^{(2)},\dots, P_0^{(N)})$,
are permuted, when $\lambda_0$ moves along the path $\gamma_k$,
according to the rule
\[
 (P_0^{(1)},P_0^{(2)},\dots, P_0^{(N)}) \longrightarrow
\mathcal{S}(\gamma_{k})(P_0^{(1)},P_0^{(2)},\dots, P_0^{(N)}),
\quad k=1,\dots,2m+2.
\]
Therefore $\mathcal{C}_{N,m}$ given in (\ref{zvercurve})
is the Riemann surface associated with the
permutation representation  (\ref{perm0}).
In general the derivation of an algebraic expression for the cover
from the permutation representation is a hard task.

The genus of the surface $\mathcal{C}_{N,m}$ obtained from (\ref{RHu}) is equal
to $g=(N-1)m$.
We observe that in our case, the
complex dimension of the space of quasi-permutation
monodromy matrices $M_k=G_kG_{k-1}^{-1}\in SL(N,\C)$, $k=1,\dots, 2m+1$  is
equal to $(N-1)(2m+1)$. In order to solve the R-H problem by using only
the Szeg\"o kernel as suggested  in \cite{kor01},
the complex dimension of the space of monodromy matrices
must be at most equal to  $2g=2(N-1)m$. For this reason one of the monodromy matrices
must be fixed as a suitable permutation or quasi-permutation matrix.

Our derivation of the solution of the R-H problem {\bf (i)-(iv)},
incorporates both the method of  \cite{dikz99}, implemented for
hyperelliptic curves  and the general treatment of \cite{kor01}. First,
we  solve the so-called canonical R-H problem, namely the
problem {\bf (i)-(iv)} when all the matrices $G_{2k}$ are set
equal to the identity and all the matrices $G_{2k+1}$ are set
equal to the quasi-permutation $\mathcal{P}_N$, where
\begin{equation}
\label{PP0}
\mathcal{P}_N=\begin{pmatrix}
0&0&\dots&0&0&(-1)^{N-1}\\
1&0&0&\dots&0&0\\
0&1&0&\dots&0&0\\
\dots&&\dots&\dots&0&0\\
0&0&\dots&1&0&0\\
0&0&\dots&0&1&0\\
\end{pmatrix}.
\end{equation}
More precisely the canonical R-H problem consists of finding a
matrix valued function $X(\lb)$ analytic in the complex plane off
the segment $\li_0=\cup_{k=1}^{m+1}  [\lambda_{2k-1},\lambda_{2k}]$  such that
\begin{equation}
\label{BX}
\begin{split}
X_-(\lb)&=X_+(\lb)\mathcal{P}_N,\quad
\lb\in \li_0,\\
X(\lb_0)&=1_N, \quad \lb_0\in C_+.
\end{split}
\end{equation}
The solution of the R-H problem (\ref{BX}) can be obtained in
 an elementary way by
diagonalising the matrix $\mathcal{P}_N=U\e^{2\pi \i\sigma_N}U^{-1}$,
where the matrix $\sigma_N$ reads
\begin{equation}
\label{sigmaN0}
\sigma_{N}=\mbox{Diag}\left(\dfrac{-N+1}{2N},\dfrac{-N+3}{2N},\dots,
\dfrac{N-3}{2N},\dfrac{N-1}{2N}\right),
\end{equation}
and the matrix $U$ is chosen so that the entries  $U_{1k}=1,\;k=1,\dots,N$ and $\mbox{Det}(U)\neq 0$.
Then it is quite immediate to verify that
\begin{equation}
\label{solX0}
X(\lambda)=
U\left(\dfrac{p(\lb)}{q(\lb)}\dfrac{q(\lb_0)}{p(\lb_0)}\right)^{\sigma_N}U^{-1}
\end{equation}
solves the R-H problem (\ref{BX}).
The entries of the  matrix $X(\lb)$ can be expressed
in terms of the Szeg\"o kernel with  zero characteristics, $S[0](P,Q)$,
defined on $\mathcal{C}_{N,m}$. We show that
\begin{equation}
\label{szegozero} S[0](P,Q)=\dfrac{1}{N}\dfrac{ \sqrt{\d z(P)\d z (Q)}}
{z(P)-z(Q)}\sum\limits_{k=0}^{N-1} \left(
\dfrac{q(z(P))}{p(z(Q))}  \dfrac{p(z(Q))}{q(z(P))}
\right)^{-\frac{k}{N}+\frac{N-1}{2N}},\quad
P,Q\in\mathcal{C}_{N,m},
\end{equation}
where $z (P)$ is a local coordinate near the point $P$ and the
polynomials  $p$ and $q$ have been defined in (\ref{pq}).
Then the entries of  the matrix $X(\lb)$ in (\ref{solX0})
can be written in the form
\begin{align*}
X_{rs}(\lb)=& S[0](P^{(s)},P_0^{(r)})\dfrac
{z(P)-z(Q)}{ \sqrt{\d z(P)\d z (P_0)}}\\
=&\dfrac{1}{N}\sum\limits_{k=0}^{N-1}
\left( \e^{2\pi\i\frac{(s-r)}{N}}\sqrt[N]
{\dfrac{p(\lb)}{q(\lb)}\dfrac{q(\lb_0)}{p(\lb_0)}}
\right)^{-k
+\frac{N-1}{2}},\quad
\lb_0\notin D,
\end{align*}
where $P^{(s)}=(\lb,\rho^{s-1} y)$ and  $P_0^{(r)}=(\lb_0,\rho^{r-1} y_0)$,
$r,s=1,\dots,N$,
denote the points on the $s$-th  and $r-$th sheet of
$\mathcal{C}_{N,m}$ respectively.
When $N=2$ and $\sqrt[4]{\dfrac{q(\lb_0)}{p(\lb_0)}}=1$,
such a formula coincides with the canonical solution obtained in 
\cite{diz97},\cite{dikz99}.

The solution $Y(\lb)$, of the full R-H problem {\bf (i)-(iv)},
where the constant matrices $G_k$, $k=1,\dots,2m+1$, are
parametrised by $2(N-1)m$ arbitrary complex constants,  is
obtained, following \cite{kor01}, using the Szeg\"o kernel with
non-zero characteristics. From the relation (\ref{szegozero}),
we are able to write  the global solution
$Y(\lb)=(Y_{rs}(\lambda))_{r,s,=1,\dots,n}$ of the  R-H problem {\bf (i)-(iv)} in the form
\begin{equation}
\label{sol0}
Y_{rs}(\lb)=X_{rs}(\lb)
\dfrac{\theta\pq\left(\int\limits_{P_0^{(r)}}^{P^{(s)}}\d\boldsymbol{v}
;\Pi\right)}
{\theta\left(\int\limits_{P_0^{(r)}}^{P^{(s)}}\d\boldsymbol{v};\Pi\right)}
\dfrac{\theta(\boldsymbol{0};\Pi)}{\theta\pq(\boldsymbol{0};\Pi)},
\quad r,s=1,\dots,N,
\end{equation}
where $\d\boldsymbol{v}$ is the vector of normalized
 holomorphic differentials  on $\mathcal{C}_{N,m}$, $\Pi$ is the period matrix
with respect to  $\d\boldsymbol{v}$,   $\theta\pq$ is
the canonical  $\theta$-function with characteristics
 $\boldsymbol{\epsilon}$ and $\boldsymbol{\delta}$
 determined from the non-zero entries of the matrices $G_k$,
$k=1,\dots,2m+1$.
The solution (\ref{sol0}) exists if
\[
\theta\pq(\boldsymbol{0};\Pi)\neq 0,
\]
that is if $\Pi\boldsymbol{\delta}+\boldsymbol{\epsilon}\notin(\Theta)$,
where $(\Theta)$ is the $\theta$-divisor in the Jacobian variety $\mathrm{Jac}(\mathcal{C}_{N,m})$  of
the Riemann surface $\mathcal{C}_{N,m}$.
 This solution coincides with the solution obtained in \cite{dikz99},\cite{kk98}
for $N=2$. The formula (\ref{sol0}) 
permits us to evaluate {\it explicitly} the characteristics $\boldsymbol{\delta}$ and
$ \boldsymbol{\epsilon}$ in terms of the monodromy matrix entries
thus solving the R-H problem effectively.

The Fuchsian system (\ref{Fuchsian0}) is recovered from the solution
(\ref{sol0}) by evaluating
the residue
\begin{equation}
\label{A00}
A_k=A_k(\lambda_1,\dots,\lambda_{2m+1}|M_1,\dots,M_{2m+1})=\res[\lb=\lambda_k]
\left[\dfrac{\d Y(\lb)}{\d\lb}Y^{-1}(\lb)\right].
\end{equation}
If none of the monodromy matrices $M_i$, $i=1,\dots,2m+2$, depends on the position of the
singular  points $\lambda_k$, $k=1,\dots,2m+1$, then the matrices $A_k$
satisfy the  Schlesinger system \cite{schl12} (see below
\ref{Schlesinger}).
The Jimbo-Miwa-Ueno  $\tau$-function \cite{JMU81}
\[
\frac{\partial}{\partial \lambda_k}\mathrm{log}\,\tau =\frac12
\res[\lambda=\lambda_k]\,\mathrm{Tr}\,\left(
\frac{\mathrm{d}Y(\lambda)} {\mathrm{d}\lambda} Y(\lambda)^{-1}
\right)^2,\]
corresponding to the particular solution
(\ref{A00})  of the Schlesinger system, has the form
\begin{equation}
\tau(\lambda_1,\dots,\lambda_{2m+1})=\dfrac{\theta\pq(\boldsymbol{0};\Pi)}{\theta(\boldsymbol{0};\Pi)}\dfrac
{\prod\limits_{\substack{k<i\\i,k=0}}^m(\lambda_{2k+1}-\lambda_{2i+1})^{\frac{N^2-1}{6N}}
\prod\limits_{\substack{k<i\\k,i=1}}^m(\lambda_{2k}-\lambda_{2i})^{\frac{N^2-1}{6N}} }
{\prod\limits_{\substack{i<j\\i,j=1}}^{2m+1}(\lambda_{i}-\lambda_{j})^{\frac{N^2-1}{12N}}}.
\end{equation}
For $N=2$ the above expression has been obtained in \cite{kk98}. 
The $\tau$-function can be written in a different form
by using the Thomae-type formula which we derive for the families of
curves $\mathcal{C}_{N,m}$
\[
\theta^8(\boldsymbol{0};\Pi)=\dfrac{\prod_{s=1}^{N-1}\mathrm {det}
\A^4_s}{(2\pi\i)^{4m(N-1)} }
\prod_{i<j}(\lambda_{2i}-\lambda_{2j})^{2(N-1)}
\prod_{k<l}(\lambda_{2k+1}-\lambda_{2l+1})^{2(N-1)}.
\]

The form (\ref{sol0}) of the solution of the R-H problem  enables
us to show the following:
\begin{enumerate}
\item  if the non-singular characteristics
$\boldsymbol{\delta},\;\boldsymbol{\epsilon}$ correspond to
 a non-special divisor supported on the branch points, then
$\boldsymbol{\delta},\;\boldsymbol{\epsilon}\in
(\Z/N\Z)^{(N-1)m}$ and  the solution of the R-H problem corresponds to a reducible monodromy representation;
\item when  two solutions $Y(\lb)$ and $\tilde{Y}(\lb)$
 have their corresponding  characteristics  equivalent  modulo
$(\Z/N\Z)^{(N-1)m}$, the matrix
entries   $Y_{rs}(\lb)$ and $\tilde{Y}_{rs}(\lb)$
are related by an algebraic transformation.
The corresponding monodromy representations $\mathcal{M}=\{M_1,M_2,\dots,M_{2m+1},M_{\infty}\}$ and  $\mathcal{\tilde{M}}=
\{\tilde{M}_1,\tilde{M}_2,\dots,\tilde{M}_{2m+1},\tilde{M}_{\infty}\}$
are equivalent up to  multiplication
by $N$-roots of unity. That is $\tilde{M}_k=\e^{\frac{2\pi i}{N}j_k}M_k$, $j_k$ integer, $\sum_{k=1}^{2m+2} j_k=0\;\text{mod}\,N$.
\end{enumerate}
We remark that the result in (2) has been  suggested by Dubrovin and Mazzocco
\cite{duma03} following their  investigations of the symmetries
of the Schlesinger system. These symmetries generalise the Okamoto symmetries derived in the $2\times 2$
case \cite{ok87}.

Finally we study in detail the solution of  the rank $3$  problem
with four singular points $(\lb_1,\lb_2,\lb_3,\infty)$.
The monodromy matrices read
\begin{align}
\label{M1}
M_1&=\begin{pmatrix}
0&0&c_1\\
\dfrac{c_2}{c_1}&0&0\\
0&\dfrac{1}{c_2}&0
\end{pmatrix},\quad
M_{2}=
\begin{pmatrix}
0&\dfrac{c_1d_1}{c_{2}}&0\\
0&0&c_{2}d_{2}\\
\dfrac{1}{c_1 d_1d_{2}}&0&0
\end{pmatrix},\quad\\
 M_{3}&=\begin{pmatrix}
0&0&d_1d_{2}\\
\dfrac{1}{d_{1}}&0&0\\
0&\dfrac{1}{d_{2}}&0
\end{pmatrix},\,\quad M_{\infty}=\begin{pmatrix}
0&1&0\\
0&0&1\\
1&0&0
\end{pmatrix},\notag
\end{align}
where $c_1, c_2, d_1,d_2$ are non-zero constants.
The solution of the R-H problem is defined in terms of
the Szeg\"o kernel of the  genus two Riemann
surface
\[
\mathcal{C}_{3,1}:\,y^3=(\lambda-\lb_1)(\lambda-\lb_2)^2)(\lambda-\lb_3).
\]
The period matrix of the surface has the symmetric form
\[
\Pi=\begin{pmatrix} 2\T&\T\\\T&2\T\end{pmatrix},\quad \text{Im}\,T>0,
\]
with
\[
\T=\dfrac{\i\sqrt{3}}{3}
\dfrac{F\left(\frac13,\frac23,1;1-t\right)}
{F\left(\frac13,\frac23,1;t\right)}, \quad t=\dfrac{\lb_2-\lb_1}{\lb_3-\lb_1}
\]
where $F\left(\frac13,\frac23,1;1-t\right)$ and $F\left(\frac13,\frac23,1;t\right)$
are two independent solutions of the Picard-Fuchs equation
\[
t(1-t)\dfrac{d^2}{dt^2} F +(1-2t)\dfrac{d}{dt} F -\frac{2}{9} F=0.\]
The inverse function $t=t(T)$ is in general not single valued.
For $\T$ belonging to  Siegel half-space ${\mathcal H}_1$
modulo the sub-group $\Gamma_0(3)$ of the modular group,  the function $t=t(T)$ 
is single-valued  and reads
\begin{equation}
\label{tmodular1}
  t=27\vartheta_3^4(0; 3\T)
\dfrac{(\vartheta_3^4(0; 3\T)-\vartheta_3^4(0; \T))^2}{(3\vartheta^4_3(0; 3\T)+\vartheta^4_3(0; \T))^3}.
\end{equation}
Clearly, the above expression is automorphic under the action of the group
$\Gamma_0(3)$ and can be expressed in terms of the Dedekin $\eta$-function \cite{hm00}.
 From the classical theory of the hypergeometric equation
it follows that the function $t=t(T)$ satisfies the Schwarz equation
(see for example \cite{fo29})
\[
\{t,\T\}+\dfrac{\dot{t}^2}{2}\left(\dfrac{1}{t^2}+\dfrac{1}{(t-1)^2}-
\dfrac{10}{9t(t-1)}\right)=0, 
\]
where  $\dot{t}=\dfrac{d t}{d\T}$ abd  $\{\;,\;\}$ is the Schwarzian
derivative
\begin{equation}
\label{schder}
\{t,\T\}:=
\dfrac{\dddot{t}}{\dot{t}}-\dfrac{3}{2}\left(\dfrac{\ddot{t}}{\dot{t}}\right)^2.
\end{equation}
From the function $t=t(T)$ it is possible to derive an
expression for the solution of the corresponding general Halphen system equivalent to
the one derived in \cite{hm00}.

\noindent
The surface $\mathcal{C}_{3,1}$ is a two-sheeted cover of  two elliptic
curves that are $3$-isogenous. As a result, the solution of the R-H
problem and of the Schlesinger equations can be expressed explicitly
in terms of Jacobi's $\vartheta$-functions. The corresponding
$\tau$-function of the Schlesinger system reads
\begin{equation}
\begin{split}
 \tau(\lb_1,\lb_2,\lb_3)
&=\left(\dfrac{\lb_1-\lb_3}{(\lb_1-\lb_2)(\lb_2-\lb_3)}\right)^{\frac29}
\e^{2\pi\i [\T(\delta_1^2+\delta_1\delta_2+\delta_2^2)
+\epsilon_1\delta_1+\epsilon_2\delta_2]}\\
&\times\dfrac{\sum_{k=2}^3\vartheta_k(\epsilon_1+\epsilon_2+3\T(\delta_1+\delta_2);\,6\T)\vartheta_k(\epsilon_1-\epsilon_2+\T(\delta_1-\delta_2);\,2\T)}{\vartheta_3(0;\,6\T)\vartheta_3(0;\,2\T)+\vartheta_2(0;\,6\T)\vartheta_2(0;\,2\T)},
\end{split}
\end{equation}
where $\vartheta_i$, $i=2,3$ are  the Jacobi's
$\vartheta$-functions and
\[
\epsilon_i=\dfrac{1}{2\pi\i}\log c_i,\quad\delta_i=\dfrac{1}{2\pi\i}\log d_i,\;\;i=1,2.
\]

This paper is organized as follows. In the  Section 2 we
 give some general backgrounds about
 the theory of R-H problems and we describe
the R-H problem we are going to solve.
In  the    Section 3  some backgrounds about classical algebraic
geometry of Riemann surfaces and kernel forms are given.
We describe in detail the curve $\mathcal{C}_{N,m}$ in the Section 4,
 namely its homology basis, the characteristics supported on branch points,
the Szeg\"o kernel for $1/N$ characteristics and the projective connection.
This section contains mainly new material.
In the  Section 5 we solve the R-H problem for quasi-permutation monodromy
matrices and we study the symmetry  properties of the solution
which are inherited from the symmetries of the curve. We derive the $\tau$-function
for the Schlesinger system and Thomae-type formula for the $Z_n$ curve.
We describe extensively an example for a $3\times 3$ matrix
R-H problem with four singular points in the sixth Section  and we derive
the solution of the corresponding $3\times 3$ Schlesinger system. Interesting
relations with the modular surface $\mathcal{H}_1/\Gamma_0(3)$ are pointed out.
We draw our conclusion in the last section.

\section*{Acknowledgments}
The authors are grateful to B.Dubrovin for many useful conversations and for
letting us know \cite{duma03} before publication. We are grateful to Korotkin
for the fruitful conversations and for the many suggestions which improve the paper.
They also wish to thank P. Deift, J.Harnad,  A.Its,
 J.McKay, M. Mazzocco, K. McLaughlin, M.Narasimhan  and   F.Nijhoff for the
discussion of the results. We are also grateful to Yu.Brezhnev for
pointing out  the paper \cite{hut02}. During
the preparation of the  manuscript the authors  used the Maple
software by B.Deconnink and M. van Hoeij \cite{dh01} to compute
algebraic curves.

\section{The $N\times N$ matrix  Riemann-Hilbert problem}
 The method of  \cite{pl64} to solve the R-H problem consists
of reducing it to the so-called homogeneous Hilbert boundary value
problem of the theory of singular equations \cite{muskhel72}. The
reduction is carried out in the following way. Let us  assume that
the  set of points $\lambda_1,\dots,\lambda_{2m+1}$ satisfy the relation
\[
\mathrm{Re}\,\lambda_1<\mathrm{Re}\,\lambda_3
<\mathrm{Re}\,\lambda_3<\dots<\mathrm{Re}\,\lambda_m<\mathrm{Re}\,\lambda_{2m+1}.
\]
Let  $\li$ be   the oriented polygonal line which connects
this set of points  and infinity
\[
\li=(\infty,\lambda_1)\cup(\lambda_1,\lambda_{2})\cup(\lambda_{2},\lambda_3)\cup\dots\cup(\lambda_{2m},\lambda_{2m+1})
\cup(\lambda_{2m+1},\infty).
\]
We  denote by $C_+$ and $C_-$ the positive and negative parts
of the plane $\C$ with respect to $\li$ (see Figure~\ref{fig1}).

Let us consider the set of $2(N-1)m$ non-zero complex constants
$c_1,\dots,c_{(N-1)m}$ and $d_1,\dots,d_{(N-1)m}$ and define the
$N\times N$ quasi-permutation matrices $G_k\in SL(N,\C)$  as
\begin{equation}
\label{Gkm1}
G_{2k-1}=\begin{pmatrix}
0&0&\dots&0&0&(-1)^{N-1}c_k\\
\dfrac{c_{k+m}}{c_k}&0&0&\dots&0&0\\
0&\dfrac{c_{k+2m}}{c_{k+m}}&0&\dots&0&0\\
\dots&&\dots&\dots&0&0\\
0&0&\dots&\dfrac{c_{k+(N-2)m}}{c_{k+(N-3)m}}&0&0\\
0&0&\dots&0&\dfrac{1}{c_{k+(N-2)m}}&0\\
\end{pmatrix}
\end{equation}
for $k=1,\dots,m$ and $G_{2m+1}=\mathcal{P}_N$,
 where
$\mathcal{P}_N$ has been defined in (\ref{PP0}); the diagonal
matrix $G_{2k}$ reads
\begin{equation}
\label{Dk}
G_{2k}=\mbox{diag}\left(d_k,d_{k+m},\dots,
d_{k+(N-2)m}, \prod_{j=0}^{N-2}\dfrac{1}{d_{k+jm}}\right),
\end{equation}
for $k=1,\dots,m$ and $G_{0}=G_{2m+1}=1_{N}$.
We define the $N\times N$ matrix
 function $Y(\lb)$ as the solution of the
following R-H problem:
\begin{align}
\label{RH0}
&Y(\lb)\;\; \text{is analytic in }\;\;\mathbb{CP}^1\backslash \li, \\
\nonumber
&\text{The}\;\; L_2-\text{limits}\;\; Y_{\pm}(\lb)\;\;\text{ as}\;\; \lb\rightarrow \li_{\pm}\;\;\text{ satisfy the jump conditions:}\\
\label{RH1}
&Y_-(\lb)=Y_+(\lb)G_{k},
\quad \lb \in(\lambda_{k},\lambda_{k+1}),~~k=0,\dots,2m+1,~~\lambda_0=\lambda_{2m+2}=\infty\\
\label{RH3}
&Y(\lb_0)=1_N,\quad  \lambda_0\in C_+\backslash  D.
\end{align}
Assuming the existence of the solution of the R-H problem (\ref{RH0})-(\ref{RH3}), one can find that the monodromy matrices are obtained from (\ref{Gk})
 by
\[
Y(\gamma_k(\lb))=Y(\lb)M_k,
\]
where
\begin{equation}
\label{Mk}
M_{k}=G_{k}\,(G_{k-1})^{-1},\quad k=1,\dots,2m+2,
\end{equation}
\[
Y(\lb)_{\frac{1}{\lb}\rightarrow\frac{1}{\lb}\e^{2\pi
i}}=Y(\lb)M_{\infty},
\]
where
\begin{equation}
\label{Minfinity}
M_{\infty}=\mathcal{P}_N^{-1}.
\end{equation}
\begin{remark}\label{remarkreduc}
The monodromy representation described by the matrices (\ref{Mk})
is irreducible if
\[
c_{k+sm}\neq\xi_k^{s+1},\quad d_{k+sm}\neq\zeta_k, \quad s=0,\dots,N-2,
\]
 where $\xi_k$ and $\zeta_k$,  $k=1,\dots,m$,  are  any $N$-th root
of unity and   $\xi_{m+1}=\zeta_{m+1}=1$.
Indeed on the contrary, the matrices $G_k$ read
\[
G_{2k}=\zeta_k\, 1_N,\quad G_{2k-1}=\xi_k\,\mathcal{P}_N,\quad k=1,\dots,m.
\]
The corresponding    reducible monodromy representation is given by the matrices
\begin{equation}
\begin{split}
\label{Mreducible}
&M_{2k}=G_{2k}(G_{2k-1})^{-1}=\dfrac{\zeta_k}{\xi_k}\mathcal{P}^{-1}_N,\quad k=1,\dots,m+1,\\
&M_{2k-1}=G_{2k-1}(G_{2k-2})^{-1}=\dfrac{\xi_{k}}{\zeta_{k-1}}\mathcal{P}_N,
\quad k=1,\dots,m+1.
\end{split}
\end{equation}
\end{remark}
The matrices $M_k$ can be written in the form
\begin{equation}
\label{Ck}
M_{k}=U^{-1}_k\e^{2\pi \i\sigma_N}U_k,\quad k=1,\dots,2m+1,\;\;U_k\in GL(N,\C),
\end{equation}
where the matrix $\sigma_N$ reads
\begin{equation}
\label{sigmaN}
\sigma_{N}=\mbox{Diag}\left(\dfrac{-N+1}{2N},\dfrac{-N+3}{2N},\dots,
\dfrac{N-3}{2N},\dfrac{N-1}{2N}\right).
\end{equation}

The function $Y(\lb)$ has regular singularities
of the following form near the points $\lambda_k$
\begin{equation}
\label{Yexp}
Y(\lb)=\hat{Y}_k(\lb)(\lb-\lambda_k)^{\sigma_N}U^{\pm}_k,\;\;\lb\in C_{\pm},
\end{equation}
where  the matrices
$\hat{Y}_k(\lb)$ are holomorphic and invertible
 at $\lb=\lambda_k$, $U^+_k=U_k$ and $U_k^-=U_k G_{k-1}$ and the matrices
$G_k$ and $U_k$, $k=1,\dots,2m+1$,
 have been defined in (\ref{Gkm1}), (\ref{Dk}) and (\ref{Ck}) respectively.

It follows from the above expansion that $
\frac{dY(\lb)}{d\lb}Y^{-1}(\lb) $ is  meromorphic in
$\mathbb{CP}^1$  with simple poles at $\lambda_1,\lambda_2,\dots
\lambda_{2m+1}$ and $\infty$. Therefore $Y(\lb)$
satisfies the Fuchsian equation
\begin{equation}
\label{Fuchsian} \dfrac{\d Y(\lb)}{\d\lb}=
\sum_{k=1}^{2m+1}\dfrac{A_k}{\lb-\lambda_k} \;Y(\lb),
\end{equation}
where
\begin{equation}
\begin{split}
\label{Ak}
A_k&=A_k(\lambda_1,\dots,\lambda_{2m+1}|M_1,\dots,M_{2m+1})
=\res[\lb=\lambda_k]\left[\dfrac{\d Y(\lb)}{\d \lb}Y^{-1}(\lb)\right]=\\
&=\hat{Y}_k(\lambda_k)\sigma_N
\hat{Y}^{-1}_k(\lambda_k), \;\;k=1,\dots,2m+1,
\end{split}
\end{equation}
which 	follows from  (\ref{Yexp}).
If none of the monodromy matrices depend on the position of
the singular  points $\lambda_k$, $k=1,\dots,2m+1$,
 the function $Y(\lb;\lambda_1,\dots,\lambda_{2m+1})$ in addition to
(\ref{Fuchsian})  satisfies the following equations
\begin{equation}
\label{Fuchsian1}
\dfrac{\partial}{\partial \lambda_k}Y(\lb)=\left(\dfrac{A_k}{\lb_0-\lambda_k}-\dfrac{A_k}{\lb-\lambda_k}\right)Y(\lb),\quad k=1,\dots,2m+1.
\end{equation}
Compatibility conditions of (\ref{Fuchsian}) and (\ref{Fuchsian1})
 are described by the system of Schlesinger equations \cite{schl12}
\begin{equation}
\label{Schlesinger}
\begin{split}
&\dfrac{\partial}{\partial \lambda_j}A_k=\dfrac{[A_k,A_j]}{\lambda_k-\lambda_j}-
\dfrac{[A_k,A_j]}{\lb_0-\lambda_j}, \quad j\neq k,\quad\\
&\dfrac{\partial}{\partial \lambda_k}A_k= -\sum_{\substack{j\neq
k\\j=1}}^{2m+1}
\left( \dfrac{[A_k, A_j]}{\lambda_k-\lambda_j}-\dfrac{[A_k,A_j]}{\lb_0-\lambda_j}\right).
\end{split}
\end{equation}
Thus the solution of the R-H problem (\ref{RH0})-(\ref{RH3})
leads immediately to the particular solution (\ref{Ak})  of
the Schlesinger system (\ref{Schlesinger}).

From the  solution of the Schlesinger equation (\ref{Ak})
one can define the corresponding holomorphic $\tau$-function
given by the formula \cite{JMU81}
\begin{equation}
\label{tau}
\frac{\partial}{\partial \lambda_k}\mathrm{log}\,\tau =\frac12
\res[\lambda=\lambda_k]\,\mathrm{Tr}\,\left(
\frac{\mathrm{d}Y(\lambda)} {\mathrm{d}\lambda} Y(\lambda)^{-1}
\right)^2. 
\end{equation}
The set of zeros of the $\tau$-function in the space of singularities of
the R-H problem is called the Malgrange divisor $(\theta)$ \cite{ma83}.
It plays a crucial role in the discussion of the solvability of the
R-H problem with the given monodromy data.

\section{Riemann surface of an algebraic curve}
In order to solve the R-H problem  (\ref{RH1})-(\ref{RH3}),
we first need to introduce some basic objects on Riemann surfaces.

\subsection{The curve and differentials} 
Let   $\mathcal C$ be the Riemann
surface of    the algebraic equation
$$ y^N+p_1(\lambda)y^{N-1}+\ldots+p_N(\lambda)=0, $$ where
$p_1,\ldots,p_N$ are polynomials in $\lambda$.
 In a neighbourhood $U_R$ of  the point
$R=(\eta,w)\in\mathcal C$, a local coordinate $z(P)$, $P=(\lb,y)\in U_R$,
 is  the function
defined by
 \begin{equation}
z(P)=\begin{cases} \lambda-\eta &\text{if}\quad R\quad \text{is an ordinary point, }\\
                   \sqrt[l]{ \lambda-\eta} &\text{if}\quad R\quad
                   \text{is a  finite branch point of order $l$},\\
                   \frac{1}{\lambda }&\text{if}\quad R\quad
                   \text{is an ordinary point at infinity,}\\
                   \frac{1}{\sqrt[m]{\lambda}}&\text{if}\quad R\quad
                   \text{is  a branch point at infinity of order}\quad
                  m.
\end{cases}\label{local}
\end{equation}
Let $\mathrm{d}\boldsymbol{v}(P)=(\mathrm{d}v_1(P),\ldots,
\mathrm{d}v_1(P)) $ be the  basis of normalized
holomorphic differentials, with respect to the canonical homology basis in $H_1(\mathcal{C},\Z)$
of $\alpha$ and $\beta$-cycles, 
$(\alpha_1,\ldots,\alpha_g;\beta_1,\ldots,\beta_g)$. The matrix of $\beta$-periods
\begin{equation}\Pi=\left(\oint_{\beta_i}\mathrm{d}v_k(P)
\right)_{i,k=1,\ldots,g}
\label{matrixpi}
\end{equation}
 belongs to the  Siegel half space, ${\mathcal
H}_g=\{\Pi|\Pi^t=\Pi, \mathrm{Im}\,\Pi>0\}. $ The
Jacobian variety of the curve ${\mathcal C}$ is  denoted by
$\mathrm{Jac}({\mathcal C}) ={\mathbb C}^g/(1_g\oplus \Pi)$.

We also mention the  variation
formulas  which describe  the dependence of the period matrix $\Pi$
on the branch points. These formula can be  already found in the
hyperelliptic case in Thomae \cite{th69}. For general surfaces the infinitesimal
variation of the period matrix with respect to a Beltrami differential is due to 
 Rauch \cite{ra59}  (see also Fay \cite{fa92}) and Korotkin
reduced this deformation to the useful form \cite{kor01}:
\begin{equation}
\label{Rauch} \dfrac{\partial}{\partial \lambda_k}\Pi_{ij}=2\pi \i
\res[\lb=\lambda_k]\left\{ \dfrac{1}{(\mathrm{d} z(P))^2}\sum_{s=1}^N\d
v_i(P^{(s)}) \d v_j(P^{(s)})\right\},\;\;
\end{equation}
where $i,j=1,\dots,2m,\;\; k=1,\dots,2m+1$ and $P^{(s)}$ is a point on
the sheet $s$ of $\mathcal{C}$.

\subsection{$\theta$-function}
Any point $\boldsymbol{e}\in\C^{g}$ can be written uniquely as
$\boldsymbol{e}=(\be,\bd)(^{1_g}_\Pi)$,
where $\be,\bd\in\Rn^g$ are the characteristics
of $\boldsymbol{e}$. We use the  notation $[\boldsymbol{e}]=[^{\bd}_{\be}]$.

If $\be$ and $\bd$ are half integer,
then we say that the corresponding characteristics
$[\mathbf{e}]$ are half-integer.
The half-integer  characteristics are odd or even, whenever
 $4\langle\bd,\be\rangle$ is equal
to $1$ or $0$ modulo $2$.
The bracket $\langle\;,\;\rangle$  denotes the standard
 Euclidean scalar product.

The Riemann $\theta$-function with characteristics $\pq$ is given
on ${\mathcal H}_g\times \mathrm{Jac}({\mathcal C})$ as the Fourier
series
\begin{equation}
\theta\pq(\boldsymbol{z};\Pi)=\sum_{\boldsymbol{n}\in \mathbb{Z}^g}
\exp(\pi \i\,\langle \Pi
\boldsymbol{n}+\Pi\bd,\boldsymbol{n}+\bd\rangle+2\pi
\i\,\langle
\boldsymbol{z}+ \be, \boldsymbol{n}+\bd\rangle).
\end{equation}
The $\theta$-function is an entire function in the variable
$\boldsymbol{z}$ with  periodicity properties:
\begin{align}
\label{periodicity1}
&\theta\pq(\boldsymbol{z}+\boldsymbol{e}_k;\Pi)=\e^{2\pi \i \delta_k}
\theta\pq(\boldsymbol{z};\Pi),\\
\label{periodicity2}
&\theta\pq(\boldsymbol{z}+\boldsymbol{e}_k\Pi;\Pi)=\e^{-2\pi \i
\epsilon _k} \e^{-2\pi \i z_k}\e^{-\pi \i
\Pi_{kk}}\theta\pq(\boldsymbol{z};\Pi),\\
&\theta[^{\boldsymbol{\delta}+\boldsymbol{n}^{\prime}}_
{\boldsymbol{\epsilon}+\boldsymbol{n}^{\prime\prime}}](\boldsymbol{z};\Pi)=
\e^{2\pi\i\langle\boldsymbol{\epsilon},\boldsymbol{n}^{\prime}\rangle}
\theta\pq(\boldsymbol{z};\Pi),
\end{align}
where
$\boldsymbol{e}_k=(0,\dots,\stackrel{k\downarrow}{1},\dots,0)$ is
the standard basis in $\C^g$, $\boldsymbol{n}^{\prime}$ and $\boldsymbol{n}^{\prime\prime}$ integer vectors.
When $\pq$ is equal to zero we write
$\theta[{}^{\boldsymbol{0}}_{\boldsymbol{0}}](\boldsymbol{z};\Pi)
=\theta(\boldsymbol{z};\Pi).$ The function
$\theta(\boldsymbol{z};\Pi)$ is even and clearly satisfies the
relation
\begin{equation}
\frac{\partial}{\partial z_i}
\theta(\boldsymbol{z};\Pi)\big|_{\boldsymbol{z}=0}=0,\quad
i=1,\ldots,g. \label{thetaderzero}
\end{equation}

The $\theta$-function with arbitrary
characteristics satisfies  the heat equation
\begin{equation}
\frac{\partial^2}{\partial z_k\partial
z_l}\theta\pq(\boldsymbol{z};\Pi)=
2\imath\pi(1+2\delta_{k,l})\frac{\partial}{\partial
\Pi_{kl}}\theta\pq(\boldsymbol{z};\Pi), \quad k,l=1,\ldots,g.
\label{heat}\end{equation}

The zeros of the $theta$-function are described by the  Riemann vanishing theorem.
\begin{theorem}
Let $\boldsymbol{e}\in \mathrm{Jac}({\mathcal C})$ be an arbitrary
vector. Then the multi-valued function
\[
P\rightarrow \theta\left(\int_{Q_0}^P\d\boldsymbol{v}-\boldsymbol{e};\Pi\right)
\]
  has on $\mathcal{C}$ exactly $g$ zeros $Q_1,Q_2,\dots,Q_g$ provided
it does not vanish identically.  There is a one-to-one
correspondence between $\boldsymbol{e}\in  \mathrm{Jac}({\mathcal C}) $
and the non-special divisor $\sum_{i=1}^g Q_i$
\[
\boldsymbol{e}=\sum_{i=1}^g\int_{Q_0}^{Q_i}
\d\boldsymbol{v}-\boldsymbol{K}_{Q_0},
\]
where $\boldsymbol{K}_{Q_0}$ is the vector of Riemann constants
\begin{equation}
\label{RC}
(\boldsymbol{K}_{Q_0})_j=\dfrac{1+\Pi_{jj}}{2}
-\sum_{i=1,i\neq j}^g\oint_{\alpha_j}\d v_i(P)\int_{Q_0}^P\d v_j.
\end{equation}

\end{theorem}
\begin{remark}
The vector of Riemann constants depends on the homology basis
$(\alpha_1,  \dots,  \alpha_g;\newline  \beta_1,  \dots,  \beta_g)
\in    H_1(\mathcal{C},\mathbb{Z})$ and the base point
$Q_0\in \mathcal{C}$.
\end{remark}
For  a point $P\in \mathcal {C}$, we define the Abel map
$\boldsymbol{\mathfrak{A}}: \mathcal{C}\longrightarrow
\mathrm{Jac}(\mathcal {C})$ by setting
\begin{equation}
\label{AbelQ0}
\boldsymbol{\mathfrak{A}}(P)=\int_{P_0}^P\d\boldsymbol{v},
\end{equation}
for some base point $P_0\in\mathcal{C}$. For a positive divisor $\mathcal{D}$ of 
degree $n$ the Abel map reads
\[
\boldsymbol{\mathfrak{A}}(\mathcal{D})=\int_{nP_0}^{\mathcal{D}}\d\boldsymbol{v}.
\]
There exists a non-positive  divisor $\Delta$ of degree $g-1$ such that
\begin{equation}
\label{VRC}
\boldsymbol{\mathfrak{A}}(\Delta-(g-1)Q_0)=\boldsymbol{K}_{Q_0},
\end{equation}
where $\boldsymbol{K}_{Q_0}$ has been defined in (\ref{RC}).
The divisor $\Delta$ is called the Riemann divisor and satisfies the condition
$2\Delta={\mathcal K}_{\mathcal C}$, where ${\mathcal K}_{\mathcal C}$ is the canonical
class (that is the class of divisors of Abelian differentials).
\begin{definition}
The characteristic $\pq$ of a point $\boldsymbol{e}=
\boldsymbol{\epsilon}+\boldsymbol{\delta}\Pi$
is called singular if
\[
\theta(\boldsymbol{e};\Pi)=0.
\]
The odd half-integer characteristics
$[^{\hat{\boldsymbol{\delta}}}_{\hat{\boldsymbol{\epsilon}}}]$
 of a point $\boldsymbol{\gamma}=\boldsymbol{\hat{\epsilon}}+\boldsymbol{\hat{\delta}}\Pi$ is non-singular if among the derivatives
\[
\left .\dfrac{\partial }{\partial z_j}
\theta[\boldsymbol{\gamma}](\boldsymbol{z};\Pi)\right|_{\boldsymbol{z}=\boldsymbol{0}},\quad j=1,\dots,g,
\]
there is at least one non-vanishing.
\end{definition}

\subsection{Kernel-forms}
The Schottky-Klein prime form $E(P,Q)$, $P,Q\in \mathcal{C}$ is a
skew-symmetric $(-\frac12,-\frac12)$-form on $\mathcal C\times\mathcal C$
\cite{fa73}
\begin{equation}
\label{prime} E(P,Q)=\frac{\theta[\gamma]\left(\int\limits_Q^{P}\d
\boldsymbol{v};\Pi\right)}{ h(P) h(Q) },
\end{equation}
where $[\gamma]$ is a non-singular  odd half-integer  characteristics and
\[ h^2(P)=\sum_{j=1}^g\frac{\partial }{\partial z_j}
\theta[\boldsymbol{\gamma}](\boldsymbol{0};\Pi) \d v_j(P).
\]
The prime form does not depend on the point $\boldsymbol{\gamma}$.
The automorphic factors of the prime form along all cycles
$\alpha_k$ are trivial; the automorphic factor along each
$\beta_k$ cycle in the $Q$ variable equals $ \exp\{-\pi \i \Pi_{kk}-2\pi
\i\int_{P}^Q\d v_k\}$. If the points $P$ and $Q$ are placed in the
vicinity of the point $R$ with local coordinate $z$, $z(R)=0$,
then the prime form has the following local behaviour as
$P\rightarrow Q$
\begin{equation}
E(P,Q)=\frac{z(P)-z(Q)}{\sqrt{\mathrm{d}z(P)}\sqrt{\mathrm{d}z(Q)}
}
 \left(1+ O(1)) \right).
\end{equation}

The prime form $ E(P,Q)$ is the  generating form of the Bergmann
and Szeg\"o  kernels. Let $P=(\lambda,y)$ and
$Q=(\mu,w)$. Then the Bergmann kernel $\omega(P,Q)$ is
defined as a symmetric 2-differential,
\begin{equation}
\omega(P,Q)=\mathrm{d}_{\lambda}\mathrm{d}_{\mu}\,\mathrm{log}\,
E(P,Q).\label{Bergmann}
\end{equation}
All the $\alpha$-periods of $\omega(P,Q)$ with respect to any
of its  two variables vanish. The  period of the Bergmann kernel
with respect to the variable $P$  or $Q$,
along the $\beta_k$ cycle,
 is equal to $2\pi\i \d v_k(Q)$ or
$2\pi \i\d v_k(P)$ respectively. The Bergmann kernel has a double
pole along the diagonal with the following local behaviour
\begin{align}
\omega(P,Q)&=\left(\frac{1}{(z(P)-z(Q))^2}+H(z(P),z(Q))
+\text{higher order terms}\right)\mathrm{d}z(P)\mathrm{d}z(Q),
\end{align}
where $H(z(P),z(Q))\mathrm{d}z(P)\mathrm{d}z(Q)$ is the non-singular part
of $\omega(P,Q)$ in each coordinte chart. The restriction
of $H$ on the diagonal is the projective connection (see for example \cite{ti78})
\begin{equation}
\label{pcon}
R(z(P))=6H(z(P),z(P))
\end{equation}
which  depends non-trivially  on the chosen  system of local coordinates.
Namely the projective connection transforms as follows  with respect to a
change of local coordinates $z\rightarrow f(z)$
\[
R(z)\rightarrow R(f(z))[f'(z)]^2+\{f(z),z\},
\]
where $\{\,,\,\}$ is the Schwarzian derivative (\ref{schder}).

The Szeg\"o kernel $S\pq(P,Q)$ is defined for all
non-singular characteristics
$\pq$ as the
$(\frac12,\frac12)$-form on $\mathcal{C}\times \mathcal{C}$ \cite{fa73}
\begin{equation}
\label{Szego} S\pq(P,Q)=\frac{\theta\pq\left(\int\limits^P_{Q} \d
\boldsymbol{v};\Pi \right)}{\theta\pq(\boldsymbol{0};\Pi)E(P,Q) }.
\end{equation}
The local behaviour of the Szeg\"o kernel when
$P\rightarrow Q$  is
\begin{equation}
\label{szegoexp}
S\pq(P,Q)=\dfrac{\sqrt{\mathrm{d}z(P)}\sqrt{\mathrm{d}z(Q)}}{z(P)-z(Q)}
\left[1+T(z(P)) (z(P)-z(Q)) + O ((z(P)-z(Q)^2)\right],
\end{equation}
where 
\begin{equation}
\label{fayszego}
T(z(P))dz(P)=\sum_{k=1}^g\frac{\partial}{\partial z_k}
\log\theta\pq(\boldsymbol{0};\Pi)\mathrm{d}v_k(P).
\end{equation}
The Szeg\"o kernel transforms when the variable
$P$ goes around $\alpha_k$ and $\beta_k$-cycles as follows
\begin{align}
\label{Szego1}
&S\pq(P+\alpha_k,\,Q)=\e^{2\pi \i \delta_k}S\pq(P,Q), \\
\label{Szego2}
&S\pq(P+\beta_k,\,Q)=\e^{-2\pi \i \epsilon_k}
S\pq(P,Q),\quad k=1,\dots,g.
\end{align}
The Riemann divisor $\Delta$ is the divisor
class of the Szeg\"o kernel  with zero characteristics $[^{\boldsymbol{0}}_{\boldsymbol{0}}]$ (\cite{fa73}, p. 7).

Another important relation \cite{fa73}, Cor. 2.12, connects
the Szeg\"o and Bergmann kernels
\begin{equation}
S\pq(P,Q)S\mpq(P,Q)=\omega(P,Q)
+\sum_{k,l=1}^g\frac{\partial^2}{\partial z_k\partial z_l}
\log\theta\pq(\boldsymbol{0};\Pi)\mathrm{d}v_k(P)\mathrm{d}v_l(Q).
\label{fay212}
\end{equation}
Finally we point the following equality,  \cite{fa73}, Cor. 2.19,
\begin{equation}
\label{det}
\det\left(\left(S\pq(P_j,Q_k)\right)_{j,k=1,\ldots,n}\right)=
\dfrac{\theta\pq\left(\sum\limits_{j=1}^n\int\limits^{P_j}_{Q_j}\d\boldsymbol{v};\Pi)\right)}{\theta\pq(\boldsymbol{0};\Pi)}
\dfrac{\prod_{1\leq j<k\leq n}E(P_j,P_k)E(Q_k,Q_j)}
{\prod_{j,k=1}^nE(P_j,Q_k)},
\end{equation}
 for any two sets of points $P_1,\dots,P_n$, and $Q_1,\dots,Q_n$,
 $n\geq g$,
and non-singular characteristics $\pq$.

\section{$Z_N$ curves}
In order to solve the R-H problem (\ref{RH0})-(\ref{RH3}) explicitly
 we need to study in detail the Riemann surface
$\mathcal{C}_{N,m}$ of the curve
\begin{equation}
\label{CNM}
y^N=p(\lb)q(\lb)^{N-1},
\end{equation}
where $p(\lb)$ and $q(\lb)$ have been defined in (\ref{pq}).
 The curve (\ref{CNM})   has  singularities
at the points $(\lambda_{2k},0)$, $k=1,\dots,m$. These singularities 
can be  resolved \cite{mi95} to give rise to a compact Riemann
surface which we  denote by $\mathcal{C}_{N,m}$.
The genus $g$ of the curve (\ref{CNM}) can be computed from (\ref{RHu}) and is equal to
$(N-1)m$.

The branch points of the curve are $(\lambda_1,0),\dots,(\lambda_{2m+1},0)$ 
and $(\infty,\infty)$.
 The projection $\pi:(\lb,y)\ra \lb$,
defines $\mathcal{C}_{N,m}$ as a $N-$sheeted
covering of the complex plane   $\mathbb{CP}^1$. Therefore the
pre-image of an ordinary point $\lb\in \mathbb{CP}^1$
consists of $N$ points. The $N$-cyclic automorphism $J$ of
$\mathcal{C}_{N,m}$ is given by the action 
$J:(\lb,y)\ra (\lb,\rho y)$, where $\rho$ is the $N$-primitive root of unity, 
namely $\rho=\e^{\frac{2\pi \i}{N} }$.
In a neighbourhood $U_R$ of  the point
$R=(\eta,w)\in\mathcal{C}_{N,m}$, a local coordinate $z(P)$, $P=(\lb,y)\in U_R$,
 is  the function defined by
\begin{equation}
\label{localN}
z(P)=\begin{cases} \lambda-\eta, &\text{if}\quad R\quad
\text{is an ordinary point, }\\
            \sqrt[N]{ \lambda-\eta}, &\text{if}
\quad R=(\lambda_k,0),\;k=1,\dots,2m+1,\\
            \frac{1}{\sqrt[N]{\lambda}},&\text{if}\quad R=(\infty,\infty).
\end{cases}
\end{equation}

\subsection{Homologies and periods of $Z_N$-curves}

\begin{figure}[htb]
\centering
\mbox{\epsfig{figure=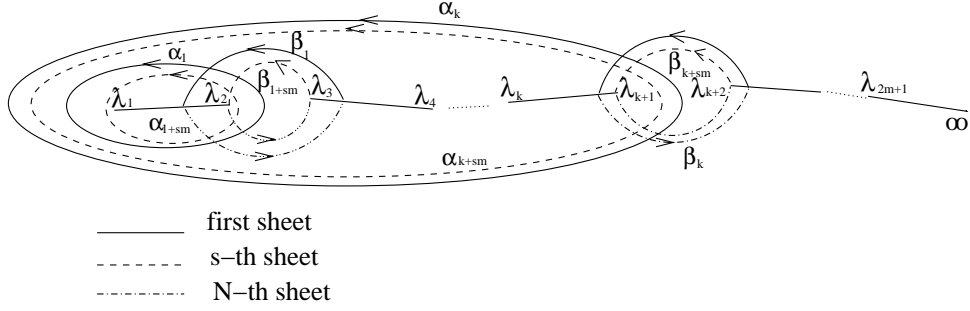,width=0.8\textwidth}}
     \caption{The homology basis. \label{fig2}}
\end{figure}

\noindent
The canonical homology basis,

$$(\alpha_1,\ldots, \alpha_{(N-1)m}; \beta_1,\ldots,
\beta_{(N-1)m})\in H(\mathcal{C},\mathbb{Z})$$
of $\mathcal{C}_{N,m}$ is
shown in the Figure\ref{fig2}.
Namely the cycles  $\alpha_{j+km}$,  $j=1,\dots,m$ lie on the
$k+1$ sheet, $k=0,\dots,N-2$. The cycles $\beta_{j+km}$,
$j=1,\dots, m$, $k=0,\dots,N-2$,  emerges on the $(k+1)$th sheet on the cut
$(\lambda_{2j-1},\lambda_{2j})$, pass anti-clockwise to the $N$th
sheet through the cut
$(\lambda_{2j+1},\lambda_{2j+2})$  and return to the initial point
through the $N$th sheet.
\begin{remark}
We remark that on  Figure 3, when $N>3$, the  $\beta$-cycles placed
from the  second  to the $(N-2)$th sheet  should intersect the cuts only
on the  branch  points. If we drop this requirement we need to draw a more
complicated but equivalent homology basis.
\end{remark}

The action of the automorphism  $J$ on the basis of cycles is given by
\begin{align}
\label{aa1}
&J\alpha_{i+sm}=\alpha_{i+(s+1)m},~~~~i=1,\dots,m,~~~s=0,\dots,N-3,\\
\label{aa2}
&J\alpha_{i+(N-2)m}=-\sum_{s=0}^{N-2}\alpha_{i+sm},\quad i=1,\ldots,m,\\
\label{aa3}
&J\beta_{i+sm}=\beta_{i+(s+1)m}-\beta_i, \;\;s=0,\dots,N-3,\quad J\beta_{i+(N-2)m}=-\beta_i,\quad i=1,\ldots,m.
\end{align}
The  basis of canonical holomorphic differentials reads
\begin{equation}
\label{holo}
\mathrm{d}u_{j+sm}(P)=\dfrac{\lb^{j-1}q(\lb)^{s}}{y^{s+1}}\mathrm{d}\lb,
\quad j=1,\dots,m,\;\;s=0,\dots,N-2.
\end{equation}
The $(N-1)m\times(N-1)m $ matrices  $\A$ of  $\alpha$-periods and
$\mathcal B$ of  $\beta$-periods are expressible in terms of
$m\times m$-matrices
\begin{equation}
\label{As} (\A_{s+1})_{kj}=\oint_{\alpha_{j}}\d u_{k+ms}, \;\quad
(\mathcal{B}_{s+1})_{kj}=\oint_{\beta_j}\d u_{k+ms}, \;\quad
j,k=1,\dots,m,\;s=0,\dots,N-2,
\end{equation}
in the following way. Let us introduce the $(N-1)m\times (N-1)m$ dimensional
 matrices
\begin{align}
\label{RA}
&\mathcal{R}_{A}=\left(\dfrac{\rho^{-i(k-1)}-\rho^{-ik}}{1-\rho^{-i}}\right)_{i,k=1,\dots,N-1}\otimes 1_{m},\\
\label{RB}
&\mathcal{R}_{B}=\left(\dfrac{\rho^{-i(k-1)}-\rho^{-i(N-1)}}{1-\rho^{-(N-1)i}}\right)_{i,k=1,\dots,N-1}\otimes 1_{m}.
\end{align}
Then
\begin{align}
\A&=\left(\oint\limits_{\alpha_j}
\mathrm{d}u_k\right)_{k,j=1,\ldots,(N-1)m}=
\mbox{Diag}(\mathcal{A}_1,\mathcal{A}_2,\dots,
\mathcal{A}_{N-1})\mathcal{R}_{\mathcal{A}},\\
\mathcal{B}&=\left(\oint\limits_{\beta_j}
\mathrm{d}u_k\right)_{k,j=1,\ldots,(N-1)m}=\mbox{Diag}(\mathcal{B}_1,
\mathcal{B}_2,\dots, \mathcal{B}_{N-1})\mathcal{R}_{\mathcal{B}},
\end{align}
where
\[
\mbox{Diag}(\mathcal{A}_1,\mathcal{A}_2,\dots, \mathcal{A}_{N-1}),
\quad\mbox{Diag}(\mathcal{B}_1,\mathcal{B}_2,\dots, \mathcal{B}_{N-1})
\]
are  the block diagonal $(N-1)m\times (N-1)m$ dimensional matrices
having as entries the matrices $A_s$ and $B_s$, $s=1,\dots,N-1$, respectively.

The basis of normalized holomorphic differentials
$$\d\boldsymbol{v}=(\d v_1,\dots,\d v_{(N-1)m}),\;\;
\oint_{\alpha_j}\d v_k=\delta_{jk},$$ is written as
\[
\d v_j=\sum_{k=1}^{g}(\A^{-1})_{jk}\d u_k,\quad j=1,\dots,(N-1)m.
\]
The period matrix $\Pi$, $$\Pi_{j,k}=\oint_{\beta_{k}}\d v_j,\quad j,k=1,\dots,(N-1)m$$ is
given by
\begin{equation}
\label{periodmatrix}
\Pi=\mathcal{R}_{\mathcal{A}}^{-1}\mbox{Diag}(\mathcal{A}^{-1}_1\mathcal{B}_1,
\mathcal{A}^{-1}_2\mathcal{B}_2,\dots, \mathcal{A}^{-1}_{N-1}
\mathcal{B}_{N-1})\mathcal{R}_{\mathcal{B}}
\end{equation}
with $\mathcal{R}_A$ and $\mathcal{R}_B$ defined in (\ref{RA})
and (\ref{RB}) respectively.

\subsection{Characteristics supported on branch points\label{charac}}

\noindent
In this section we are going to compute the integrals of the form
\[
\int_{P_{\infty}}^{P_k}\d v_j,\quad j,k=1,\dots,(N-1)m,
\]
in terms of the period matrix $\Pi$, where $P_k=(\lb_k,0)$ and 
$P_{\infty}=(\infty,\infty)$. Below we shall omit the second coordinate of the
points $P_k$ and $P_{\infty}$. 
\begin{lemma}\label{character}
The following relations are satisfied for $k=1,\dots,m,\;s=0,\dots,N-2$,
\begin{align}
\label{car1}
&\int_{\lambda_{2k}}^{\lambda_{2k-1}}\d v_{k+sm}=\dfrac{N-1-s}{N},\\
\label{car2}
&\int_{\lambda_{2k+2}}^{\lambda_{2k+1}}\d v_{k+sm}=-\dfrac{N-1-s}{N},\\
\label{car3}
&\int_{\lambda_{2k+2}}^{\lambda_{2k+1}}\d v_{j+sm}=0,\;\;j\neq k,k+1,\;j=1,\dots,m\;\;
\end{align}
and
\begin{equation}
\label{car4}
\int_{\lambda_{2j}+1}^{\lambda_{2j}}\d v_{k+sm}=\dfrac{N-1}{N}\Pi_{k+sm,j}-\dfrac{1}{N}
\sum_{r=1}^{N-2}\Pi_{k+sm,j+rm},
\end{equation}
for $\quad k,j=1,\dots,m$, $s=0,\dots,N-2$.
\end{lemma}
\begin{proof}
To prove (\ref{car1}) we observe that for $r,s=0,\dots,N-2$,
\[
\oint_{\alpha_{j+rm}}\d v_{k+sm}=0=\sum_{l=1}^{j}\int_{\lambda_{2l}}^{\lambda_{2l-1}}
(J^{(r)}(\d v_{k+sm})-J^{r+1}(\d v_{k+sm})),\;\;j<k.
\]
Since
\begin{equation}
\label{holoId}
\sum_{r=0}^{N-1}J^{(r)}(\d\boldsymbol{v})=0,
\end{equation}
the above two equations imply that
\begin{equation}
\label{CH1}
\int_{\lambda_{2j-1}}^{\lambda_{2j}}J^{(r)}(\d v_{k+sm})=0,\;\;j<k,\;\;r,s=0,\dots,N-2.
\end{equation}
Therefore for $k=1,\dots,m$ and $s=0,\dots,N-2$
\begin{equation}
\begin{split}
\label{CH2}
&\oint_{\alpha_{k+sm}}\d v_{k+sm}=1=\int_{\lambda_{2k-1}}^{\lambda_{2k}}
\left(J^{(s)}(\d v_{k+sm})-J^{(s+1)}(\d v_{k+sm})\right),\\
&\oint_{\alpha_{k+rm}}\d v_{k+sm}=0=\int_{\lambda_{2k-1}}^{\lambda_{2k}}
\left(J^{(r)}(\d v_{k+sm})-J^{(r+1)}(\d v_{k+sm})\right),\;\;r\neq s.
\end{split}
\end{equation}
Combining (\ref{holoId}) and (\ref{CH2})
 we can write the system
\begin{equation}
\begin{split}
\label{CH3}
&\begin{pmatrix}
1&-1&0&\dots&0&0\\
0&1&-1&\dots&0&0\\
\dots&\dots&&\dots&&\dots\\
0&0&0&\dots&1&-1\\
1&1&1&\dots&1&2
\end{pmatrix}
\begin{pmatrix}
\int_{\lambda_{2k-1}}^{\lambda_{2k}} \d v_{k+sm}\\
\int_{\lambda_{2k-1}}^{\lambda_{2k}} J(\d v_{k+sm})\\
\dots\\
\int_{\lambda_{2k-1}}^{\lambda_{2k}} J^{(N-3)}(\d v_{k+sm})\\
\int_{\lambda_{2k-1}}^{\lambda_{2k}} J^{(N-2)}(\d v_{k+sm})
\end{pmatrix}=
\begin{pmatrix}
0\\
\dots\\
1\\
\dots\\
0
\end{pmatrix}
\begin{matrix}
\\
\\
\stackrel{s+1}{\leftarrow}\\
\\
\\
\end{matrix}\\
&\text{for}\quad s=0,\dots,N-2\quad \text{and}\quad k=1,\dots,m,
\end{split}
\end{equation}
which leads to  (\ref{car1}).
The relation (\ref{car2}) follows from the combination of (\ref{CH1}),
(\ref{CH3}) and the fact that
\[
\oint_{\alpha_{k+1+rm}}\d v_{k+sm}=0 \;\;\text{for}\;\; \;r,s=0,\dots,N-2.
\]
The relation (\ref{car3}) follows from (\ref{car1}), (\ref{car2}) and
the fact that  $\d\boldsymbol{v}$ are  normalized differentials.

Finally to prove  (\ref{car4}) we observe that
\[
\int_{\beta_{j+rm}}\d v_{k+sm}=\Pi_{k+sm,\,j+rm}=
\int_{\lambda_{2j+1}}^{\lambda_{2j}}J^{(r)}(\d  v_{k+sm})-J^{(N-1)}(\d v_{k+sm}), \;\;r=0,\dots,N-2.
\]
Writing the above equation in matrix form and using (\ref{holoId})  we obtain
\begin{equation}
\begin{split}
&\begin{pmatrix}
2&1&1&\dots&1&1\\
1&2&1&\dots&1&1\\
\dots&\dots&&\dots&&\\
1&1&1&\dots&2&1\\
1&1&1&\dots&1&2
\end{pmatrix}
\begin{pmatrix}
\int_{\lambda_{2j+1}}^{\lambda_{2j}}\d  v_{k+sm}\\
\int_{\lambda_{2j+1}}^{\lambda_{2j}}J(\d v_{k+sm})\\
\dots\\
\int_{\lambda_{2j+1}}^{\lambda_{2j}}J^{(N-3)}(\d v_{k+sm})\\
\int_{\lambda_{2j+1}}^{\lambda_{2j}}J^{(N-2)}( \d v_{k+sm})
\end{pmatrix}=
\begin{pmatrix}
\Pi_{k+sm,\,j}\\
\Pi_{k+sm,\,j+m}\\
\dots\\
\Pi_{k+sm,\,j+(N-3)m}\\
\Pi_{k+sm,\,j+(N-2)m}
\end{pmatrix}\\
&\qquad\qquad\qquad\text{for}\quad s=0,\dots,N-2\quad \text{and}\quad k=1,\dots,m,
\end{split}
\end{equation}
which is equivalent to (\ref{car4}).
\end{proof}
We observe that the quantities in (\ref{car2}) and (\ref{car4}) satisfy
\[
\dfrac{N-1}{N}\Pi_{k+sm,j}=-\dfrac{1}{N}\Pi_{k+sm,j}\;\;\text{modulo lattice},\]\[
-\dfrac{N-1-s}{N}=\dfrac{s+1}{N}\;\;\text{modulo lattice}.
\]
From the relations (\ref{car1})-(\ref{car4}) and the above
observation we are able to write the
characteristics $[\Uc_k]$ of the vectors
\[
\Uc_k=\int_{\infty}^{\lambda_k}\d\boldsymbol{v}
\]
in the form
\begin{eqnarray*}
\label{U2m1}
&[\Uc_{2m+1}]&=\underbrace{\left[
\begin{array}{cccc}0&\ldots&0&\stackrel{m\downarrow}{0}\\
                   0&\ldots&0&\frac{1}{N}
\end{array}\right.}_m\dots
\underbrace{
\begin{array}{cccc}0&\ldots&0&\stackrel{sm\downarrow}{0}\\
                   0&\ldots&0&\frac{s}{N}\end{array}}_{m}\dots
\underbrace{\left.
\begin{array}{cccc}0&\ldots&0&\stackrel{(N-1)m\downarrow}{0}\\
                   0&\ldots&0&\frac{N-1}{N}\end{array}\right]}_{m},\\
\end{eqnarray*}
which immediately follows from (\ref{car2}). To pass from $[\Uc_{2m+1}]$ to
$[\Uc_{2m}]$, and in general from $[\Uc_{2k+1}]$ to $[\Uc_{2k}]$
we use (\ref{car3}), while for passing from $[\Uc_{2k}]$ to
$[\Uc_{2k-1}]$
we use (\ref{car1}) and (\ref{car2}), thus obtaining
\begin{align*}
\label{U2m}
&[\Uc_{2m}]=\underbrace{\left[
\begin{array}{cccc}0&\ldots&0&\stackrel{m\downarrow}{-\frac{1}{N}}\\
                   0&\ldots&0&\frac{1}{N}
\end{array}\right.}_m\dots
\underbrace{
\begin{array}{cccc}0&\ldots&0&\stackrel{sm\downarrow}{-\frac{1}{N}}\\
                   0&\ldots&0&\frac{s}{N}\end{array}}_{m}\dots
\underbrace{\left.
\begin{array}{cccc}0&\ldots&0&\stackrel{(N-1)m\downarrow}{-\frac{1}{N}}\\
                   0&\ldots&0&\frac{N-1}{N}\end{array}\right]}_{m},\\
\nonumber
&\\
&~~~~~~~~~~~~~~~~~\qquad\qquad\qquad\vdots\\
&\\
&[\Uc_{2k+1}]=\underbrace{\left[
\begin{array}{cccccc}
0&\dots&\stackrel{k\downarrow}{0}&-\frac{1}{N}&\dots&-\frac{1}{N}\\
0&\ldots&\frac{1}{N}&0&\dots&0
\end{array}\right.}_m\dots
\underbrace{
\begin{array}{cccccc}
0&\dots&\stackrel{k+(s-1)m\downarrow}{0}&-\frac{1}{N}&\dots&-\frac{1}{N}\\
0&\ldots&\frac{s}{N}&0&\dots&0\end{array}}_{m}\dots\\
&~~~~~~~~~~~~~~~~~~~~~~~~~~~~~~~~~\qquad\qquad\quad\dots
\underbrace{\left.
\begin{array}{cccccc}
0&\dots&\stackrel{k+(N-2)m\downarrow}{0}&-\frac{1}{N}&\dots&-\frac{1}{N}\\
0&\ldots&\frac{N-1}{N}&0&\dots&0
\end{array}\right]}_{m},\\
&[\Uc_{2k}]=\underbrace{\left[
\begin{array}{cccccc}0&\dots&\stackrel{k\downarrow}{-\frac{1}{N}}&-\frac{1}{N}&\dots&-\frac{1}{N}\\
                   0&\ldots&\frac{1}{N}&0&\dots&0
\end{array}\right.}_m\dots
\underbrace{
\begin{array}{cccccc}0&\dots&\stackrel{k+(s-1)m\downarrow}{-\frac{1}{N}}&-\frac{1}{N}&\dots&-\frac{1}{N}\\
                   0&\ldots&\frac{s}{N}&0&\dots&0\end{array}}_{m}\dots\\
&~~~~~~~~~~~~~~~~~~~~~~~~~~~~~~~~~~~~~~~~~~~~~~~~\qquad\qquad\quad\dots
\underbrace{\left.
\begin{array}{cccccc}0&\dots&\stackrel{k+(N-2)m\downarrow}{-\frac{1}{N}}&-\frac{1}{N}&\dots&-\frac{1}{N}\\
                   0&\ldots&\frac{N-1}{N}&0&\dots&0\end{array}\right]}_{m},\\
&~~~~~~~~~~~~~~~~~~~~~~~~~~~~~~~~~~~~~~~~~~~~~~~~~~~\qquad\qquad\qquad\vdots\\
&[\Uc_{2}]=\underbrace{\left[
\begin{array}{cccc}-\frac{1}{N}&-\frac{1}{N}&\ldots&-\frac{1}{N}\\
                   \frac{1}{N}&0&\ldots&0
\end{array}\right.}_m\dots
\underbrace{
\begin{array}{cccc}-\frac{1}{N}&-\frac{1}{N}&\ldots&-\frac{1}{N}\\
                   \frac{s}{N}&0&\ldots&0\end{array}}_{m}\dots
\underbrace{\left.
\begin{array}{cccc}-\frac{1}{N}&-\frac{1}{N}&\ldots&-\frac{1}{N}\\
                   \frac{N-1}{N}&0&\ldots&0\end{array}\right]}_{m},\\
&[\Uc_{1}]=\underbrace{\left[
\begin{array}{cccc}-\frac{1}{N}&-\frac{1}{N}&\ldots&-\frac{1}{N}\\
                   0&0&\ldots&0
\end{array}\right.}_m\dots
\underbrace{
\begin{array}{cccc}-\frac{1}{N}&-\frac{1}{N}&\ldots&-\frac{1}{N}\\
                   0&0&\ldots&0\end{array}}_{m}\dots
\underbrace{\left.
\begin{array}{cccc}-\frac{1}{N}&-\frac{1}{N}&\ldots&-\frac{1}{N}\\
                   0&0&\ldots&0\end{array}\right]}_{m}.
\end{align*}
These formulas will be useful for  the construction of $1/N$ non-singular
characteristics.


\subsection{Szeg\"o kernel for $\frac{1}{N}$-periods }
In this section we construct  the Szeg\"o kernel for $1/N$ characteristics.
For the purpose we first need to determine the vector of Riemann constants
of the curve $\mathcal{C}_{N,m}$.
\begin{lemma}
The vector of Riemann constants computed in the homology basis described
in Figure~\ref{fig2} and  with  base point $\infty$
equals
\begin{equation}
\boldsymbol{K}_{\infty}=(N-1)\sum_{k=1}^m\int\limits_{\infty}^{P_{2k}}\d
\boldsymbol{v},\label{Rconstant}
\end{equation}
where $P_{2k}=(\lb_{2k},0)$.
\end{lemma}
The proof of the above relation  is obtained by direct calculations
from the definition (\ref{RC}) following the lines of the proof of
the Lemma~\ref{character}.

\begin{lemma}
The Riemann divisor $\Delta$  of the curve $\mathcal{C}_{N,m}$ in the homology basis described in Figure~\ref{fig2} is equivalent to
\begin{equation}
\Delta= (N-1)\sum_{k=1}^mP_{2k}- \infty.\label{riemanndiv}
 \end{equation}
\end{lemma}
\begin{proof}
The above relation follows immediately from (\ref{Rconstant}).
\end{proof}
 The canonical divisor ${\mathcal K}_{\mathcal C}$ is the  divisor class of any Abelian differential on $\mathcal{C}_{N,m}$.
Choosing $\d \lb$ as representative differential,
we have according to (\ref{localN}) that
\begin{equation}
\label{canonicalD}
{\mathcal K}_{\mathcal C}=(N-1)\sum_{i=1}^{2m+1}P_{i}-(N+1)\,\infty.
\end{equation}
From (\ref{riemanndiv}) and (\ref{canonicalD}) it is possible to verify that
 $2\Delta ={\mathcal K}_{\mathcal C}$.

According to the results of Section~\ref{charac},  the formula (\ref{AbelQ0})
when $Q_0$ is a branch point,  put into correspondence
divisors  $\mathcal{D}$ consisting of branch points with  $1/N$-periods.

Following Diez \cite{diez91}, we  describe a family
of non-special divisors on $\mathcal{C}_{N,m}$ supported on the branch points.
For $m\leq l \leq 2m+1$, let  $s_1,\ldots, s_l$
be   positive integers  such that
\begin{align}
\sum_{i=1}^l s_i=(N-1)m,\quad s_i\leq N-1,
\end{align}
i.e. when $l=m$ all $s_i=N-1$.  For each $l$ let us define the  divisor class
$\mathcal{D}_l$ supported on the  branch points
\begin{equation}
\label{Dl}
\mathcal{ D}_l=s_1P_{i_1}+\ldots+s_lP_{i_l}, \quad
\mathbb{ I}_l=\{i_1,\ldots,i_l\}\in \{1,\ldots, 2m+1\}.
\end{equation}
In particular, the  divisor class $\mathcal{ D}_m$ contains
$\left(\begin{array}{c}2m+1\\m\end{array}\right)$  divisors
\begin{equation}
\mathcal{D}_m=(N-1)P_{i_1}+\ldots+(N-1)P_{i_{m-1}}+(N-1)P_{i_m}.
\label{diez1}
\end{equation}
Among the divisors with $m+1$ branch points we consider the divisor class
$\mathcal{ D}_{m+1,1}$ which contains
$\frac12(m+1)m\left(\begin{array}{c}2m+1\\m+1\end{array}\right)$ divisors
\begin{equation}
\mathcal{D}_{m+1,1}=(N-1)P_{i_1}+
\ldots+(N-1)P_{i_{m-1}}+(N-2)P_{i_m}+P_{i_{m+1}}.
\label{diez2}
\end{equation}
It is out of the scope of the present manuscript to classify all the  
non-singular divisors of the form (\ref{Dl}). 
However we can single out two families of non-special divisors.
\begin{lemma} \label{diezlemma}
The divisors $\mathcal{D}_{m}$
defined in (\ref{diez1}) are non-special and the divisors
 $\mathcal{D}_{m+1,1}$ defined in (\ref{diez2}) are non-special for $N>3$.
At $N=3$ the divisors
\begin{equation}
\begin{split}
&\mathcal{D}_{m+1,1}=2P_{i_1}+
\ldots+2P_{i_{m-1}}+P_{i_m}+P_{i_{m+1}},\\
&\;\;i_m\in\{1,3,5,\dots,2m+1\},\quad i_{m+1}\in\{2,4,6,\dots,2m\},
\label{diez22}
\end{split}
\end{equation}
are non-special.\label{non-special}
\end{lemma}
The proof is given in the Appendix.

\begin{remark}
The importance of the divisor classes $\mathcal{D}_{m}$ and
$\mathcal{D}_{m+1,1}$ is due to the fact  that one can construct meromorphic
functions with zeros and poles in prescribed branch points. Indeed let
  $\mathcal{D}_{m}$ and $\mathcal{D}_{m+1,1}$ be the divisors
defined in (\ref{diez1}) and (\ref{diez2}).
Then the function
\begin{equation} f(P)= C\left(\frac{\theta\left(
\int\limits_{\infty}^P\mathrm{d}\boldsymbol{v}-
\int\limits_{g\infty}^{\mathcal{D}_{m}}\mathrm{d}\boldsymbol{v}
+\boldsymbol{K}_{\infty};     \Pi\right)} {\theta\left(
\int\limits_{\infty}^P\mathrm{d}\boldsymbol{v}-
\int\limits_{g\infty}^{\mathcal{D}_{m+1,1}}\mathrm{d}\boldsymbol{v}
+\boldsymbol{K}_{\infty} ;    \Pi\right)}\right)^N, \label{coordinate}
 \end{equation}
with  $C$  a constant, has the only zero of $N$-th order at
the point $P_{i_m}$ and the only pole of
 $N$-th order at the point
$P_{i_{m+1}}$. When
the normalising constant $C$ is chosen in an appropriate way, the function
$f(P)$ can be identified with the coordinate $\lambda$ of the curve.
\end{remark}

Now we associate to the  non-special divisors ${\mathcal D}_m$
the Szeg\"o kernel corresponding to such divisors.
The Szeg\"o kernel for the  complete  class of divisors (\ref{Dl})
will be considered in  a separate publication.

For the purpose we define the divisor class $\mathfrak{D}$ as
\begin{equation}
\label{D}
\mathfrak{D}=P+J(P)+J^{(2)}(P)+\dots+J^{(N-1)}(P),
\end{equation}
which is independent from the point $P\in\mathcal{C}_{N,m}$.
The following relations
hold
\[
\mathfrak{D}=NP_i,\quad i=1,\dots,2m+1,\quad \mathfrak{D}=NP_{\infty}.
\]
Let us associate to the divisor $\mathcal{D}_m$  the divisor of degree
$m(N-1)-1$
\begin{equation}
\label{Dmm}
\tilde{\mathcal{D}}_m=\mathcal{D}_m+(N-1)P_{\infty}-\mathfrak{D},
\end{equation}
and let  $[\tilde{\mathcal{D}}_m]$ be the corresponding $1/N$ period
of the   divisor
$\tilde{\mathcal{D}}_m$, that is
\begin{equation}
\label{eIm}
[\tilde{\mathcal{D}}_m]=\mathfrak{A}\left(\mathcal{D}_m+(N-1)P_{\infty}-
\mathfrak{D}-\Delta\right),
\end{equation}
where $\Delta$ is the Riemann divisor.
We observe that when the base point is at infinity then (\ref{eIm}) reads
\[
[\tilde{\mathcal{D}}_m]=\int\limits^{\mathcal{D}_m}_{(N-1)mP_{\infty}}\d\boldsymbol{v}-\boldsymbol{K}_{\infty},
\]
therefore, the characteristics $[\tilde{\mathcal{D}}_m]$
is non-singular because of Lemma~\ref{diezlemma}.
Let us define the function
\begin{equation}\psi_k(P,Q)=\frac{z(P)-\lambda_k}{z(Q)-\lambda_k},\quad k=1,\ldots,2m+1.
\end{equation}
and agree to omit the arguments $(P,Q)$ if no ambiguities appear.
\begin{theorem}
\label{tszegoN1}
 The  Szeg\"o kernel associated to the
characteristics $[\tilde{\mathcal{D}}_m]$ reads
\begin{align}
\label{szegoN1}
 S[\tilde{\mathcal{D}}_m](P,Q)&
=\dfrac{1}{N}\dfrac{ \sqrt{\mathrm{d}z(P)\mathrm{d}z(Q)}}{z(P)-z(Q)}
\sum\limits_{s=0}^{N-1} \left( \frac{\prod\limits_{i_k\in \mathbb{I}_m}\psi_{i_k}(P,Q)}
 {\prod\limits_{j_k\in \mathbb{J}_{m+1}} \psi_{j_{k}}(P,Q)}
\right)^{-\frac{s}{N}+\frac{N-1}{2N}},
\end{align}
where $\mathbb{I}_m=\{i_1,\dots,i_m\}\subset \{1,2,\dots,2m+1\}$ and $\mathbb{J}_{m+1}=
\{1,2,\dots,2m+1\}\backslash \{i_1,\dots,i_m\}$.
In particular, the Szeg\"o kernel with zero characteristics reads
\begin{align}
\label{szegoN0} S[0](P,Q)&
=\dfrac{1}{N}\dfrac{ \sqrt{\mathrm{d}z(P)\mathrm{d}z(Q)}}{z(P)-z(Q)}
\sum\limits_{s=0}^{N-1} \left( \dfrac{q(z(P))}{p(z(P))}
\dfrac{p(z(Q))}{q(z(Q))} \right)^{-\frac{s}{N}+\frac{N-1}{2N}}.
\end{align}
where the polynomials $p(\lb)$ and $q(\lb)$ have been defined in (\ref{pq}).
\end{theorem}

\begin{proof}
The Szeg\"o kernel $S[\tilde{\mathcal{D}}_m](P,Q)$ is   the unique,
 up to a constant,   $(\frac12,\frac12)$-form on
$\mathcal{C}_{N,m}\times \mathcal{C}_{N,m}$ that has a simple  pole  along the
diagonal $P=Q$ and   divisor  $\mathcal{K}_{\mathcal{C}}-\tilde{\mathcal{D}}_m$
in the variable $P$
 and  $\tilde{\mathcal{D}}_m$ in the variable $Q$
 (see e.g. Narasimhan \cite{narXX}). Here
 $\mathcal{K}_{\mathcal{C}}$ is the canonical divisor and $\tilde{\mathcal{D}}_m$
has been defined in (\ref{Dmm}).
Therefore  we just need  to verify
that the right hand sides of the expressions
(\ref{szegoN1}) and  (\ref{Szego})
have the same divisor.  It is enough to show this by  setting
$Q=P_{j_1}$. Regarding the formula (\ref{Szego})
we have
\begin{align*}
&\mathrm{Div} \left( \frac{\theta[\tilde{\mathcal{D}}_m]
\left(\int\limits_{P_{j_1}}^P\mathrm{d}\boldsymbol{v}
 \right) } {E(P,P_{j_1})}\right)=(N-1)\sum_{k=2}^{m+1}P_{j_k}-P_{j_1}=\mathcal{K}_{\mathcal{C}}-\widetilde{\mathcal{D}}_m,
\end{align*}
Next putting $Q=P_{j_1}$ into the expression    (\ref{szegoN1})
we obtain
\begin{align*}
&\mathrm{div} \left.\left( \dfrac{1}{N}\dfrac{ \sqrt{\mathrm{d}z(P)\mathrm{d}z(Q)}}{z(P)-z(Q)}
\sum\limits_{s=0}^{N-1} \left( \frac{\prod\limits_{k=1}^m\psi_{i_k}(P,Q)}
 {\prod\limits_{k=1}^{m+1} \psi_{j_{k}}(P,Q)}
\right)^{-\frac{s}{N}+\frac{N-1}{2N}}\right)\right|_{Q=P_{j_1}}
=\dfrac{1}{\sqrt{N}}
\left(\dfrac{\prod\limits_{k=1}^{m}(\lambda_{j_1}-\lambda_{i_k})}{\prod_{k=2}^{m+1}
(\lambda_{j_1}-\lambda_{j_k})}\right)^{\frac{N-1}{2N}}\times\\
&\qquad\times
\mathrm{div} \left( \dfrac{ \sqrt{\mathrm{d}z(P)}}{z(P)-\lambda_{j_1}}
\left( \frac{\prod\limits_{k=1}^{m+1}(z(P)-\lambda_{j_k})}{\prod\limits_{k=1}^m(z(P)-\lambda_{i_k})}\right)^{\frac{N-1}{2N}}\right)=(N-1)\sum_{k=2}^{m+1}P_{j_k}-P_{j_1}=\mathcal{K}_{\mathcal{C}}-\widetilde{\mathcal{D}}_m,
\end{align*}
which shows that the two expressions (\ref{szegoN1}) and  (\ref{Szego})
have the same divisor class in $P$. In the same way one can check the divisor
class in $Q$. Therefore the  expressions
(\ref{Szego}) and (\ref{szegoN1})   of the Szeg\"o kernel  differ at most
from  a multiplicative constant.
This constant is equal to one because when $P\rightarrow Q$ the expression
 (\ref{szegoN1}) has the following expansion
\[
S[\tilde{\mathcal{D}}_m](P,Q)=\dfrac{\sqrt{\mathrm{d}z(P)}\sqrt{\mathrm{d}z(Q)}}{z(P)-z(Q)}
\left[1+O ((z(P)-z(Q))\right],
\]
which coincides with the leading coefficient of the expansion (\ref{szegoexp}).
\end{proof}

\begin{example}
In particular for  $N=3$, the above families of Szeg\"o kernels read
\begin{align*}
&S\{2P_{i_1}+\ldots+2P_{i_m}\}(P,Q)=\frac13\left(\sqrt[3]{
\frac{   \psi_{i_1}\cdots\psi_{i_m}  }
{\psi_{j_{1}}\cdots\psi_{j_{m+1}}    } }+1+\sqrt[3]{
\frac{\psi_{j_{1}}\cdots\psi_{j_{m+1}}     }
{\psi_{i_1}\cdots\psi_{i_m}}}
\right)\frac{\sqrt{\d z(P) \d z(Q)}}{z(P)-z(Q)},
\end{align*}
for $N=4$,
\begin{align*}
&S\{3P_{i_1}+\ldots+3P_{i_m}\}(P,Q)=\frac14\frac{\sqrt{\d z(P) \d z(Q)}}{z(P)-z(Q)}\\
&\quad\quad\left(\sqrt[8]{
\frac{   \psi_{i_1}^3\cdots\psi_{i_m}^3  }
{\psi_{j_{1}}^3\cdots\psi_{j_{m+1}}^3    } }
+\sqrt[8]{
\frac{   \psi_{i_1}\cdots\psi_{i_m}  }
{\psi_{j_{1}}\cdots\psi_{j_{m+1}}    } }+ \sqrt[8]{
\frac{\psi_{j_{1}}\cdots\psi_{j_{m+1}}     }
{\psi_{i_1}\cdots\psi_{i_m}}}+
\sqrt[8]{
\frac{\psi_{j_{1}}^3\cdots\psi_{j_{m+1}}^3     }
{\psi_{i_1}^3\cdots\psi^3_{i_m}}}
\right).
\end{align*}
\end{example}

The following corollary can be checked in a straightforward
manner.
\begin{corollary}
The expansion of the Szeg\"o kernel with zero characteristics
as $P\rightarrow Q$  reads
\begin{align}
S[0](P,Q)&=\frac{\sqrt{\mathrm{d}z(P)\mathrm{d}z(Q)
}}{z(P)-z(Q)}\notag
\\
&\times\left\{ 1+ \left(\frac16\{z(P),P\}+
\frac{N^2-1}{24 N^2}
\left[ \frac {\mathrm{ d}}{\mathrm{d} z}\,\mathrm {log}
\frac { p(z(P))}{ q(z(P))}
\right]^2\right)(z(P)-z(Q))^2 +\ldots \right\},
\label{szegoexpan1g}
\end{align}
where $\{z(P),P\}$ is the Schwarzian derivative (\ref{schder}).
\end{corollary}

\section{Solution of the Riemann-Hilbert problem for the $Z_N$-curve}
Now we are ready to solve the canonical R-H problem (\ref{BX}),
that is we determine a $N\times N$ matrix valued function  $X(\lb)$ that
satisfies
\begin{equation}
\label{BX1}
\begin{split}
&X_-(\lb)=X_+(\lb)\mathcal{P}_N,\quad
\lb\in\cup_{k=0}^m(\lb_{2k+1},\lb_{2k+2}),\\
&X(\lb_0)=1_N,\quad \lb_0\in C_+,
\end{split}
\end{equation}
where $\mathcal{P}_N$ has been defined in (\ref{PP0}).
The quasi-permutation monodromy matrix $\mathcal{P}_N$ can be diagonalised
to the form
\[
\mathcal{P}_N=U\e^{2\pi\i\sigma_N}U^{-1},\]
where the diagonal matrix $\sigma_N$ is defined in (\ref{sigmaN}) and
the matrix $U$ can be chosen with entries  $U_{1k}=1$, $k=1,\dots,N$, $\text{Det}(U)\neq 0$.
In this way the canonical R-H problem (\ref{BX1}) is reduced to the form
\begin{equation}
\nonumber
\begin{split}
&(U^{-1}XU)_-=(U^{-1}XU)_+\e^{2\pi\i\sigma_N}\quad
\lb\in\cup_{k=0}^m(\lb_{2k+1},\lb_{2k+2}),\\
&U^{-1}X(\lb_0)U=1_N,\quad \lb_0\in C_+.
\end{split}
\end{equation}
It is easy to verify that the diagonal matrix
\[
U^{-1}X(\lb)U=\left(\dfrac{p(\lb)}{q(\lb)}\dfrac{q(\lb_0)}
{p(\lb_0)}\right)^{\sigma_N}
,\]
where the polynomials $p(\lb)$ and $q(\lb)$ has been defined in (\ref{pq}),
solves the above R-H problem. Indeed choosing the  function
\[
\left(\dfrac{p(\lb)}{q(\lb)}\right)^{\frac{N-1}{2N}}\rightarrow
\lb^{\frac{N-1}{2N}},\quad \lb\rightarrow \i\infty,\;\text{arg}
\lb=\frac{\pi}{2}
\]
it follows that
\[
\left(\dfrac{p(\lb)}{q(\lb)}\right)_-^{-k+\frac{N-1}{2N}}=\e^{2\pi\i
(-k+\frac{N-1}{2N})}\left(\dfrac{p(\lb)}{q(\lb)}\right)_+^{-k+
\frac{N-1}{2N}},\quad k=0,\dots,N-1.
\]
Furthermore
\[
\det(U^{-1}X(\lb)U)=\det\left(\left(\dfrac{p(\lb)}{q(\lb)}
\dfrac{q(\lb_0)}{p(\lb_0)}\right)^{\sigma_N}\right)=1,\quad
\lb\in\C\cup\infty\]
and clearly $U^{-1}X(\lb_0)U=1_N$.
Therefore the matrix function
\[
X(\lb)=U\left(\dfrac{p(\lb)}{q(\lb)}\dfrac{q(\lb_0)}{p(\lb_0)}\right)^{\sigma_N}U^{-1}
\]
solves the canonical R-H problem (\ref{BX1}).
The entries of the  matrix $X(\lb)$ can be expressed
in terms of the Szeg\"o kernel with  zero characteristics, $S[0](P,Q)$,
defined on $\mathcal{C}_{N,m}$ and derived in (\ref{szegoN0}).
Indeed it turns out that the entries $X_{rs}(\lb)$, $r,s,=1,\dots,N,$, of $X(\lb)$ are also equal to
\begin{align*}
X_{rs}(\lb)=& S[0](P^{(s)},P_0^{(r)})\dfrac
{z(P)-z(Q)}{ \sqrt{\d z(P)\d z (P_0)}}=\dfrac{1}{N}\sum\limits_{k=0}^{N-1}
\left( \e^{2\pi\i\frac{(s-r)}{N}}\sqrt[N]
{\dfrac{p(\lb)}{q(\lb)}\dfrac{q(\lb_0)}{p(\lb_0)}}
\right)^{-k
+\frac{N-1}{2}},\;
\lb_0\notin D,
\end{align*}
where $P^{(s)}=(\lb,\rho^{s-1} y)$ and  $P_0^{(r)}=(\lb_0,\rho^{r-1} y_0)$,
$r,s=1,\dots,N$,
denote the points on the $s$-th  and $r-$th sheet of
$\mathcal{C}_{N,m}$ respectively.
When $N=2$ and $\sqrt[4]{\dfrac{q(\lb_0)}{p(\lb_0)}}=1$,
such formula coincides with the canonical solution obtained in \cite{dikz99}.

We also observe that the matrix $X(\lb)$ satisfies the differential equation
\[
\dfrac{\d
X(\lb)}{\d\lb}=\sum_{k=1}^{2m+1}\dfrac{A_k}{\lb-\lambda_k}X(\lb)
\]
where the matrices $A_k$ are given by
\begin{equation}
\label{A0char}
A_k=(-1)^{k-1}U\sigma_N U^{-1},\quad k=1,\dots,2m+1,
\end{equation}
with  $\sigma_N$ defined (\ref{sigmaN}).
Therefore the canonical R-H problem gives a constant solution of
the Schlesinger system
(\ref{Schlesinger}).

We are now ready to derive the solution of the R-H problem
(\ref{RH1})-(\ref{RH3}) for arbitrary non-zero values of the constants
$c_k$ and $d_k$, $k=1,\dots (N-1)m$.

\begin{theorem}[Main Theorem]
\label{trhs}
Let the characteristics $\boldsymbol{\epsilon},
\boldsymbol{\delta}\in\C^{(N-1)m}$ be
\begin{align}
\label{chare}
\epsilon_{k+sm}&=\dfrac{1}{2\pi \i}\log\dfrac{c_{k+sm}}{c_{k+1+sm}},\;
\;s=0,\dots,N-2,\;\;k=1,\dots,m-1,\\
\nonumber
\epsilon_{sm}&=\dfrac{1}{2\pi \i}\log c_{sm},\;\;s=1,\dots,N-1,\\
\label{chard}
\delta_k&=\dfrac{1}{2\pi \i}\log d_k\;\;k=1,\dots,(N-1)m.
\end{align}
Suppose that $\theta\pq(\boldsymbol{0};\Pi)\neq 0$.
Then the matrix valued function $Y(\lb)=(Y_{rs}(\lb))_{r,s,=1,\dots,N}$
\begin{equation}
\label{solN}
Y_{rs}(\lb)=X_{rs}(\lb)\dfrac{\theta\pq
\left(\int\limits_{P_0^{(r)}}^{P^{(s)}}\d\boldsymbol{v};\Pi
\right)}{\theta\left(\int\limits_{P_0^{(r)}}^{P^{(s)}}
\d\boldsymbol{v};\Pi\right)}
\dfrac{\theta(\boldsymbol{0};\Pi)}{\theta\pq(\boldsymbol{0};\Pi)}
, \quad r,s=1,\dots,N
\end{equation}
solves the R-H problem (\ref{RH0})-(\ref{RH3}) and $\det Y(\lb)\neq 0$ for
$\lb\neq \lambda_k$, $k=1,\dots,2m+2$.
\end{theorem}
\begin{proof}
First of all we show that matrix (\ref{solN}) is holomorphic
outside the singular set $\lb\neq \lambda_1,\dots,\lambda_{2m+1},\infty$.
Indeed combining (\ref{Szego}) and (\ref{szegoN0}),  the entries of the
matrix (\ref{solN})  can be written in the form given in \cite{kor01}
\begin{equation}
\label{solN2}
Y_{rs}(\lb)=S\pq(P_0^{(r)},P^{(s)})\dfrac{z(P^{(s)})-z(P^{(r)}_0)}
{\sqrt{\d z(P^{(s)})\d z(P^{(r)}_0)}},\quad r,s=1,\dots,N.
\end{equation}
From the properties of the Szeg\"o kernel,  the matrix  (\ref{solN2})
is clearly holomorphic for $\lb\neq \lambda_1,\dots,\lambda_{2m+1},\infty$.
Furthermore, using the  formula (\ref{solN2}) for the entries of
$Y(\lb)$ and applying  the relation (\ref{det}) and (\ref{holoId})
we  conclude that
\begin{equation}
\det Y(\lb)=\left( \dfrac{z(P^{(1)})-z(P^{(1)}_0)}
{\sqrt{\d z(P^{(1)})\d z(P^{(1)}_0)}}\right)^N
\dfrac{\prod\limits_{1\leq r<s\leq N}E(P_0^{(r)},P_0^{(s)})E(P^{(r)},P^{(s)})}
{\prod\limits_{r,s=1}^N E(P_0^{(r)},P^{(s)})}\neq 0,
\end{equation}
for $P\neq (\lambda_k,0)$ or $(\infty,\infty)$.
In the above formula we have used  the relation
$z(P^{(r)})=z(P^{(1)})$ and $z(P_0^{(r)})=z(P_0^{(1)})$, $r=1,\dots,N$.
Evidently we have that
\[
\det Y(\lb_0)=1_N.
\]
In order to prove that (\ref{solN}) does indeed satisfy the R-H
problem (\ref{RH1})-(\ref{RH3}) the following considerations are
needed.
The action of the automorphism $J$ on $\d v_j$ is given by the relation
\begin{equation}
\label{Jdv}
J(\d v_j(\lb,y)) =\sum_{k=1}^m\sum_{r=1}^{N} \gamma^r_{j,k}
\lb^{m-k}    \dfrac{  ( q(\lb) )^{r-1}  }{  \rho^r y^r  } \d\lb,
\end{equation}
where $\gamma^r_{jk}$ are the normalisation constants
of the holomorphic differentials  and $\rho$ is the $N$th root of unity.
Let us consider the Abelian integral
\begin{equation}
\label{Abi}
\boldsymbol{v}(P)=\int_{\infty}^{P}\d\boldsymbol{v}.
\end{equation}
The action of the automorphism $J$ on $\boldsymbol{v}(P)$
is naturally given by
\begin{equation}
\label{Abb}
J(\boldsymbol{v}(P))=\int_{\infty}^{J(P)}\d\boldsymbol{v}=\int_{\infty}^{P}J
(\d\boldsymbol{v}).
\end{equation}
When $P^{(s)}=(\lb,\rho^{s-1}y)$
is  on the $s$-th sheet, we denote by $J^{(s-1)}(\boldsymbol{v}(\lb))$
the natural restriction of the integral $\boldsymbol{v}(P^{(s)})$
on $C_+\cup C_-$:
\[
J^{(s-1)}(\boldsymbol{v}(\lb)):=
\int_{\infty}^{(\lb,y)}J^{(s-1)}(\d\boldsymbol{v})=
\sum_{k=1}^m\sum_{r=1}^{N} \gamma^r_{j,k}
\int_{\infty}^{(\lb,y)}\xi^{m-k}    \dfrac{  ( q(\xi) )^{r-1}  }
{  \rho^{r(s-1)} w^r  } \d\lb,\;\;(\xi,w)\in\mathcal{C}_{N,m}.
\]
The integral itself is taken  on the first sheet of $\mathcal{C}_{N,m}$ and
the integration path  lies in  $C_+$ and $C_-$ for $\lb\in C_+$
or $\lb\in C_-$ respectively. The integral in (\ref{solN}) is defined as
\[
\int\limits_{P_0^{(r)}}^{P^{(s)}}\d\boldsymbol{v}:=J^{(s-1)}(\boldsymbol{v}(\lb))-J^{(r-1)}(\boldsymbol{v}(\lb_0)).
\]
If $\lb\in\li$ the integrals $\boldsymbol{v}_{\pm}(\lb)$, are shown on
Figure~\ref{figho}, namely the  integration path of $\boldsymbol{v}_{\pm}(\lb)$,
lies in $C_{\pm}$ respectively. 
\begin{figure}[htb]
\centering
\mbox{\epsfig{figure=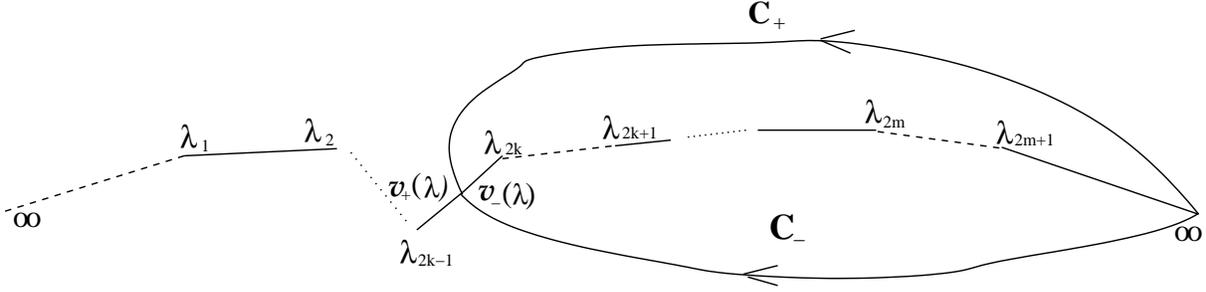,width=1.0\textwidth}}
     \caption{The different paths of integration of 
$\boldsymbol{v}_{\pm}(\lambda)=\int_{\infty}^{\lambda}\d\boldsymbol{v}$ in $C_{\pm}$. \label{figho}}
\end{figure}
From the properties of
the homology basis (\ref{aa1})-(\ref{aa3})
the following relations can be easily derived:
\begin{align}
\label{ii1}
&\left.\left[J^{(s-1)}\boldsymbol{v}_-(\lb)-J^{s}\boldsymbol{v}_+(\lb)\right]
\right|_{[\lambda_{2k-1},\lambda_{2k}]}=
\sum_{j=k}^{m}\left(\oint\limits_{\beta_{j+(s-1)m}}\d
\boldsymbol{v}-\oint\limits_{\beta_{j+sm}}\d \boldsymbol{v}\right)
\end{align}
for $s=1,\dots,N-2$ and $k=1,\dots,m$,
\begin{align}
\label{ii2}
&\left.\left[J^{(N-2)}\boldsymbol {v}_-(\lb)-J^{(N-1)}
\boldsymbol{v}_+(\lb)\right]\right|_{[\lambda_{2k-1},\lambda_{2k}]}=
\sum_{j=k}^{m}\oint\limits_{\beta_{j+(N-2)m}}\d\boldsymbol{v},\\
\label{ii3}
&\left.\left[J^{(N-1)}\boldsymbol {v}_-(\lb)-\boldsymbol{v}_+(\lb)\right]\right|_{[\lambda_{2k-1},\lambda_{2k}]}=-\sum_{j=k}^{m}\oint_{\beta_{j}}\d\boldsymbol{v}
\end{align}
for $\;\;k=1,\dots,m$ and
\begin{align}
\left.\left[J^{(s-1)}\boldsymbol{v}_-(\lb)-J^{s}\boldsymbol{v}_+(\lb)\right]
\right|_{[\lambda_{2m+1},\infty]}=0,\quad s=1,\dots,N.
\end{align}
In the same way we obtain
\begin{align}
\label{ii4}
&\left.\left[J^{s}\boldsymbol{v}_-(\lb)-J^{s}\boldsymbol{v}_+(\lb)\right]
\right|_{ [\lambda_{2k},\lambda_{2k+1}]  }=\oint_{\alpha_{k+sm}}\d\boldsymbol{v},\;\;s=0,\dots,N-2,\\
\label{ii5}
&\left.\left[J^{(N-1)}\boldsymbol{v}_-(\lb)-J^{(N-1)}\boldsymbol{v}_+(\lb)\right] \right|_{[\lambda_{2k},\lambda_{2k+1}]  }=-\sum_{s=0}^{N-2}
\oint_{\alpha_{k+sm}}\d\boldsymbol{v},
\end{align}
for $\;\;k=1,\dots,m$ and
\begin{align}
\label{ii6}
\left.\left[J^{s}\boldsymbol{v}_-(\lb)-J^{s}\boldsymbol{v}_+(\lb)\right]
\right|_{ [\infty,\lambda_{1}]  }&=0,\quad s=0,\dots,N-1.
\end{align}
Now let us suppose that $\lb\in [\lambda_{2k-1},\lambda_{2k}]$. Then for $s=1,\dots,N-2$ and $r=1,\dots, N$ we have
\begin{align}
\nonumber
(Y_{-}(\lb))_{rs}&=(X_{-}(\lb))_{rs}\dfrac{\theta\pq
\left(J^{(s-1)}\boldsymbol{v}_-(\lb)-J^{(r-1)}\boldsymbol{v}(\lb_0);\Pi
\right)}{\theta\left(J^{(s-1)}\boldsymbol{v}_-(\lb)-J^{r-1}\boldsymbol{v}(\lb_0);\Pi\right)}
\dfrac{\theta(\boldsymbol{0};\Pi)}{\theta\pq(\boldsymbol{0};\Pi)}\\
\nonumber
&=(X_{+}(\lb)\mathcal{P}_{N})_{rs}\dfrac{\theta\pq
\left(\int\limits_{P_0^{(r)}}^{P^{(s+1)}}\d\boldsymbol{v}_++\int\limits_{\infty}^{P^{(s)}}\d\boldsymbol{v}_--\int\limits_{\infty}^{P^{(s+1)}}\d\boldsymbol{v}_+;\Pi
\right)}{\theta\left(\int\limits_{P_0^{(r)}}^{P^{(s+1)}}
\d\boldsymbol{v}_+\int\limits_{\infty}^{P^{(s)}}\d\boldsymbol{v}_--\int\limits_{\infty}^{P^{(s+1)}}\d\boldsymbol{v}_+;\Pi\right)}
\dfrac{\theta(\boldsymbol{0};\Pi)}{\theta\pq(\boldsymbol{0};\Pi)}\\
\label{YPr}
&=(X_{+}(\lb)\mathcal{P}_{N})_{rs}\dfrac{\theta\pq
\left(\int\limits_{P_0^{(r)}}^{P^{(s+1)}}\d\boldsymbol{v}_+;\Pi
\right)}{\theta\left(\int\limits_{P_0^{(r)}}^{P^{(s+1)}}
\d\boldsymbol{v}_+;\Pi\right)}
\dfrac{\theta(\boldsymbol{0};\Pi)}{\theta\pq(\boldsymbol{0};\Pi)}\times\left\{
\begin{array}{lll}
\dfrac{\e^{\sum_{j=k}^m\epsilon_{j+sm} }}
{\e^{\sum_{j=k}^m\epsilon_{j+(s-1)m}}},&s=1,\dots,N-2&\\
&&\\
\e^{-\sum_{j=k}^m\epsilon_{j+(N-2)m}},&s=N-1&\\
&&\\
\e^{\sum_{j=k}^m\epsilon_{j}},&s=N,&
\end{array}\right.
\end{align}
where the quantities $(Y_-(\lb))_{rs}$,  $(X_-(\lb))_{rs}$,
and  $(X_-(\lb)\mathcal{P}_N)_{rs}$
denote the $r,s$ entry of the matrix $Y_-(\lb)$,  $X_-(\lb)$, and
$X_-(\lb)\mathcal{P}_N$ respectively,  and in the last identity we
have
 used the relation (\ref{ii1})-(\ref{ii3}) and the periodicity
 property
 (\ref{periodicity2}) of the $\theta$-function.  From (\ref{YPr})
it is immediate to verify that the constants $\epsilon_1,\dots \epsilon_{(N-1)m}$ and
$c_1,\dots, c_{(N-1)m}$ are related by (\ref{chare}) if and only if the
 matrix $Y(\lb)$ satisfies
\[
Y_-(\lb)=Y_+(\lb)G_{2k-1},
\quad \lb \in[\lambda_{2k-1},\lambda_{2k}],~~k=1,\dots,m+1,~~\lambda_{2m+2}=\infty,
\]
where the matrix $G_{2k-1}$ has been define in (\ref{Gk}).
Repeating the same procedure for $\lb\in [\lambda_{2k},\lambda_{2k+1}]$ and using  (\ref{ii4})-(\ref{ii6}) and the periodicity properties
(\ref{periodicity1}) of the $\theta$-function, we derive (\ref{chard}) if and only if
the matrix valued function $Y(\lb)$ satisfies
\[
Y_-(\lb)=Y_+(\lb)G_{2k  },
\quad \lb \in[\lambda_{2k},\lambda_{2k+1}],~~k=0,\dots,m,~~\lambda_0=\infty,
\]
where  $G_{2k}$ has been define in (\ref{Dk}).
We conclude that the matrix  (\ref{solN}) satisfies the R-H problem (\ref{RH1})-(\ref{RH3}).
\end{proof}
The form of the solution (\ref{solN}) and the $Z_N$ symmetry of the curve
$\mathcal{C}_{N,m}$ enable us to prove the following.
\begin{proposition}
\label{theoeq}
Let  $\boldsymbol{\delta}_N,\,\boldsymbol{\epsilon}_N\in (\Z/N\Z)^{(N-1)m}$ be
the characteristics   associated to the non-singular divisor  $\mathcal{D}_l$
supported on the branch points, that is
\begin{equation}
\label{deltac}
\boldsymbol{\epsilon}_N+\boldsymbol{\delta}_N\Pi=\sum_{i_l}s_{i_l}\int_{\infty}^{P_{i_l}}
\d\boldsymbol{v}-\boldsymbol{K}_{\infty},\quad \sum_{i_l}s_{i_l}=(N-1)m,
\end{equation}
where $i_l\in\{1,2,\dots,2m+1\}$ and $\boldsymbol{K}_{\infty}$ is
 the vector of Riemann constants
(\ref{Rconstant}). Then the matrix $Y(\lb)$ with entries
\begin{equation}
\label{solNr}
Y_{rs}(\lb)=X_{rs}(\lb)
\dfrac{\theta[^{\boldsymbol{\delta}_N}_{\boldsymbol{\epsilon}_N}]
\left(\int\limits_{P_0^{(r)}}^{P^{(s)}}\d\boldsymbol{v};\Pi
\right)}{\theta\left(\int\limits_{P_0^{(r)}}^{P^{(s)}}
\d\boldsymbol{v};\Pi\right)}
\dfrac{\theta(\boldsymbol{0};\Pi)}{\theta[^{\boldsymbol{\delta}_N}_{\boldsymbol{\epsilon}_N}](\boldsymbol{0};\Pi)}
, \quad r,s=1,\dots,N,
\end{equation}
solves  the R-H problem (\ref{RH1})-(\ref{RH3}) with reducible monodromy
representation (\ref{Mreducible}).
\end{proposition}
\begin{proof}
Using the notations of Section \ref{charac}
we write (\ref{deltac}) in the form
\begin{equation}
\label{deltacc}
\boldsymbol{\epsilon}_N+\boldsymbol{\delta}_N\Pi=
\sum_{i_l}s_{i_l}[\Uc_{i_l}]-(N-1)\sum_{l=1}^m [\Uc_{2l}],
\end{equation}
where $[\Uc_{k}]$ are the characteristics defined in Sect.\ref{charac}.
Because of the relations (\ref{chare}) and (\ref{chard}), the operation of summing 
two  characteristics is equivalent to the operation of multiplying two different
sets of constants $c^{(k)}_j,\;d^{(k)}_j$, $k=1,2$, $j=1,\dots,(N-1)m$. 
For this reason,  we associate to  the characteristics $[\Uc_{k}]$,
 the constants $c^{(k)}_j,\;d^{(k)}_j$, $j=1,\dots,(N-1)m$ 
according to the rule (\ref{chare}) and (\ref{chard}), that is 
\[
[\Uc_1]\longleftrightarrow  c^{(1)}_j=1,\quad d^{(1)}_j=\e^{-\frac{2\pi\i}{N}},\;
j=1,\dots (N-1)m;
\]
for $k=2,\dots,m$
\begin{align*}
&[\Uc_{2k-1}]\longleftrightarrow c^{(2k-1)}_{k-1+sm}=\e^{\frac{2\pi\i(s+1)}{N}},\;\; 
c^{(2k-1)}_{j+sm}=1,\,j\neq k-1,\;j=1,\dots,m,\\
&~~~~~~~d^{(2k-1)}_{j+sm}=\e^{-\frac{2\pi\i}{N}},\;k\leq j\leq m,\;\; d^{(2k-1)}_{j+sm}
=1,\;1\leq j<k,\;\;s=0,\dots,N-2;
\end{align*}
for $k=1,\dots,m$
\begin{align*}
&[\Uc_{2k}]\leftrightarrow c^{(2k)}_{k+sm}=\e^{\frac{2\pi\i(s+1)}{N}},\;\; 
c^{(2k)}_{j+sm}=1,\,j\neq k,\;j=1,\dots,m,\\
&~~~~~~~d^{(2k)}_{j+sm}=\e^{-\frac{2\pi\i}{N}},\;k\leq j\leq m, \;\;d^{(2k)}_{j+sm}=1,\;
1\leq j<k,\;\;s=0,\dots,N-2.
\end{align*}
Combining the above relations it is possible to verify
that the monodromy representation associated to the non-singular characteristics (\ref{deltacc}) is
\begin{align*}
&\left\{ M_{k}=\dfrac{\prod_{i_{l_k}}\xi_{i_{l_{k}}}^{s_{i_l}} }
{ \prod\limits_{n_k \text{even}}
\xi^{(N-1)}_{n_k}}\mathcal{P}^{(-1)^{k-1}}, \;k=1,\dots,2m+1, \;M_{\infty}=
\mathcal{P}_N^{-1},\;\;\;
\prod_{k=1}^{2m+1}\dfrac{\prod_{i_{l_k}}\xi_{i_{l_{k}}}^{s_{i_l}} }{ \prod\limits_{n_k \text{even}}
\xi^{(N-1)}_{n_k}}=1,\right.\\
&\;\;\left.\xi_{i_{l_{k}}}, \xi_{n_k}\in \{1,\e^{\frac{2\pi\i}{N}},\dots,\e^{\frac{2\pi\i(N-1)}{N}}\},\quad \sum_{i_l}s_{i_l}=(N-1)m\right\}.
\end{align*}
According to remark~\ref{remarkreduc}, the above monodromy representation
is reducible.
\end{proof}
In the following  we consider a couple of divisors whose difference
is a non-singular divisor supported on the branch points.
If  $\pq$ is  a non-singular characteristics corresponding to a divisor
in general position and if $\boldsymbol{\epsilon}_N,\boldsymbol{\delta}_N
\in  (\Z/N\Z)^{(N-1)m}$ is  a  characteristics corresponding to the
non-special divisors $\mathcal{D}_l$ defined in (\ref{Dl}),
 then $[^{\boldsymbol{\delta}+\boldsymbol{\delta}_N }_{
\boldsymbol{\epsilon}+\boldsymbol{\epsilon}_N}]$ is a non-singular
characteristics. Indeed the characteristics  $[^{\boldsymbol{\delta}
+\boldsymbol{\delta}_N }_{
\boldsymbol{\epsilon}+\boldsymbol{\epsilon}_N}]$ correspond to
  a divisor of degree $2g$.   But  all divisors $\mathcal{D}$ of degree
  $\mathrm{deg}\,\mathcal{D}>2g-2$ are non-special \cite{fk80}.
\begin{theorem}
\label{theosym}
Let  $\pq$ and  $[^{\boldsymbol{\delta}+\boldsymbol{\delta}_N }_{
\boldsymbol{\epsilon}+\boldsymbol{\epsilon}_N}]$ be as
above.
Then the entries $Y_{rs}(\lb)$ and $\widetilde{Y}_{rs}(\lb)$
of the solutions $Y(\lb)$ and $\widetilde{Y}(\lb)$  of the
R-H problem  (\ref{RH0})-(\ref{RH3}) with characteristics  $\pq$ and  $[^{\boldsymbol{\delta}+\boldsymbol{\delta}_N }_{
\boldsymbol{\epsilon}+\boldsymbol{\epsilon}_N}]$,
  are equivalent up to an algebraic  transformation.
The monodromy representation $\mathcal{M}=\,\{M_1,\,M_2,\,\dots,M_{2m+1},\,M_{\infty}\}$ and  $\mathcal{\widetilde{M}}=\{\widetilde{M}_1,\widetilde{M}_2,\dots,\widetilde{M}_{2m+1},\widetilde{M}_{\infty}\}$   associated to  the solutions $Y(\lb)$ and
$\widetilde{Y}(\lb)$ respectively,  are equivalent  up to multiplication by $N$th roots of unity.  That is $\widetilde{M}_k=\e^{\frac{2\pi\i}{N} j_k}M_k$, $j_k$ integer,
$\sum_{k=1}^{2m+1}j_k=0\,\text{mod}\,N$.
\end{theorem}
\begin{proof}
If $\boldsymbol{\delta}_N,\,\boldsymbol{\epsilon}_N\in  (\Z/N\Z)^{(N-1)m}$ then, by  (\ref{periodicity1}), (\ref{periodicity2}), the  ratio
\begin{equation}
\label{RR1}
\mathcal{F}(P^{(s)},P_0^{(r)}):=\left(\dfrac{\theta[^{\boldsymbol{\delta}+\boldsymbol{\delta}_N }_{
\boldsymbol{\epsilon}+\boldsymbol{\epsilon}_N}]\left(\int\limits_{P^{(r)}_0}^{P^{(s)}}\d\boldsymbol{v};\Pi\right)}
{\theta\pq
\left(\int\limits_{P^{(r)}_0}^{P^{(s)}}\d\boldsymbol{v};\Pi\right)}
\right)^N
\end{equation}
is a single-valued function on $\mathcal{C}_{N,m}$ in both the  arguments
$P^{(s)}$ and $P_0^{(r)}$. Hence  $\mathcal{F}(P^{(s)},P_0^{(r)})$  is a
meromorphic function. This means that
$\mathcal{F}(P^{(s)},P_0^{(r)})$ is a rational expression in
$\lb,  y,  \,  \lb_0,  \,y_0$, therefore
$\sqrt[N]{\mathcal{F}(P^{(s)},P_0^{(r)})}$ is algebraic.
Hence, from (\ref{solN}) and (\ref{RR1})
\[
\widetilde{Y}_{rs}(\lb)=\sqrt[N]{\mathcal{F}(P^{(s)},P_0^{(r)})}Y_{rs}(\lb),\quad r,s=1,\dots,N,
\]
which is the first statement of the theorem.
The equivalence of the corresponding  monodromy representation
$\mathcal{M}$ and $\mathcal{\widetilde{M}}$   up to multiplication
by $N$th roots of unity, follows  from the proof of Proposition~\ref{theoeq}.
\end{proof}
\begin{example}
We consider the case $N=3$ and $m=1$ when the Riemann surface
$\mathcal{C}_{3,1}:\{(\lb,y), \;\;y^3=(\lb-\lambda_1)(\lb-\lambda_2)^2(\lb-\lambda_3)\}$ is of genus $2$. Let
$\boldsymbol{\epsilon}$ and $\boldsymbol{\delta}$ be
a  non-singular characteristics. We consider the non-singular characteristics
 $\boldsymbol{\epsilon}_3,\boldsymbol{\delta}_3$ supported on the branch points
 given by
\[
\boldsymbol{\epsilon}_3+\boldsymbol{\delta}_3\Pi=
2[\Uc_3]-2[\Uc_2]=\left(-\dfrac43,\dfrac23\right)\Pi,
\]
as follows from the  relations derived in Section (\ref{charac}).
If $\{M_1,M_2,M_3,M_{\infty}\}$ are the monodromy matrices associated to the characteristics
$\pq$, the monodromy matrices associated
to the characteristics $[^{\boldsymbol{\delta}+\boldsymbol{\delta}_3 }_{
\boldsymbol{\epsilon}+\boldsymbol{\epsilon}_3}]$ are
\[
\{M_1,\e^{\frac{4\pi\i}{3}}M_2,\e^{\frac{2\pi\i}{3}}M_3,M_{\infty}\}.
\]
\end{example}
\subsection{Solution of the Schlesinger equations}
From the solution of the R-H problem (\ref{solN}) 
one can derive a particular solution  of the Schlesinger system.
In this sub-section   we denote by  $Y(\lb,\lb_0)$ the solution (\ref{solN}) of 
the R-H problem  (\ref{RH0})-(\ref{RH3}) with base point $\lb_0$, that is $Y(\lb_0,\lb_0)=1_N$.
According to the results in  \cite{kor01}, the solution 
of the Schlesinger system can be derived from the relation (\ref{Fuchsian1})
that can be re-written in the form
\[
(\lb-\lb_k)\dfrac{\partial}{\partial \lb_k}Y(\lb,\lb_0)Y^{-1}(\lb,\lb_0)=(\lb-\lb_k)\left(\dfrac{A_k}{\lb_0-\lb_k}-\dfrac{A_k}{\lb-\lb_k}\right).
\]
 Taking the derivative with respect to $\lb$  of both sides of the above relation and setting $\lb$ equal to $\lb_0$ it follows that 
\[
A_k=(\lb_0-\lb_k)^2\dfrac{\partial}{\partial \lb}\left.\left(\dfrac{\partial}{\partial \lb_k}Y(\lb,\lb_0)Y^{-1}(\lb,\lb_0)\right)\right|_{\lb=\lb_0},\quad k=1,\dots,2m+1,
\]
or, equivalently, \cite{kor01}
\[
A_k=(\lb_0-\lb_k)^2\left.
\dfrac{\partial^2}{\partial\lb\partial\lb_k}Y(\lb)\right|_{\lb=\lb_0},\quad k=1,\dots,2m+1.
\]
Using the formula (\ref{solN}) for the solution $Y(\lb,\lb_0)$  
of the R-H problem and the expansion (\ref{fayszego}) of the Szeg\"o kernel,
the above relation can  be written in the form
\begin{align}
\label{Akss}
(A_k)_{ss}&=(\lb_0-\lb_k)^2
\dfrac{\partial}{\partial\lb_k}\left(\sum_{l=1}^{(N-1)m}\dfrac{\partial}{\partial z_l}\log \theta\pq
\left(\boldsymbol{0};\Pi\right)\left.\dfrac{\d v_l(P)}{\d z(P)}
\right|_{P=P^{(s)}_0}\right),\quad s=1,\dots,N,\\
\label{Akrs}
(A_k)_{rs}&=(\lb_0-\lb_k)^2\sum_{l=0}^{N-1}\left(\dfrac{l}{N^2}\exp\left\{2\pi\i\frac{r-s}{2N}(2l-N+1)\right\}\right)\times\\
\nonumber
&\times\dfrac{\partial}{\partial\lb_k}
\left[\left(\dfrac{\partial}{\partial \lb_0}\log \dfrac{q(\lb_0)}{p(\lb_0)}\right)
\dfrac{\theta\pq
\left(\int\limits_{P_0^{(r)}}^{P_0^{(s)}}\d\boldsymbol{v};\Pi
\right)}{\theta\left(\int\limits_{P_0^{(r)}}^{P_0^{(s)}}
\d\boldsymbol{v};\Pi\right)}
\dfrac{\theta(\boldsymbol{0};\Pi)}{\theta\pq(\boldsymbol{0};\Pi)}\right]
, \quad r\neq s,\; r,s=1,\dots,N.
\end{align}
The matrix $A_{\infty}$ is obtained from the relation
\[
A_{\infty}=-\sum_{k=1}^{2m+1}A_k.
\]
From the relation (\ref{Akss}) and (\ref{holoId}) we can immediately verify that
\[
\text{Trace}(A_k)=0,\quad k=1,\dots,2m+1.
\]
Because of (\ref{Ak}), the eigenvalues of the matrices $A_k$, $k=1,\dots,2m+1$,  are
\[
\text{Eigenvalues}(A_k)=\left(\dfrac{-N+1}{2N},\dfrac{-N+3}{2N},\dots,
\dfrac{N-3}{2N},\dfrac{N-1}{2N}\right).
\]
The substitution (\ref{Akss}) and (\ref{Akrs}) into the above
equation leads to non-trivial $\theta$-function equivalence.

\subsection{$\tau$-function of the Schlesinger equations}
From the solution of the R-H problem one can derive the $\tau$
function for the Schlesinger system defined by (\ref{tau}).
\begin{theorem}
The $\tau$-function for the Schlesinger system reads
\begin{equation}
\label{tau1}
\tau(\lambda_1,\lambda_2,\dots,\lambda_{2m+1})
=\dfrac{\theta\pq(\boldsymbol{0};\Pi)}{\theta(\boldsymbol{0};\Pi)}\dfrac
{\prod\limits_{\substack{k<i\\i,k=0}}^m(\lambda_{2k+1}-\lambda_{2i+1})^{ \frac{N^2-1}{6N} }
\prod\limits_{\substack{k<i\\k,i=1}}^m(\lambda_{2k}-\lambda_{2i})^{\frac{N^2-1}{6N}}}
{\prod\limits_{\substack{i<j\\i,j=1}}^{2m+1}(\lambda_{i}-\lambda_{j})^{\frac{N^2-1}{12N}}}.
\end{equation}
\end{theorem}
\begin{proof}
We define the  $\tau$-function by the formula
 (\ref{tau}), where $Y(\lambda)$ is the solution 
 (\ref{solN}) of the R-H problem (\ref{RH1})-(\ref{RH3}).
It follows from the definition that the $\tau$-function does not
depend on the normalisation point $\lb_0$. In order to obtain the explicit
expression of the residue in the r.h.s. of (\ref{tau}), we use
the relation obtained in \cite{kor01}, namely
\begin{equation}
\label{taup1} 
\begin{split}
\frac{\partial}{\partial \lambda_k}\mathrm{log}\,\tau &=\frac12
\res[\lambda=\lambda_k]\,\mathrm{Tr}\,\left(
\frac{\mathrm{d}Y(\lambda)} {\mathrm{d}\lambda} Y(\lambda)^{-1}
\right)^2\\
&=\dfrac{\partial}{\partial
\lambda_k}\log\theta\pq(\boldsymbol{0};\Pi)-\res[P=(\lambda_k,0)]
\left\{\sum_{\substack{r<s\\r,s=1}}^N \dfrac{\d
\omega(P^{(r)},P^{(s)})}{(\d z(P))^2}\right\},
\end{split}
\end{equation}
where $\omega(P,Q)$ is the Bergmann kernel and $P^{(s)}$, is on
the $s$-th sheet of $\mathcal{C}_{N,m}$,   $s=1,\dots,N$. Since
\[
\res[P=(\lambda_k,0)] \left\{\sum_{\substack{r<s\\r,s=1}}^N \dfrac{\d
\omega(P^{(r)},P^{(s)})}{(\d z(P))^2}\right\}=-\dfrac{1}{2}
\res[P=(\lambda_k,0)] \left\{\sum_{\substack{s=1}}^N \dfrac{\d
\omega(P^{(s)},P^{(s)})}{(\d z(P))^2}\right\},
\]
we can write (\ref{taup1}) in the form
\begin{equation}
\label{taup2} \frac12\res[\lb=\lambda_k]\left\{\mathrm{Tr}\,\left(\frac{\d
Y(\lb)}{\d \lb} Y(\lb)^{-1}
\right)^2\right\}=\dfrac{\partial}{\partial
\lambda_k}\log\theta\pq(\boldsymbol{0};\Pi)+\frac12\res[P=(\lambda_k,0)]
\left\{\sum_{\substack{s=1}}^N \dfrac{\d
\omega(P^{(s)},P^{(s)})}{(\mathrm{d}z(P))^2}\right\}.
\end{equation}
From the identity  (\ref{fay212}) and the expansion (\ref{szegoexpan1g})
we express (\ref{taup2}) in the form
\begin{equation}
\nonumber
\begin{split}
\nonumber &\frac12\res[\lb=\lambda_k]\left\{\mathrm{Tr}\left(\frac{\d
Y(\lb)}{\d \lb}Y(\lb)^{-1}\right)^2\right\}=
\dfrac{\partial}{\partial \lambda_k}\log\theta\pq(\boldsymbol{0};\Pi)+
\dfrac{N^2-1}{24N}\res[\lb=\lambda_k]
\left\{\left[
\frac {\mathrm{ d}}{\mathrm{d} \lb}\,\mathrm {log} \frac { p(\lb)}{
q(\lb)}\right] \right\}^2 -\\
&\quad\quad\quad-\dfrac{1}{2}\sum_{i,j=1}^g\frac{\partial^2
}{\partial z_i\partial z_j}\mathrm{log}\,\theta(\boldsymbol{0};\Pi)
\left(\res[P=(\lambda_k,0)]\sum_{\substack{s=1}}^N
\left\{\mathrm{d}v_i(P^{(s)})\mathrm{d}v_j(P^{(s)})\right\}\right).
\end{split}
\end{equation}
Using the relation (\ref{Rauch}),  the heat equation (\ref{heat})
and the property (\ref{thetaderzero}), the above formula
 can be reduced to the form
\begin{equation}
\label{residue2}
\begin{split}
\frac12\res[\lb=\lambda_k]\left\{\mathrm{Tr}\, \left(  \frac{\d Y(\lb)}{\d
\lb}  Y(\lb)^{-1} \right)^2\right\}=&\dfrac{\partial}{\partial
\lambda_k}\log\theta\pq(\boldsymbol{0};\Pi)+
\frac{N^2-1}{24N}\res[\lb=\lambda_k]
 \left\{\left[
 \frac {\mathrm{ d}}{\mathrm{d} \lb}\,\mathrm {log} \frac { p(\lb)}{
q(\lb)}\right] \right\}^2 -\\
-&\dfrac{\partial}{\partial \lambda_k}\log \theta(\boldsymbol{0};\Pi).\\
\end{split}
\end{equation}
From the above formula, we can easily obtain the $\tau$ function
defined in (\ref{tau1}).
\end{proof}
We remark that the formula for the $\tau$-function obtained
in \cite{kor01} for the case of a general $N$-sheeted Riemann surface
reads $\tau(\lb_1,\dots,\lb_{2m+1})=F(\lb_1,\dots,\lb_{2m+1})
\theta\pq(\boldsymbol{0};\Pi)$, where the function $F$
depends  only on the
Bergmann projective connection of the Riemann surface.
In the formula (\ref{tau1}) the term derived  from the  projective connection
is explicitly evaluated.

The set of zeros of the $\tau$-function in the space of singularities of
the R-H problem, that is the set
\[
\{(\lb_1,\dots\lb_{2m+1}),\;\;\lb_i\neq \lb_j\neq \infty,\;\;\;i,j=1,\dots, 2m+1,\;\;\;\tau(\lb_1,\dots,\lb_{2m+1})=0\}
\]
is called the Malgrange divisor $(\theta)$ \cite{ma83}.
From the expression (\ref{tau1}) it follows that the $\tau$-function vanishes
 when
\[
\theta\pq(\boldsymbol{0};\Pi)= 0,
\]
that is when $\Pi\boldsymbol{\delta}+\boldsymbol{\epsilon}\notin(\Theta)$.
Therefore the set of singularities $(\lb_1,\dots,\lb_{2m+1})$ belongs to the
 Malgrange divisor $(\theta)$ if the vector $\boldsymbol{\epsilon}+\boldsymbol{\delta}\Pi$ belongs to the    $(\Theta)$-divisor \cite{kor01}.

The expression for the $\tau$ function can be written in a
different form substituting  the Thomae formula for the
$\theta$-constant.

\begin{theorem}\label{thomaetheorem}
The Thomae-type formula for $\theta(\boldsymbol{0};\Pi)$ reads
\begin{equation}
\label{tome}
\theta^8(\boldsymbol{0};\Pi)=\dfrac{\prod_{s=1}^{N-1}\mathrm {det}
\A^4_s}{(2\pi\i)^{4(N-1)m } }
\prod_{i<j}(\lambda_{2i}-\lambda_{2j})^{2(N-1)}
\prod_{k<l}(\lambda_{2k+1}-\lambda_{2l+1})^{2(N-1)},
\end{equation}
where the matrices $\A_s$, $s=1,\dots, N-1$,  are defined in (\ref{As}).
\end{theorem}
The proof of the theorem is shown in the Appendix.
\begin{remark}
We remark that the  Thomae formulae for  $Z_N$ curve
$y^N=\prod_{k=1}^{mN}(\lambda-\lambda_k) $ , $\lambda_i\neq\lambda_j$
was discovered by Bershadsky and Radul \cite{br88} and Knizhnik \cite{kn89}
and rigorously proved by Nakayashiki
\cite{na97}. The formula (\ref{tome}) is written for  singular $Z_N$ curves
and it does not follow from the  results in \cite{br88,na97}.
\end{remark}

Combining (\ref{tau1}) and (\ref{tome}) we have
\begin{align}
\label{tau2} \tau(\lambda_1,\lambda_2,\dots,\lambda_{2m+1})&=\xi
(2\pi\i)^{(N-1)\frac{m}{2} }
\dfrac{\theta\pq(\boldsymbol{0};\Pi)}{ (\prod_{s=1}^{N-1}\mathrm
{det} \A_s)^{\frac12}}
{\prod\limits_{\substack{i<j\\i,j=1}}^{2m+1}(\lambda_{i}
-\lambda_{j})^{-\frac{N^2-1}{12N}}}\times\\
 &\times
{\prod\limits_{\substack{k<i\\i,k=0}}^m(\lambda_{2k+1}-\lambda_{2i+1})^{
-\frac{(N-1)(N-2)}{12N} }
\prod\limits_{\substack{k<i\\k,i=1}}^m(\lambda_{2k}
-\lambda_{2i})^{-\frac{(N-1)(N-2)}{12N}}}, \notag
\end{align}
where $\xi^8=1$.
For $N=2$ (\ref{tau2})  coincides with the expression
derived in \cite{kk98}.

\section{Example: R-H problem with four singular points} Consider the class of curves (\ref{zvercurve}) for
$m=1$ 
\begin{equation}
{\mathcal C}_{N,1}:
\quad y^N=(\lambda-\lb_1)(\lambda-\lb_3)(\lambda-\lb_2)^{N-1}. \label{curvem1}
\end{equation}
The curve (\ref{curvem1}) is a non-ramified covering over the
hyperelliptic curve
\begin{equation}{\mathcal C}_{\rm hyperel}:
\quad w^2=\xi^{2N}+2(\lb_1+\lb_3-2\lb_2)\xi^N+(\lb_1-\lb_3)^2 .\label{hyperell}
\end{equation}
The coordinates of the cover $\psi: {\mathcal C}_{N,1}\rightarrow
{\mathcal C}_{\rm hyperel}$ are
\begin{equation}
\xi=\frac{y}{\lambda-\lb_2},\quad w= \frac{\lambda^2-2\lb_2\lambda+\lb_2(\lb_1+\lb_3)-\lb_1\lb_3}{\lambda-\lb_2}.
\label{cover}
\end{equation}
The canonical holomorphic differentials of both curves,
\begin{equation}
\d u_k(\lambda,y)=\frac{(\lambda-\lb_2)^{k-1}}{y^k}\d \lambda,
\quad\text{and} \quad
\d U_k(\xi,w)=\xi^{k-1}\frac{\d \xi}{w},\quad k=1,\ldots,N-1
\end{equation}
are linked under the action of $\psi$ as
\begin{equation}
\d u_k(\lambda,y)=N \d U_{N-k}(\xi,w).
\end{equation}
The curve (\ref{hyperell}) admits two automorphisms $f_{\pm}$ of order two
different from the hyperelliptic involution $\mathcal{I}$:
\[
f_+(\xi,w)=\left(\dfrac{(\lb_3-\lb_1)^{\frac2N}}{\xi},(\lb_3-\lb_1)\dfrac{w}{\xi^N}\right),\quad
f_-(\xi,w)=(f_+\circ\mathcal{I})(\xi,w).
\]
We observe that the  automorphism group of the surfaces (\ref{hyperell}) is
\[
\{Id,\mathcal{I},f_+,J\},\quad J(\xi,w)=(\e^{\frac{2\pi\i}{N}},w).
\]
The quotient of the above automorphism  group by $\{Id,\mathcal{Id}\}$ is isomorphic to the dihedral  group $D_N$ of symmetries of the $N$-sided regular polygon.
For $N=3$ this result was pointed out in \cite{bo88a}.

For $N$ odd each of the maps $f_+$ and $f_-$ fixes  exactly two points
of ${\mathcal C}_{\rm hyperel}$. The automorphism
$f_+$ fixes  the two points were $\xi=(\lb_3-\lb_1)^{\frac1N}$ while
$f_-$ fixes the two points were $\xi=-(\lb_3-\lb_1)^{\frac1N}$.
According to Riemann-Hurwitz formula the quotient surfaces:
\[
\mathcal{C}_{\pm}={\mathcal C}_{\rm hyperel}/\{Id,f_{\pm}\}
\]
have genus equal to $\frac{N-1}{2}$.

For $N$ even, the map $f_+$ fixes  the four points were $\xi=\pm (\lb_3-\lb_1)^{\frac1N}$ while
$f_-$ has no fixed points. Therefore the quotient
surfaces
$\mathcal{C}_{\pm}={\mathcal C}_{\rm hyperel}/\{Id,f_{\pm}\}$
have genus equal to $\frac{N}{2}-1$ and  $\frac{N}{2}$ respectively.

The Jacobian varieties  $\mathrm{Jac}({\mathcal C}_{N,1})$
and $ \mathrm{Jac}(\mathcal{C}_{+})\times \mathrm{Jac}(\mathcal{C}_{-})$
are complex tori of dimension $N-1$. Following \cite{earleXX}
it is possible to show that
these tori are isomorphic.
From the   factorisation of the Jacobian variety $\mathrm{Jac}
({\mathcal C}_{\rm hyperel})$,
the   $\theta$-functions defined on the surface of genus $N-1$
 can be expressed in terms of $\theta$-functions
defined on two surfaces of genus $\frac{N-1}{2}$ for $N$ odd and of genus
$N/2-1$ and $N/2$  for $N$ even. The procedure for obtaining the period matrix
of the two quotient surfaces and the factorisation of the $\theta$-function
is illustrated, for automorphisms of order two, in \cite{farkas74} and
\cite{fa73}.

We are going to study in detail  the case  $N=3$ when  $\theta$-functions
decomposes as  product of Jacobi's $\vartheta$-functions.

\begin{remark} We remark that the curves of the form (\ref{CNM}) are
hyperelliptic only in the case $m=1$.
\end{remark}

\subsection{Decomposition of two-dimensional
Jacobian to elliptic curves: $N=3$ and $m=1$ }

\noindent
We restrict  ourselves to the curve
\begin{equation} {\mathcal C}_{3,1}:
\quad y^3=(\lambda-\lb_1)(\lambda-\lb_3)(\lambda-\lb_2)^2.
\label{curvem13}
\end{equation}
Its  holomorphic differentials are
\[
\d u_1(\lambda,y)=\dfrac{\d \lambda}{y},\quad
\d u_2(\lambda,y)=\dfrac{(\lambda-\lb_2)\d \lambda}{y^2}.
\]
The  matrices of $\alpha$ and $\beta$-periods in the
 homology basis given on the Figure 3, are
\begin{equation}
\label{AA1}
{\mathcal A}=\left(\begin{array}{cc}
{\mathcal A}_1&\rho^2{\mathcal A}_1\\
{\mathcal A}_2&\rho{\mathcal A}_2
\end{array}\right),\quad  {\mathcal B}=\left(\begin{array}{cc}
{\mathcal B}_1&-\rho{\mathcal B}_1\\
{\mathcal B}_2&-\rho^2{\mathcal B}_2
\end{array}\right),\;\;\rho=\e^{\frac{2\pi\i}{3}},
\end{equation}
where $\oint_{\alpha_1}\d u_i=\mathcal{A}_i,\,
\oint_{\beta_1}\d u_i=\mathcal{B}_i$,
$i=1,2$. Evaluating explicitly the integrals we obtain
\begin{equation}
\label{AA2}
\begin{split}
\mathcal{A}_1&=\int_{\alpha_1}\d u_1=(1-\rho^2)\int\limits^{\lb_2}_{\lb_1}
\frac{\d \xi }{\sqrt[3]
{(\xi-\lb_1)(\lb_3-\xi)(\lb_2-\xi)^2}}=\dfrac{(1-\rho^2)}
{\sqrt[3]{\lb_3-\lb_1}}\int\limits_0^1\frac{\d \xi
}{\sqrt[3]{\xi(1-\xi)^2(1-\frac{\lb_2-\lb_1}{\lb_3-\lb_1}\xi)}}=\\
&=\dfrac{2\pi}{\sqrt 3}\dfrac{(1-\rho^2)}
{\sqrt[3]{\lb_3-\lb_1}}F\left(\frac13,\frac23,1;t\right)=
\dfrac{2\pi}{\sqrt 3}\dfrac{(1-\rho^2)}
{\sqrt[3]{\lb_3-\lb_1}}F\left(\frac23,\frac13,1;t\right)=
-\rho^2\mathcal{A}_2\sqrt[3]{\lb_3-\lb_1},
\end{split}
\end{equation}
where  $F(a,b,c,\lb)$ is the standard hypergeometric function and 
\begin{equation}
\label{t}
t=\frac{\lb_2-\lb_1}{\lb_3-\lb_1},
\end{equation}
and, analogously,  
\begin{equation}
\label{AA3}
\mathcal{B}_1=\dfrac{2\pi\i}{\sqrt[3]{\lb_3-\lb_1}} F\left(\frac13,\frac23,1;1-t\right)=\mathcal{B}_2\sqrt[3]{\lb_3-\lb_1}.
\end{equation}

From (\ref{AA1}) and (\ref{AA2}) the normalized holomorphic
differentials read
\begin{equation}
\label{dv}
\d v_1=\dfrac{1}{\mathcal{A}_1(1-\rho)}(\d u_1+\sqrt[3]{\lb_3-\lb_1}\d u_2),\quad \d v_2=\dfrac{1}{\mathcal{A}_1(1-\rho^2)}(\rho\d u_1+\sqrt[3]{\lb_3-\lb_1}\d u_2).
\end{equation}
From (\ref{AA1})-(\ref{AA3}),  the Riemann $\Pi$-matrix has the form
\begin{equation}
\Pi=\left(\begin{array} {cc} 2\T& \T\\ \T&2 \T
 \end{array}\right),\quad \mathrm{Im} \,\T>0,
\label{tauhutchinson}
\end{equation}
where
\begin{equation}
\label{T}
\T=\dfrac{1}{1-\rho}\dfrac{\mathcal{B}_1}{\mathcal{A}_1}=\dfrac{\i}{\sqrt 3}
\dfrac{F\left(\frac13,\frac23,1;\frac{\lb_3-\lb_2}{\lb_3-\lb_1}\right)}
{F\left(\frac13,\frac23,1;\frac{\lb_2-\lb_1}{\lb_3-\lb_1}\right)}.
\end{equation}

The curve $\mathcal{C}_{3,1}$ covers the hyperelliptic curve of genus two
\begin{equation}{\mathcal C}_{\rm hyperel}:
\quad w^2=\xi^{6}+2(\lb_1+\lb_3-2\lb_2)\xi^3+(\lb_1-\lb_3)^2. \label{hyperell2}
\end{equation}
 Bolza \cite{bo88a}, Igusa \cite{ig60} and Lange \cite{lange76}
 have classified the curves of genus two with automorphism and in
particular the curves with involutions. The moduli of such a curves
describe a $2$-dimensional sub-variety of the moduli space of curves 
of genus two.  The automorphism group of the
curve (\ref{hyperell2}) is generated by  $\{Id,\,\mathcal{I},\,J,\,\,f_+\}$
 where  now $J(\xi,w)=(\e^{\frac{2\pi\i}{3}}\xi,w)$ and
\[
f_+(\xi,w)=\left(\dfrac{(\lb_3-\lb_1)^{\frac23}}{\xi},(\lb_3-\lb_1)\dfrac{w}{\xi^3}\right).
\]
The reduced group of automorphism is isomorphic to the dihedral group $D_3$ \cite{bo88a}
and such curves describes a one-dimensional variety in the moduli space 
of curves of genus two.
The quotient surfaces $\mathcal{C}_{\pm}={\mathcal C}_{\rm hyperel}/\{Id,f_{\pm}\}$ with $f_-=f_+\circ\mathcal{I}$, are elliptic surfaces.

We are going to construct the covering maps $\phi_{\pm}:=h_{\pm}\circ \psi$
\cite{fa73},\cite{farkas74}:
\[
\mathcal{C}_{N,1}\stackrel{\psi} {\longrightarrow}{\mathcal C}_{\rm hyperel}
\stackrel{h_{\pm}} {\longrightarrow}\mathcal{C}_{\pm}
\]
where $\psi$ is defined in (\ref{cover}).
 Let   $(a_1,a_2;b_1,b_2)$ be
the canonical homology basis defined on ${\mathcal C}_{\rm hyperel}$
so  that $f_+(a_1)=a_2$ and $f_+(b_1)=b_2$. Then  $f_-(a_1)=-a_2$ and $f_-(b_1)=-b_2$.
It is easy to verify that  $a_1=\psi(\alpha_1),\;\;a_2=-\psi(\alpha_1)-\psi(\alpha_2)$
and  $b_1=\psi(\beta_1)-\psi(\beta_2)$, $b_2=-\psi(\beta_2)$,  where $\{\alpha_1,\alpha_2,\beta_1,\beta_2\}$ is  the canonical homology basis defined on $\mathcal{C}_{3,1}$.
We fix $h_{\pm}(a_1)=\alpha_{\pm}$ and $h_{\pm}(b_1)=\beta_{\pm}$
where $\{\alpha_{\pm}, \beta_{\pm}\}$ is the canonical homology basis on
$\mathcal{C}_{\pm}$ respectively.
It follows that $h_{\pm}(a_2)=\pm\alpha_{\pm}$ and $h_{\pm}(b_2)=\pm\beta_{\pm}$ so that
\begin{align}
\label{Nor1}
&\phi_+(\alpha_1)=\alpha_+,\;\phi_+(\alpha_2)=-2\alpha_+,\;\;\phi_+(\beta_1)=0,\;\;\phi_+(\beta_2)=-\beta_+,\\
\label{Nor2}
&\phi_-(\alpha_1)=\alpha_-,\;\phi_+(\alpha_2)=0,\;\;\phi_+(\beta_1)=0,\;\;\phi_+(\beta_2)=\beta_-.
\end{align}
The action of $f_{\pm}$ on the holomorphic differentials $\d U_{k}$, $k=1,2$, of the curve ${\mathcal C}_{\rm hyperel}$ is given by
\[
f_{\pm}(\d U_1(\xi,w))=\dfrac{\mp 1}{\sqrt[3]{\lb_3-\lb_1}} \d U_2(\xi,w),\quad f_{\pm}\d U_2(\xi,w))=\mp\sqrt[3]{\lb_3-\lb_1} \d U_1(\xi,w).
\]
Therefore the differentials $\sqrt[3]{\lb_3-\lb_1}\d U_1\mp\d U_2$ are invariant under the action
of $f_{\pm}$. It follows that the differentials
$\sqrt[3]{\lb_3-\lb_1}\d U_1\mp\d U_2=\dfrac{1}{3}(\sqrt[3]{\lb_3-\lb_1}\d u_2\mp \d u_1) $ project to the holomorphic
differentials  of the curve $\mathcal{C}_{\pm}$.
From the above considerations and from (\ref{dv}) we conclude that
\begin{equation}
\label{dvpm}
\d v_+=\d v_1-2\d v_2,\quad \d v_-=\d v_1,
\end{equation}
where $\d v_{\pm}$ are  the holomorphic  differentials  of the curve $\mathcal{C}_{\pm}$.
From the relations (\ref{Nor1}) and (\ref{Nor2}) we deduce that
 $\d v_{\pm}$ are the normalized holomorphic differentials
of $\mathcal{C}_{\pm}$ with periods
\[
\oint_{\beta_-}\d v_-=\T,\quad \oint_{\beta_+}\d v_+=3\T.
\]
Therefore the elliptic curves  $\mathcal{C}_{\pm}$  are $3$-isogenous. 
Let us write the equations of the two elliptic curves  $\mathcal{C}_{\pm}$ in the Legendre form:
\[
{\mathcal C}_{\pm}:\quad
z_{\pm}^2=\eta(1-\eta)(1-k_{\pm}^2\eta).
\]
The Jacobi's moduli $k_{\pm}$ are related by a third order transformation and
are parametrised as
\begin{equation} k^2_-=\frac{1}{16p}(p+1)^3(3-p),\quad
k_+^2=\frac{1}{16 p^3}(p+1)(3-p)^3,
\label{moduli}\end{equation}
where the  parameter $p$ can  be expressed in terms of $\vartheta$-constants
(see e.g. \cite{lawd89} )
\begin{equation}
\label{ppar}
p=\frac{3\vartheta_3^2(0; 3\T)}{\vartheta_3^2(0; \T)}.
\end{equation}
The holomorphic differentials $\d v_{\pm}$ reads
\begin{align}
\label{dvpm1}
&\d v_{\pm}=\dfrac{1}{4 K_{\pm}}\dfrac{\d\eta}{z_{\pm}},\;\;\;K_{+}=\dfrac{\pi}{2}\vartheta_3^2(0;
3\T),\quad K_{-}=\dfrac{\pi}{2}\vartheta_3^2(0;\T).
\end{align}
From the relation (\ref{dvpm}) and  (\ref{dvpm1})  we  construct 
the coordinates of  the covers
$\phi_{\pm}:\mathcal{C}_{3,1}\rightarrow\mathcal{C}_{\pm}$
\begin{equation*}
\begin{split}
&\eta= \dfrac{p^2(y-(\lambda-\lb_2)\sqrt[3]{\lb_3-\lb_1})^2+3(y+(\lambda-\lb_2)\sqrt[3]{\lb_3-\lb_1})^2}{p^2k_+^2(y-(\lambda-\lb_2)\sqrt[3]{\lb_3-\lb_1})^2+3k_-^2(y+(\lambda-\lb_2)\sqrt[3]{\lb_3-\lb_1})^2},\\
&z_{\pm}=\dfrac{\mathcal{A}_1(1\pm\rho)}{4 K_{\pm}}
\dfrac{y^2}{y\mp(\lb-\lb_2)\sqrt[3]{\lb_3-\lb_1}}\dfrac{d\eta}{d\lambda}.
\end{split}
\end{equation*}
Constructing the covering maps $\phi_{\pm}=h_{\pm}\circ \psi$
as in \cite{bo88a}, \cite{bbeim94},  by mapping the  branch points
\[
\xi_0=\sqrt[3]{2\lb_2-\lb_1-\lb_3+2\sqrt{\lb_2^2-\lb_2(\lb_1+\lb_3)+\lb_1\lb_3}},\;\rho\xi_0,\quad\rho^2\xi_0,\;\;\dfrac{1}{\xi_0},\;\;
\dfrac{\rho}{\xi_0},\;\;\dfrac{\rho^2}{\xi_0},
\]
of the hyperelliptic curve $\mathcal{C}_{hperel}$  to $(0,0),\;(\infty,\infty),\;(1,0)\in\mathcal{C}_{\pm}$ according to the rule
\begin{equation}
\label{ximage}
((\xi_0)^{\pm 1},0)\rightarrow (0,0),\quad ((\rho\xi_0)^{\pm 1},0)
\rightarrow(\infty,\infty),\quad ((\rho^2\xi_0)^{\pm 1},0)\rightarrow (1,0),
\end{equation}
 we derive the algebraic dependence of the parameter $p$ on $t=\frac{\lb_2-\lb_1}{\lb_3-\lb_1}$,
that is
\begin{equation}
\label{tmodular}
  t=\frac{p^2(p^2-9)^2}{(p^2+3)^3},
\end{equation}
and we deduce that
\begin{equation}
\label{A12}
\mathcal{A}_1=\dfrac{2\pi}{\sqrt 3}\dfrac{(1-\rho^2)}{\sqrt[3]{\lb_3-\lb_1}}F\left(\frac13,\frac23,1;\frac{\lb_2-\lb_1}{\lb_3-\lb_1}\right)=\frac{\pi}{2}\dfrac{(1-\rho^2)}{\sqrt[3]{\lb_3-\lb_1}}\dfrac{p^2+3}{ p^{\frac{3}{2}} }
F\left(\frac12,\frac12,1;k^2_+\right).
\end{equation}
Namely the holomorphic integrals $\mathcal{A}_1$ and
$\mathcal{B}_1$
defined   on the trigonal curve $\mathcal{C}_{3,1}$ are
reducible to elliptic integrals.

The equality  (\ref{A12}) can be interpreted as the
hypergeometric equalities of Goursat  \cite{goursat81} employed
by Harnad and McKay \cite{hm00} to describe the  solution of
the Halphen system in terms of automorphic functions
for groups  commensurable with the modular group.  
We are pointing here  the  link between higher order Goursat hypergeometric identities 
and the reduction of Abelian integrals to lower genera.
Indeed the equality (\ref{A12})  follows  from the superposition 
of the transformations \cite{goursat81},  (126), p. 140, (118), p. 138,  (35), \text{p}. 119.

\begin{remark}
The parameter $\T$ defined in (\ref{T}) reads
\begin{equation}
\label{TTmodular}
\T=\dfrac{\i}{\sqrt{3}}
\dfrac{F\left(\frac13,\frac23,1;1-t\right)}
{F\left(\frac13,\frac23,1;t\right)},\quad t=\frac{\lb_2-\lb_1}{\lb_3-\lb_1}
\end{equation}
where $F\left(\frac13,\frac23,1;1-t\right)$
and $F\left(\frac13,\frac23,1;t\right)$
are two independent solutions of the Gauss hypergeometric equation
\[
t(1-t) F'' +(1-2t) F' -\frac{2}{9} F=0.\]
For $\T$ belonging to  Siegel half-space ${\mathcal H}_1$
modulo the sub-group $\Gamma_0(3)$ \cite{shiga88},\cite{gonrod99},
the  expression (\ref{TTmodular}) is invertible
and the inverse function is given in (\ref{tmodular}) and reads
\begin{equation}
\label{tmodular2}
  t=27\vartheta_3^4(0; 3\T)
\dfrac{(\vartheta_3^4(0; 3\T)-\vartheta_3^4(0; \T))^2}{(3\vartheta^4_3(0; 3\T)+\vartheta^4_3(0; \T))^3}.
\end{equation}
We recall that the  sub-group  $\Gamma_0(3)$ of the modular group is defined by
the matrices $\left(\begin{array} {cc} a& b\\ c&d
 \end{array}\right)\in SL(2,\Z)$ with $c=0\;\mbox{mod}\,\,3$. 
One can show, by comparing $q$-expansions,  that 
\[
t(T/2)f(T)=1,
\]
where $f(T)$ is the automorphic function of 
$\Gamma_0(3)$ found in \cite{hm00}, Table on p. 12,
\[
f(T)= 1+\frac{1}{27}\frac{\eta(T)}{\eta(3T)},
\]
and $\eta$ is the Dedekind $\eta$-function \cite{fo29}.
Alternatively, one can express $t$ in terms of $\theta$-functions with
 $1/3$-characteristics as
\[
t=1-\left(\dfrac{\theta\left[^{0\;0}_{\frac23\frac13}\right](\boldsymbol{0};\Pi)}
{\theta(\boldsymbol{0};\Pi)}\right)^3.
\]
The equivalence of the above expression and (\ref{tmodular2}) involves non-trivial
$\theta$-function identities.

The function $t(\T)$ in (\ref{tmodular2})  gives   a solution of
the Schwarzian equation \cite{Neh52,in56}
\[
\{t,\T\}+\dfrac{\dot{t}^2}{2}V(t)=0,\quad 
\]
where $\{\;,\;\}$ is the Schwarzian derivative (\ref{schder})
 and the potential $V(t)$
is given by (see for example \cite{Neh52})
\[
V(t)=\dfrac{1-\beta^2}{t^2}+\dfrac{1-\gamma^2}{(t-1)^2}+\dfrac{\beta^2+\gamma^2-\alpha^2-1}{t(t-1)},\quad \alpha=\frac{1}{3},\;\beta=\gamma=0.
\]
It follows that the functions
\[
\omega_1=-\dfrac{1}{2}\dfrac{d}{d\T}\ln\dfrac{\dot{t}}{t(t-1)},\quad
\omega_2=-\dfrac{1}{2}\dfrac{d}{d\T}\ln\dfrac{\dot{t}}{t-1},\quad
\omega_3=-\dfrac{1}{2}\dfrac{d}{d\T}\ln\dfrac{\dot{t}}{t},
\]
solve the general Halphen system
\begin{align*}
&\dot{\omega_1}=\omega_2\omega_3-\omega_1(\omega_2+\omega_3)+R,\\
&\dot{\omega_2}=\omega_1\omega_3-\omega_2(\omega_1+\omega_3)+R,\\
&\dot{\omega_3}=\omega_1\omega_2-\omega_1(\omega_1+\omega_2)+R,
\end{align*}
where
\[
R=\alpha^2(\omega_1-\omega_2)(\omega_3-\omega_1)+\beta^2(\omega_2-\omega_3)(\omega_1-\omega_2)+\gamma^2(\omega_3-\omega_1)(\omega_2-\omega_3).
\]
When $R=0$ the above system coincides with the classical Halphen system.
The solution of the classical and general Halphen system has been
investigated by many authors \cite{hm00,AbCh96,Hi98,tak92,chha03}.
The expression (\ref{tmodular2}) gives a
formula for the solution of the general Halphen system with parameters
$\alpha=\frac13$, $\beta=\gamma=0$ equivalent to the
one derived in \cite{hm00}.
\end{remark}

In the following we derive the decomposition of the genus two
$\theta$-functions
in terms of Jacobi's $\vartheta$-functions.

\begin{lemma}
The $\theta$-function of the curve $\mathcal C_{3,1}$
is decomposed in terms of Jacobi's $\vartheta$-functions of the  curves
$\mathcal{C}^{\pm}$ as
\begin{equation}
\label{decompose}
\theta\pq(z_1,z_2;\Pi)=\e^{\pi\i\langle\boldsymbol{\delta},
\Pi\boldsymbol{\delta}\rangle+2\pi\i\langle\boldsymbol{z}+
\boldsymbol{\epsilon},\boldsymbol{\delta}\rangle}
\left[\vartheta_3(e_1;6\T)
\vartheta_3(e_2;2\T)+\vartheta_2(e_1;6\T)
\vartheta_2(e_2;2\T)\right]
\end{equation}
where
\[
e_1=z_1+z_2+\epsilon_1+\epsilon_2+3\T(\delta_1+\delta_2),\;\;
e_2=z_1-z_2+\epsilon_1-\epsilon_2+\T(\delta_1-\delta_2).
\]
\end{lemma}
\begin{proof}
By definition of $\theta$-function we obtain
\begin{equation*}
\begin{split}
\theta\pq(z_1,z_2;\Pi)&=\e^{\pi\i\langle\boldsymbol{\delta},
\Pi\boldsymbol{\delta}\rangle+2\pi\i\langle\boldsymbol{z}
+\boldsymbol{\epsilon},\boldsymbol{\delta}\rangle}
\sum_{n_1,n_2\in\Z}\exp\left[\pi\i\left
(2\T(n_1^2+n_2^2+n_1n_2)+\right.\right.\\
&\left.\left.+3\T(\delta_1+\delta_2)(n_1+n_2)
+\T(\delta_1-\delta_2)
(n_1-n_2)+2(z_1+\epsilon_1)n_1+2(z_2+\epsilon_1)n_2\right)\right].
\end{split}
\end{equation*}
Substituting in the above $m_1=n_1+n_2$ and $m_2=n_1-n_2$ where $m_i=2k_i+r$, $\,i=1,2$, $r=0,1$ we obtain
\begin{equation*}
\begin{split}
\theta\pq(z_1,z_2;\Pi)&=\e^{\pi\i\langle\boldsymbol{\delta},
\Pi\boldsymbol{\delta}\rangle+2\pi\i\langle\boldsymbol{z}+
\boldsymbol{\epsilon},\boldsymbol{\delta}\rangle}\sum_{r=0,1}
\sum_{k_1,k_2\in\Z}\exp\left[\pi\i\left
(6\T(k_1+\frac{r}{2})^2+2\T(k_2+
\frac{r}{2})^2+\right.\right.\\
&+6\T(\delta_1+\delta_2)(k_1+\frac{r}{2})
+2\T(\delta_1-\delta_2)(k_2+\frac{r}{2})+2(k_1+\frac{r}{2})(z_1+\epsilon_1+z_2+\epsilon_2)+\\
&\left.\left.+2(k_2+\frac{r}{2})(z_1-z_2+\epsilon_1
-\epsilon_2)\right)\right]=\\
&=\e^{\pi\i\langle\boldsymbol{\delta},\Pi
\boldsymbol{\delta}\rangle+2\pi\i\langle\boldsymbol{z}
+\boldsymbol{\epsilon},\boldsymbol{\delta}\rangle}
\sum_{k=2}^3\vartheta_k(z_1+z_2+\epsilon_1+\epsilon_2
+3\T(\delta_1+\delta_2);\,6\T)\times\\
&~~~~~~~~~~~~~~~~~~~~~~~~~~~~~~~~~~~~~~~~~
\times\vartheta_k(z_1-z_2+\epsilon_1-\epsilon_2
+\T(\delta_1-\delta_2);\,2\T),
\end{split}
\end{equation*}
which is equivalent to (\ref{decompose}).
\end{proof}

\subsection{Solution of the $3\times 3$ matrix R-H problem with four singular points}
\noindent
Let us consider the R-H problem with  four singular points
$\lambda_1,\,\lambda_2,\lambda_3,\,\lambda_4=\infty$ and with monodromy matrices
\begin{align}
\label{M11}
M_1&=\begin{pmatrix}
0&0&c_1\\
\dfrac{c_2}{c_1}&0&0\\
0&\frac{1}{c_2}&0
\end{pmatrix},\quad
M_{2}=\begin{pmatrix}
0&\dfrac{c_1d_1}{c_{2}}&0\\
0&0&c_{2}d_{2}\\
\dfrac{1}{c_1 d_1d_{2}}&0&0
\end{pmatrix},\quad\\
 M_{3}&=\begin{pmatrix}
0&0&d_1d_{2}\\
\dfrac{1}{d_{1}}&0&0\\
0&\dfrac{1}{d_{2}}&0
\end{pmatrix},\quad M_{\infty}=\begin{pmatrix}
0&1&0\\
0&0&1\\
1&0&0
\end{pmatrix},\notag
\end{align}
where $c_1, c_2, d_1,d_2$ are non-zero constants.
The solution of this R-H problem  is given in (\ref{solN})  and read
\begin{equation}
\label{solN3}
Y_{rs}(\lb)=X_{rs}(\lb)
\dfrac{\theta\pq\left(\int\limits_{P_0^{r}}^{P^s}\d \boldsymbol{v};\Pi\right)}
{\theta\left(\int\limits_{P_0^{r}}^{P^s}\d \boldsymbol{v};\Pi\right)}
\dfrac{\theta(\boldsymbol{0};\Pi)}{\theta\pq(\boldsymbol{0};\Pi)},
\end{equation}
with  $\d \boldsymbol{v}$ and $\Pi$ defined in (\ref{dv}) and
(\ref{tauhutchinson}) respectively and
\[
\delta_i=\dfrac{1}{2\pi \i}\log d_i,\quad\epsilon_i=\dfrac{1}{2\pi \i}\log c_i, \;\;i=1,2.\]
The entries of the matrix $X(\lb)$ in the expression (\ref{solN3}) read
\begin{align*}
X_{rs}(\lb)=&\dfrac{1}{3}\left(\e^{2\pi\i\frac{s-r}{3}}\sqrt[3]
{\dfrac{(\lb-\lb_1)(\lb-\lb_3)}{\lb-\lb_2}}\sqrt[3]{\dfrac{\lb_0-\lb_2}{(\lb_0-\lb_1)
(\lb_0-\lb_3)}}+1+\right.\\
&\left.\e^{-2\pi\i\frac{s-r}{3}}\sqrt[3]
{\dfrac{\lb-\lb_2}{(\lb-\lb_1)(\lb-\lb_3)}}\sqrt[3]{\dfrac{(\lb_0-\lb_1)(\lb_0-\lb_3)}
{\lb_0-\lb_2}}\right).
\end{align*}
Using the reduction formula  (\ref{decompose})  it is possible
to write the  solution (\ref{solN3}) in terms of Jacobi's $\vartheta$-functions
\begin{equation}
\label{reductionRH}
\begin{split}
&Y_{rs}(\lb)=X_{rs}(\lb)
\dfrac{\e^{2\pi\i\langle\boldsymbol{z},\boldsymbol{\delta}\rangle}
\vartheta_3(0;6\T)\vartheta_3(0;2\T)
+\vartheta_2(0;6\T)\vartheta_2(0;2\T)}{
\sum_{k=2}^3(\vartheta_k(\dfrac{1}{2\pi\i}\log\dfrac{c_1}{c_2^2}
-\dfrac{3\T}{2\pi\i}\log d_2;\;6\T)
\vartheta_k(\dfrac{1}{2\pi\i}\log c_1+\dfrac{\T}{2\pi\i}
\log d_1^2 d_2;\;2\T))}\\
&\times\sum_{k=2}^3\vartheta_k
\left(\int\limits_{\phi_+(P_0^r)}^{\phi_+(P^s)}
\d v_++\dfrac{1}{2\pi\i}\log\dfrac{c_1}{c_2^2}
-\dfrac{3\T}{2\pi\i}\log d_2;\;6\T\right)
\vartheta_k\left(\int\limits_{\phi_-(P_0^r)}^{\phi_-(P^s)}
\d v_-+\dfrac{1}{2\pi\i}\log c_1+\dfrac{\T}{2\pi\i}\log d_1^2 d_2;\;2\T\right)\\
&\times\left[
\sum_{k=2}^3\vartheta_k\left(\int\limits_{\phi_+(P_0^r)}^{\phi_+(P^s)}
\d v_+;\;6\T\right)\vartheta_k\left(\int\limits_{\phi_-(P_0^r)}^{\phi_-(P^s)}\d v_-;\;2\T\right)\right]^{-1},
\end{split}
\end{equation}
where $\d v_{\pm}$ have been defined in (\ref{dvpm}),  the covering maps $\phi_{\pm}$ have been described in the previous section and
\[
z_1=\int\limits_{\phi_-(P_0^r)}^{\phi_-(P^s)}\d v_,\quad z_2=\dfrac{1}{2}\int\limits_{\phi_-(P_0^r)}^{\phi_-(P^s)}\d v_--\dfrac12\int\limits_{\phi_+(P_0^r)}^{\phi_+(P^s)}\d v_+.
\]
The expression (\ref{reductionRH}) has been obtained after
performing a modular transformation of the $\theta$-function under
the action of the following symplectic transformation
\[
\begin{pmatrix}
C^2&0_2\\
0_2&C^t
\end{pmatrix},\quad
C=\begin{pmatrix}
0&1\\-1&-1\end{pmatrix},\;\;C^3=1,
\]
induced by  the automorphism  $J^{2}$. 


\subsection{Solution of the $3\times 3$  Schlesinger system }
From  (\ref{Akss}) and (\ref{Akrs}) we obtain the following expressions for the 
 solution   of the $3\times 3$ Schlesinger system (\ref{Schlesinger})
\begin{align*}
&(A_k)_{ss}=(\lb_0-\lb_k)^2\dfrac{\partial}{\partial \lb_k}\left(
\sum_{l=1}^{2}\dfrac{\partial}{\partial z_l}\log \theta\pq
\left(\boldsymbol{0};\Pi\right)\left.\dfrac{\d v_l(P)}{\d z(P)}
\right|_{P=P^{(s)}_0}\right),
\quad s=1,2,3,\\
&(A_k)_{rs}=(s-r)(-1)^{(s-r)}
\dfrac{\sqrt{3}\i}{9}(\lb_0-\lb_k)^2
\dfrac{\partial}{\partial\lb_k}
\left[\left(\sum_{l=1}^3\dfrac{(-)^{l}}{\lb_0-\lb_l}\right)
\dfrac{\theta\pq
\left(\boldsymbol{e}_{sr};\Pi
\right)}{\theta\left(\boldsymbol{e}_{sr};\Pi\right)}
\dfrac{\theta(\boldsymbol{0};\Pi)}{\theta\pq(\boldsymbol{0};\Pi)}\right]
\end{align*}
where $s\neq r$, $r,s=1,2,3$ and the vectors 
\[\boldsymbol{e}_{sr}
=\int\limits_{P_0^{(r)}}^{P_0^{(s)}}\d\boldsymbol{v},\] 
satisfy the relation $\boldsymbol{e}_{12}+
\boldsymbol{e}_{23}+\boldsymbol{e}_{31}=0$.

The matrix $A_{\infty}$ is determined from the condition
\[
A_{\infty}=-A_1-A_2-A_3.
\]
The $\tau$ function corresponding to $3\times 3$  solution
 of the Schlesinger system can be written  in terms of
Jacobi's $\vartheta$-functions.
According to the formula (\ref{tau2}), (\ref{decompose}) and (\ref{A12})
we obtain
\begin{equation}
\begin{split}
 \tau(\lb_1\lb_2,\lb_3)&
=\left(\dfrac{\lb_1-\lb_3}{(\lb_1-\lb_2)(\lb_2-\lb_3)}\right)^{\frac29}\e^{\pi\i\langle\boldsymbol{\delta},\Pi\boldsymbol{\delta}\rangle+2\pi\i\langle\boldsymbol{\epsilon},\boldsymbol{\delta}\rangle}
\times\\
&\times\dfrac{\sum_{k=2}^3\vartheta_k(\dfrac{1}{2\pi\i}\log c_1c_2+\dfrac{3\T}{2\pi\i}
\log d_1d_2;\,6\T)\vartheta_k(\dfrac{1}{2\pi\i}\log \dfrac{c_1}{c_2}+\dfrac{\T}{2\pi\i}
\log \dfrac{d_1}{d_2};\,2\T)}{\vartheta_3(0;\,6\T)\vartheta_3(0;\,2\T)+\vartheta_2(0;\,6\T)\vartheta_2(0;\,2\T) }.
\end{split}
\end{equation}
If the non-singular characteristics $\boldsymbol{\epsilon}$ and $\boldsymbol{\delta}$ 
are shifted by $1/3$ periods, the corresponding constants
$d_i$ and $c_i$ are shifted by third roots of unity and
the corresponding $\tau$ function is expressed by the above
formula with the Jacobi's $\vartheta$-function shifted by $1/6$ periods.
As an example we consider the shift
\[
\boldsymbol{\epsilon}+\boldsymbol{\delta}\Pi\rightarrow\boldsymbol{\epsilon}+\boldsymbol{\delta}\Pi+\mathfrak{A}(P_2+P_3-2P_2),\;\;P_2=(\lb_2,0),\;P_3=(\lb_3,0),
\]
where the vector  $\mathfrak{A}(P_2+P_3-2P_2)=\left(-\frac23,\frac13\right)\Pi$ is
non-singular. The corresponding constants $d_i$ and $c_i$, $i=1,2$ transform to
\[
d_1\rightarrow d_1\e^{-\frac{4\pi\i}{3}},\quad d_2\rightarrow d_2\e^{\frac{2\pi\i}{3}},\;\;c_i\rightarrow c_i,\;i=1,2.
\]
If $\{M_1,M_2,M_3,M_{\infty}\}$ are the monodromy matrices 
associated to the characteristics $\boldsymbol{\epsilon}$ and 
$\boldsymbol{\delta}$, the monodromy matrices associated
to the characteristics $\boldsymbol{\epsilon}$ and 
$\boldsymbol{\delta}+(-\frac23,\frac13)$ are
\[
\{M_1,\e^{-\frac43\pi\i}M_2,\e^{-\frac23\pi\i}M_3,M_{\infty}\}.
\]
\section{Conclusion}
In this manuscript we have studied  the solution of the R-H problem for
a particular class of quasi-permutation monodromy matrices and for a 
given set of $2m+2$ singular points. 
The dimension of the space of monodromy matrices is $2m(N-1)$. 
Inspired by \cite{kor01} we have solved the problem using the Szeg\"o
kernel of a Riemann surface. The monodromy of 
 Riemann surface is  obtained from the
reduction of the monodromy representation of the R-H problem to a permutation representation
of the symmetric group $S_N$.
The form of the monodromy matrices considered, is such that the permutation
representation obtained, generates the cyclic subgroup $Z_N$ of the permutation
group. For this reason the family of Riemann surfaces $\mathcal{C}_{N,m}$
  have $Z_N$ symmetry and  genus $N(m-1)$.
The symmetry in our problem  has enabled us to write the
entries of the $N\times N$ matrix solution of the R-H problem
as a product of an algebraic function and
$\theta$-quotients. The algebraic function turns out
to be related to the Szeg\"o kernel with zero characteristics.
The $2N(m-1)$ monodromy parameters are in one to one
correspondence with the $2N(m-1)$ characteristics of the $\theta$-quotients.
The R-H problem is solvable if the corresponding
 characteristics are non-singular.

We have studied  the  set of non-singular divisors supported
on the branch points and we have shown that the corresponding
non-singular characteristics are rational numbers
of the form $k/N$, $k=1,\ldots,N-1$.
We have shown that the solution of the R-H problem
for reducible monodromy representation is
expressed in terms of $\theta$- quotients with $k/N$ characteristics.
Furthermore we have shown that if two monodromy representations are
equivalent up to multiplication by $N$-th roots of unity,
then the corresponding solutions of the R-H problem have
characteristics that differ by $1/N$.

From the solution of the R-H problem we have straightforwardly
obtained a particular solution of the Schlesinger equations.
The Jimbo-Miwa-Ueno $\tau$-function
corresponding to this particular solution of the Schlesinger system is derived
in a complete form by the explicit evaluation of the projective connection
associated to the Riemann surfaces $\mathcal{C}_{N,m}$.

Finally we have investigated in detail the case of a $3\times 3$ matrix
R-H problem with four singular points,
 $\lb_1,\;\lb_2,\;\lb_3\;\lb_4=\infty$.
The  monodromy matrices are parametrized by  four parameters. 
The R-H problem is solved
in terms of the Szeg\"o kernel defined  on a trigonal curve of genus
two admitting the
dihedral group  $D_3$ of automorphisms. For this reason the trigonal curve
is a covering over two elliptic curves which are $3$-isogenous.
This fact enables  us to write the solution of the R-H problem in terms
of  Jacobi's $\vartheta$-functions with modulus $T=T(t)$, 
$t=\frac{\lb_2-\lb_1}{\lb_3-\lb_1}$.
The inverse function $t=t(T)$ is in general not single valued.
For $T$ belonging to  Siegel half space ${\mathcal H}_1$
modulo the group $\Gamma_0(3)$,
the function   $t=t(T)$ is single valued and the explicit
formula is given  in (\ref{tmodular2}). From this formula we have derived
an expression  for the solution  of the corresponding general
Halphen system equivalent to the one derived in \cite{hm00}.
From the solution of the R-H problem we have derived a
four parameter family of solutions of the Schlesinger system.
We suppose that these solutions would be the analogous of the elliptic
solution of the Painlev\'e VI equation \cite{Hi95}.
The study of the analytic continuation of the solutions of the above
$3\times 3$ Schlesinger system in the spirit of \cite{itno86,flnew80}
remains one of the subjects of our further investigation.
Our first observations show that the analytic continuation of the
solution of the Schlesinger system is induced by the action of $\Gamma_0(3)$ on
the characteristics $\delta_1,\delta_2,\epsilon_1,\epsilon_2$.
 We are  also interested to single out algebraic solutions and derive explicit algebraic
expression for the $3\times 3$ Schlesinger system as in 
\cite{duma00},\cite{mazz01}.

Finally, 
the R-H problem corresponding to the non-singular $Z_N$ curves
\[
y^N=\prod_{k=1}^{mN}(\lb-\lb_k),
\]
should be investigated.
This case is similar to the one treated in the present 
manuscript. The main techincal difficulty of the above
case is the determination of the explicit correspondence
between monodromy data and $\theta$-characteristics.
However, for the family of non-singular $Z_N$ surfaces, 
the fundamental quantities defined on the surfaces
like Bergmann kernel, projective connection,  Szeg\"o kernel
and Thomae type formula for $1/N$ characteristics  can be found in the literature  \cite{br88},\cite{na97}.

\section{Appendix}
\subsection{Proof of Lemma~\ref{diezlemma}} 
Assume the opposite: suppose that the divisor
$\mathcal{D}_m$ or $\mathcal{D}_{m+1,1}$
are  special,  this means that  there exists  a non-constant
meromorphic function $f(\lambda,y)$
whose divisor of poles is $\mathcal{D}_m$ or $\mathcal{D}_{m+1,1}$.
Then the function
$$\phi(\lambda,y)=f(\lambda,y)\prod_{i_j
\in\mathbb{I}_l}(\lambda-\lambda_{i_j}),\quad l=m,m+1,$$
has poles only at infinity. It follows from the Weierstrass gap
theorem, that the ring of meromorphic functions with
poles at infinity is generated in the case of the curve
$y^N=p(\lambda)q^{N-1}(\lambda)$ by powers of $\lambda$
and functions $y^i/q(\lambda)^{i-1}$, $i=0,\ldots,N-1$.
Therefore the function  $\phi(\lambda,y)$  can be written  in the form
\begin{equation}\phi(\lambda,y)=R_0(\lb)+\sum_{i=1}^{N-1}R_i(\lambda)
\dfrac{y^i}{q^{i-1}(\lambda)},\label{crucial}
 \end{equation}
where $R_i(\lambda)$ are polynomials in $\lambda$ and $q(\lambda)=\prod_{k=1}^m(\lb-\lb_k)$.\footnote{In this point our proof differs from
  that  given in \cite{diez91} which is working for Galois covers of
  the form $y^N=\prod_{i=1}^{mN}(\lambda-\lambda_i) $ where the ansatz for
  the function (\ref{crucial}) can be written as $\sum R_iy^i$  }

We remark that  $\mathrm{ord}_{\infty}\left(R_i(\lambda)
\dfrac{y^i}{q^{i-1}}\right)\neq
\mathrm{ord}_{\infty}\left(R_j(\lambda)\dfrac{y^j}{q^{j-1}}\right)$
for $i\neq j$ because otherwise
\begin{equation} N \mathrm{ord}_{\lambda} R_i(\lambda)+i\;\mathrm{deg} y -Nim  = N\mathrm{ord}_{\lambda}R_j(\lambda)+j\;\mathrm{deg} y -Njm,
 \end{equation}
and $N$ and $\mathrm{deg}\,y $ would not be relatively prime.
This observation implies that
\begin{equation}
\mathrm{ord}_{\infty}(f(\lambda,y)\prod_{i_l\in \mathbb{I}_l}
(\lambda-\lambda_{i_l})  ) =
\mathrm{ord}_{\infty}\left(R_j(\lambda)\dfrac{y^j}{q^{j-1}(\lambda)} \right )
 \label{degree} \end{equation}
for some $0\leq j \leq N-1$.   Moreover
\[
\mathrm{ord}_{\infty}(f(\lambda,y)\prod_{i_j\in \mathbb{I}_l}(\lambda-\lambda_{i_j})  )
=-N |\mathbb{I}_l|+k_l,\quad l=m,m+1,
  \]
where $k_l$, $l=m,m+1$ is the order at infinity of $f(\lambda,y)$ and
 the number of elements  $ |\mathbb{I}_m|=m$,  $ |\mathbb{I}_{m+1}|=m+1$.
From the equation of the curve we get  $\mathrm{deg}\,
y=mN+1$. Therefore the equality (\ref{degree}) can be
written as
\[
N |\mathbb{I}_l|-k_l=N(r_j+m)+j
\]
so that
\[
j=N (|\mathbb{I}_l|-r_j-m)-k_l\geq 0
.\]
When $l=m$ that is  $ |\mathbb{I}_l|=m$
it follow that $r_j=0, \,j=0,\, k_m=0$ and
\[
f(\lb,y)=\dfrac{1}{\prod_{i_n\in\mathbb{I}_m}(\lambda-\lambda_{i_n})}
\] and contradicts the assumption that $f(\lb,y)$ has divisor $\mathcal{D}_m$.

When $l=m+1$, that is  $ |\mathbb{I}_l|=m+1$ two possibility occurs:
(i) $r_j=0,\;j=N-k_{m+1},\;0\leq k_{m+1}<N$
and
(ii) $r_j=1,\,k_l=0,\;j=0$. This latter  case can be easily excluded while for the former one we have
\[
f(\lb,y)=\dfrac{1}{\prod_{i_j\in\mathbb{I}_{m+1} }(\lambda-\lambda_{i_j})}
\dfrac{ y^{N-k_{m+1} }}{q^{N-k_{m+1}-1}(\lambda)} ,\quad
\mathrm{ord}_{\infty}(f(\lambda,y))=k_{m+1}
,\]
which has divisor
\[
\text{Div}f(\lb,y)=-N\sum_{i_n\in\mathbb{I}_{m+1}}P_{i_n}+
(N-k_{m+1})\sum_{j=1}^{m+1}P_{2j+1}+k_{m+1}\sum_{j=1}^mP_{2j}.
\]
Namely the divisors of poles of $f(\lb,y)$ is
\[
\text{Div}_{\text{poles}}f(\lb,y)=(N-k_{m+1})\sum_{i_n\in
\mathbb{I}_{m+1},i_n\text{ even}}P_{i_n}+k_{m+1}
\sum_{i_n\in\mathbb{I}_{m+1},i_n\,\text{odd}}P_{i_n}
\]
and for $N>3$, differs from $\mathcal{D}_{m+1,1}$.
This contradicts  the assumption unless $f$ is constant.
For $N=3$ the divisor of poles of $f(\lb,y)$ coincides with
$\mathcal{D}_{m+1,1}$ in the following two cases:
\[
\mathcal{D}_{m+1,1}=2\sum_{k=1}^{m-1}P_{i_k}+P_{i_m}+P_{i_{m+1}},
\]
with
\[
\;\;i_m,\,i_{m+1}\in\{2,4,6\dots,2m\},
\;\;\;i_k\in\{1,3,5,\dots,2m+1\},\;k=1,\dots,m-1
\]
or
\[i_m,\,i_{m+1}\in\{1,3,5\dots,2m+1\},
\;\;\;i_k\in\{2,4,6,\dots,2m\},\;k=1,\dots,m-1.
\]
We conclude that the divisors (\ref{diez22}) where $i_m$ and $i_{m+1}$ have
different parity are non-special.
\hfill$\square$
\subsection{ Derivation of the Thomae formula}
We prove here  Theorem \ref{thomaetheorem}, that is the formula
\begin{equation}
\label{tome1}
\theta^8(\boldsymbol{0};\Pi)=\dfrac{\prod_{s=1}^{N-1}\mathrm {det}
\A^4_s}{(2\pi\i)^{4m(N-1)} }
\prod_{i<j}(\lambda_{2i}-\lambda_{2j})^{2(N-1)}
\prod_{k<l}(\lambda_{2k+1}-\lambda_{2l+1})^{2(N-1)},
\end{equation}
where the matrices $\A_s$, $s=1,\dots, N-1$,  are defined in (\ref{As}).

\begin{proof}
The proof of the theorem, consists of several steps. First we use
 Fay relation (\ref{fay212}) for zero characteristics, namely
\begin{equation}
\label{fay0} S(P,Q)^2=\omega(P,Q)
+\sum_{k,l=1}^g\frac{\partial^2}{\partial z_k\partial z_l}
\log\theta(\boldsymbol{0};\Pi)\mathrm{d}v_k(P)\mathrm{d}v_l(Q).
\end{equation}
We derive Thomae formula by evaluating the residues of
(\ref{fay0}) at $P=Q=(\lambda_i,0)$. The residue of the term containing
the derivatives of $\theta(\boldsymbol{0};\Pi)$, can be obtained
combining the heat equation (\ref{heat}),  the variation
formula (\ref{Rauch}) and the fact that the
function $\theta(\boldsymbol{0};\Pi)$ is even, which gives
\begin{equation}
\begin{split}
\label{aexp0} & \res[P=(\lambda_i,0)]\left[
\sum_{s=1}^N\sum_{k,l=1}^{(N-1)m}\frac{\partial^2}{\partial z_k\partial
z_l} \log\theta(\boldsymbol{0};\Pi)\dfrac{\mathrm{d}v_k(P^{(s)})
\mathrm{d}v_l(P^{(s)})}
{(\d z(P))^2}\right]\\
&= \sum_{k,l=1}^{(N-1)m}(1+2\delta_{kl})\dfrac{\partial}{\partial
\Pi_{k,l}}\log\theta(\boldsymbol{0};\Pi)\dfrac{\partial
\Pi_{k,l}}{\partial \lambda_i}=2\dfrac{\partial}{\partial \lambda_i}\log\theta(\boldsymbol{0};\Pi).
\end{split}
\end{equation}
From the expansion of the Szeg\"o  kernel given in
(\ref{szegoexp}) we obtain
\begin{equation}
\label{aexp1}
\res[P=Q=(\lambda_i,0)]\left[\sum_{s=1}^N
\dfrac{(S[0](P^{(s)},Q^{(s)}))^2}{dz(P)dz(Q)}\right]
=\dfrac{N^2-1}{12N}\res[\lb=\lambda_i]\left[\dfrac{
p^{\prime}(\lb)}{p(\lb)}- \dfrac { q^{\prime}(\lb)}{q(\lb)}
\right]^2.
\end{equation}

Now let us consider the Bergmann kernel $\omega(P,Q)$.
In order to write the  explicit expression for $\omega(P,Q)$,
 we follow \cite{hs66}.
The first step consists of constructing  the normalized
meromorphic differential
of the third kind $\omega_{Q,Q_0}(P)$
with simple poles at the points
$Q=(\nu, w)$ and $Q=(\nu_0, w_0)$,  with residues $\pm 1$
respectively, that is, for $P=(\lb,y)$,
\begin{equation}
\label{othird}
\begin{split}
&\Omega_{Q,Q_0}(P)=\dfrac{d\lb}{N(\lb-\nu)}\left(1+\sum_{s=1}^{N-1}
\dfrac{  w^s q(\lb)^{s-1} }{ y^s q(\nu)^{s-1} } \right)
-\dfrac{d\lb}{N(\lb-\nu_0)}\left(1+\sum_{s=1}^{N-1}
\dfrac{  w_0^s q(\lb)^{s-1} }{ y^s q(\nu_0)^{s-1} } \right)\\
&-\dfrac{1}{N}\sum_{j=1}^{(N-1)m}\d v_j(\lb)\oint_{\alpha_j}
d\xi\left[\dfrac{\left(1+\sum_{s=1}^{N-1} \dfrac{  w^s
q(\xi)^{s-1} }{ y^s_0q(\nu)^{s-1} } \right)}{(\xi-\nu)}-
\dfrac{\left(1+\sum_{s=1}^{N-1} \dfrac{  w_0^s
q(\xi)^{s-1} }{ y^s_0q(\nu_0)^{s-1} } \right)}{(\xi-\nu_0)}\right]
\end{split}
\end{equation}
where $\d v_j$, $j=1,\ldots,(N-1)m$ is the basis of normalized
holomorphic differentials and the point $(\xi,y_0)\in\mathcal{C}_{N,m}$.
The differential $\Omega_{Q,Q_0}(P)$ as a function of $Q$
is an Abelian integral with periods given  by the relations
\begin{equation}
\label{omperiods} \oint_{\alpha_j}\d_{\nu}
\Omega_{Q,Q_0}(P)=0,\quad
\oint_{\beta_j}\d_{\nu}\Omega_{Q,Q_0}(P)=2 \pi \i\,\d
v_j(P),\;\; j=1,\dots,(N-1)m.
\end{equation}
Furthermore the differential $\Omega_{Q,Q_0}(P)$ satisfies the
symmetry property $ \d_{\nu}\Omega_{Q,Q_0}(P)=
\d_{\lb}\Omega_{P,P_0}(Q)$, for $P_0\neq P$. Therefore the 2-differential,
$ \omega(P,Q):=\d_{\nu}\Omega_{Q,Q_0}(P),$
\begin{enumerate}
\item is symmetric in $P$ and $Q$;
\item is holomorphic everywhere except for a double pole along
$P=Q$, where
\[
\omega(P,Q)=\d\lb\,\d \nu \left(\dfrac{1}{(\lb-\nu)^2}+\,\text{regular terms}\right);
\]
\item  for any fixed $P$,  it satisfies (\ref{omperiods}).
\end{enumerate}
Therefore, $\omega(P,Q)$ is the Bergmann kernel given
alternatively in the form (\ref{Bergmann}).

In order to write more explicitly the Bergmann kernel, let us introduce the Abelian differentials $\sigma_{r,j}(\nu,w)$
of the second kind having the only  pole at
infinity of order $N(j+1)-r+1$, that is
\begin{equation}
\label{sigmarj}
\sigma_{r,j}(\nu,w)=\dfrac{ q(\nu)^{r-1}}
{ w^{r} }\mathcal{Q}_{r,j}(\nu)\d\nu,\;\;r=1,\dots,N-1,\;\;j\geq 0,
\end{equation}
where $\mathcal{Q}_{r,j}(\nu)$  are
polynomials in $\nu $ of degree $m+j$.
The coefficients of the polynomials
$\mathcal{Q}_{r,j}(\nu)$, $r=1,\dots,N-1$, $j\geq 0$, are uniquely determined
by the conditions
\begin{equation*}
\begin{split}
&\int_{\alpha_s}\sigma_{r,j}(\nu,w)=0,\;\;s=1,\dots,m,\\
&\sigma_{r,j}(\nu,w)\simeq
\left(\nu^{j-\frac{r}{N}}+O\left(\frac{1}{\nu^{1+\frac{r}{N}}}\right)\right)\d\nu,\;\;(\nu, w)\rightarrow(\infty,\infty).
\end{split}
\end{equation*}
From   the Riemann bilinear
relations we obtain the identities
\[
\int_{Q_0}^{Q}\sigma_{r,j}(P)+\res[P=(\infty,\infty)]\left[
\Omega_{Q,Q_0}(P)\d^{-1}\sigma_{r,j}(P)\right]=0,\quad r=1,\dots,N-1,\;j=0,\dots m-1,
\]
  so that we can  reduce the expression of $\omega(P,Q)=\d_{\nu}\Omega_{Q,Q_0}(P)$
to the form
\begin{equation}
\label{othird1}
\begin{split}
\omega(P,Q) =
\dfrac{\d\lb\d\nu}{N(\lb-\nu)^2}&\left(1+\sum_{s=1}^{N-1}
\dfrac{  w^s q(\lb)^{s-1} }{ y^s q(\nu)^{s-1} } \right)+
\dfrac{\d\lb\d\nu}{N(\lb-\nu)}\dfrac{d}{d \nu}\left(\sum_{s=1}^{N-1}
\dfrac{  w^s q(\lb)^{s-1} }{ y^s q(\nu)^{s-1} } \right)+\\
&-\dfrac{1}{N}\sum_{s=1}^{N-1}\sum_{j=1}^{m}\lb^{j-1}\dfrac{ q(\lb)^{s-1}}
{ y^{s}}\dfrac{q^{N-s-1}(\nu)\widetilde{\mathcal{Q}}_{s,j}(\nu)}{ w^{N-s}}\d\lb\d\nu,
\end{split}
\end{equation}
where $\widetilde{\mathcal{Q}}_{s,j}(\nu)$ is a polynomial depending on $\mathcal{Q}_{N-s,0}(\nu),\mathcal{Q}_{N-s,1}(\nu),\dots,\mathcal{Q}_{N-s,m-j}(\nu)$, $j=1,\dots,m$, $s=1,\dots,N-1$.
\begin{proposition}
For $s=1,\dots, N-1$,  the following identities are satisfied:
\begin{equation}
\label{detA}
\dfrac{\partial}{\partial \lambda_i}\log\det \A_s=
\dfrac{1}{\prod\limits_{\substack{l=1\\l\neq i}}^{2m+1}(\lambda_i-\lambda_l)}\sum_{j=1}^{m}\lambda_i^{j-1} \widetilde{\mathcal{Q}}_{s,j}(\lambda_i), \;\;i=1,\dots,2m+1,
\end{equation}
where the matrix $\A_s$ is defined in (\ref{As}).
\end{proposition}
{\bf Sketch of the proof.}
The integral of
$ \omega(P,Q)$ in the $P $ variable
 along the $\alpha_j$ periods is identically zero. Therefore,
substituting the local coordinate $\nu-\lambda_i=t^N$ in $\omega(P, Q)$
 and imposing that the terms
of order $dt,\,tdt,\dots,t^{N-2}dt$ of the integral
\[
\oint\limits_{\alpha_j}\omega(P,Q)=0,
\]
are identically zero, we obtain the statement.\\
Combining all the above relations we can  derive the explicit expression of the projective connection (\ref{pcon})
\[
\dfrac{1}{6}R(z(P))=\lim_{P\rightarrow Q}\left[\dfrac{\omega(P,Q)}{\d z(P)\,\d z(Q)}-\dfrac{1}{(z(P)-z(Q))^2}\right].
\]
\begin{proposition}
The projective connection $R(z(P))$ can be obtained
from (\ref{othird1}) and  reads
\begin{equation}
\label{omexp}
\begin{split}
\dfrac{1}{6}R(z(P))&=-\dfrac{ 1}{N }
\sum_{s=1}^{N-1}\sum_{j=1}^{m}\dfrac{z(P)^{j-1}\widetilde{\mathcal{Q}}_{s,j}(z(P))}{p(z(P))q(z(P))}+
 \dfrac{N^2-1}{12N^2}
 \left[\dfrac{  p^{\prime}(z(P))}{p(z(P))}- \dfrac { q^{\prime}(z(P))}{q(z(P))} \right]^2-\\
&-\dfrac{N-1}{4N}
\left[
\dfrac{ q^{\prime\prime}(z(P)) }{q(z(P))}+
\dfrac{ p^{\prime\prime}(z(P)) }{p(z(P))}
\right],
\end{split}
\end{equation}
where the prime denotes the derivative $\dfrac{d}{dz(P)}$.
\end{proposition}
Combining (\ref{detA}) and  (\ref{omexp}),  we  evaluate the residue of
 the Bergmann kernel at the branch points, namely,
\begin{equation}
\label{aexp2}
\begin{split}
\res[P=Q=(\lambda_i,0)]&\left[\sum_{s=1}^N\dfrac{\omega(P^{(s)},Q^{(s)})}{\d z(P)\d z(Q)}\right]=
\dfrac{N^2-1}{12N}\res[\lb=\lambda_i]\left[\dfrac{
p^{\prime}(\lb)}{p(\lb)}- \dfrac { q^{\prime}(\lb)}{q(\lb)}
\right]^2
-\\
&-\dfrac{N-1}{4}\res[\lb=\lambda_i]
\left[ \dfrac{ q^{\prime\prime}(\lb) }{q(\lb)}+
\dfrac{ p^{\prime\prime}(\lb) }{p(\lb)}
\right]-\sum_{k=1}^{N-1}
\dfrac{\partial}{\partial \lambda_i} \log\det \A_k.
\end{split}
\end{equation}
Substituting  (\ref{aexp0}), (\ref{aexp1}) and (\ref{aexp2}) in (\ref{fay0})
and simplifying, we obtain
\begin{equation}
2\dfrac{\partial}{\partial \lambda_i}\log\theta(\boldsymbol{0};\Pi)=
\dfrac{N-1}{4}\res[\lb=\lambda_i]
\left[
\dfrac{ q^{\prime\prime}(\lb) }{q(\lb)}+
\dfrac{ p^{\prime\prime}(\lb) }{p(\lb)}
\right]+\sum_{k=1}^{N-1}\dfrac{\partial}{\partial \lambda_i} \log\det \A_k,
\end{equation}
which gives (\ref{tome1}) up to a constant $C$.

To compute $C$  we pinch the branch points in the following way
\[
\lambda_{2k}=e_k+\epsilon,\quad \lambda_{2k-1}=e_k
-\epsilon \quad k=1,\dots,m,\quad 0<\epsilon\ll 1.
\]
In this case the l.h.s of (\ref{tome1})
becomes equal to one as $\epsilon \rightarrow 0$, more precisely
 $\theta(\boldsymbol{0};\Pi)=1+O(\epsilon)$.
Regarding the r.h.s the following relations are needed:
\[
\lim_{\epsilon\rightarrow 0}(\mathcal{A}_s)_{ij}=
2\pi\i\dfrac{e_i^{j-1}}{\prod\limits_{\substack{k\neq i
\\k=1}}^m(e_i-e_k) (e_i-\lambda_{2m+1})^{\frac{s}{N}}}
\]
so that
\begin{equation}
\label{lim0}
\lim_{\epsilon\rightarrow 0}(\det \mathcal{A}_s)=(2\pi \i)^m\dfrac{1}{\prod
\limits_{\substack{k< j\\k,j=1}}^m(e_k-e_j)}
\dfrac{1}{\prod\limits_{k=1}^m (e_k-\lambda_{2m+1})^{\frac{s}{N}}}
.\end{equation}
Substituting (\ref{lim0}) into (\ref{tome1}) and
letting  $\epsilon\ra 0$ in  all the terms of (\ref{tome1}),  we obtain
\begin{equation}
1=C \,(2\pi \i)^{4m(N-1)},
\end{equation}
and the expression for $C$ follows.
\end{proof}


\begin{thebibliography}{10}
\bibitem{kk98}
A~Kitaev and D.~Korotkin.
\newblock On solutions of {S}chlesinger equations in terms of theta-functions.
\newblock {\em Intern. Math. Res. Notices}, 17:877--905, 1998.


\bibitem{dikz99}
P.~A. Deift, A.~R. Its, A.~Kapaev, and X.~Zhou.
\newblock On the algebro-geometric integration of the {S}chlesinger equations.
\newblock {\em Commun. Math. Phys.}, 203:613--633, 1999.


\bibitem{kor01}
D.~Korotkin.
\newblock {\em Matrix {R}iemann-{H}ilbert problems related to branched coverings of
  $\mathbb {CP} 1$}.
\newblock In N.Manojlovic I.~Gohberg, A.F. dos~Santos, editor, {\em Operator
  {T}heory: {A}dvances and {A}pplications, {P}roceedings of the {S}ummer
  {S}chool on {F}actorization and {I}ntegrable {S}ystems, {A}lgarve,
  {S}eptember 6-9, 2000}, Boston, 2002. Birkh{\"a}user.
\newblock  xxx.lanl.gov/math-ph/0106009.\\
\newblock{ \em Solution of matrix Riemann-Hilbert problems with quasi-permutation monodromy matrices}, xxx.lanl.gov/math-ph/0306061.

\bibitem{aril85}
V.~I. Arnol'd and Yu.~S. Il'yashenko.
\newblock Ordinary differential equations.
\newblock In V.I.Arnold D.V.Anosov, editor, {\em Encyclopaedia of
  {M}athematical {S}ciences}, pages 7--149, Berlin, 1985. Springer Verlag.
\newblock Title of the Russian edition: Itogi nauki i tekhniki, Sovremennye
  problemy matematiki, Fundamrntal'nye napravlenia, Vol.1, Dinamicheskie
  sistemy I.

\bibitem{tr83}
A.~Treibich Kohn.
\newblock Un result de {P}lemelj.
\newblock {\em Progr. Math.}, 37:307--312, 1983.


\bibitem{an90} 
A.A. Bolibruch,\newblock {\em  The Riemann-Hilbert problem },
\newblock {\em  Russian.Math.Surveys}, 45(2):1-47, 1990.

\bibitem{anbo94}
D.V. Anosov and A.A. Bolibruch.
\newblock {\em The {R}iemann-{H}ilbert problem}, volume E 22 of {\em Aspect of
  Mathematics}.
\newblock Vieweg and Sohn, Braunschweig, 1994.

\bibitem{gla00}
A.~I. Gladyshev.
\newblock On the {R}iemann-{H}ilbert problem in dimension 4.
\newblock {\em Journ. of Dynam. Control Sys.}, 6(2):219--264, 2000.

\bibitem{dek79}
W.~Dekkers.
\newblock The matrix of a connection having regular singularities on a vector
  bundle of rank 2 on {$P^1(C)$}.
\newblock {\em Lecture Notes in Mathematics}, 712:33--43, 1979.

\bibitem{ko92}
V.~P. Kostov.
\newblock Fuchsian system on $\mathbb{CP}^1$ and the {R}iemann-{H}ilbert
  problem.
\newblock {\em C.R. Acad. Sci. Paris Ser I Math}, 315:207--238, 1992.

\bibitem{ok87}
K.~Okamoto.
\newblock Studies on the {P}ainlev\'e equations. {I}. {S}ixth {P}ainlev\'e
  equation {$P_{VI}$}.
\newblock {\em Annali Mat. Pura Appl.}, 146:337--381, 1987.

\bibitem{um90}
H.~Umemura.
\newblock Irreducibility of the {F}irst {D}ifferential {E}quation of
  {P}ainlev\'e.
\newblock {\em Nagoya Math. J.}, pages 231--252, 1990.

\bibitem{pl64}
J.~Plemelj.
\newblock {\em Problems in the sense of {R}iemann and {K}lein}.
\newblock Interscience, Mew York-London-Sydney, 1964.

\bibitem{zmnp80}
V.~E. Zakharov, S.~V. Manakov, S.~P. Novikov, and L.~P. Pitaevskii.
\newblock {\em Soliton theory: inverse scattering method}.
\newblock Nauka, Moscow, 1980.


\bibitem{bbeim94}
E.~D. Belokolos, A.~I. Bobenko, V.~Z. Enolskii, A.~R. Its, and V.~B. Matveev.
\newblock {\em Algebro {G}eometrical {A}pproach to {N}onlinear {I}ntegrable
  {E}quations}.
\newblock Springer, Berlin, 1994.


\bibitem{diz97}P.~A. Deift, A.~R. Its  and X.~Zhou,
{\em A Riemann-Hilbert approach to asymtotic problems arising in the
theory of random matrix models, and also in the theory of integrable
statistical mechanics}. {\em Ann. of Math.}, 146:149-235, 1997.

\bibitem{dkmvz99}  P.Deift,  T.Kriecherbauer, K.McLaughlin,  S. Venakides, X.Zhou,  \newblock {\em Uniform asymptotics for polynomials orthogonal with respect to varying exponential weights and
   applications to universality questions in random matrix theory}.
 {\em Comm. Pure Appl. Math.} 52 (1999), no. 11, 1335--1425.

\bibitem{dvz97} P.Deift, S.Venakides, X.Zhou, \newblock {\em New results in small dispersion KdV by an extension of the steepest descent method for Riemann-Hilbert problems}. {\em Internat. Math. Res.
   Notices} 1997, no. 6, 286--299.  


\bibitem{zv71}
E.~I. Zverovich.
\newblock Boundary problems of the theory of analytic functions.
\newblock {\em Uspekhi. Matem. Nauk}, 31(5):113--181, 1971.


\bibitem{mi95}
R.~Miranda.
\newblock {\em Algebraic Curves and Riemann Surfaces}, volume~5 of {\em
  Graduate Studies in Mathematics}.
\newblock Amer. Math. Soc., Providence, R.I., 1995.

\bibitem{hur91}
A.~Hurwitz.
\newblock {\"U}ber {R}iemann'sche {F}l{\"a}chen mit gegeben
  {V}erzweigungspunkten.
\newblock {\em Math.Ann.}, 39:1--61, 1891.



\bibitem{schl12}
L.~Schlesinger.
\newblock {\"U}ber eine {K}lasse von {D}ifferentialsystemen beliebiger
  {O}rdnung mit festen kritischen {P}unkten.
\newblock {\em J. reine angew. Math.}, 141:96--145, 1912.

\bibitem{JMU81}
M.~Jimbo, T.~Miwa, and K.~Ueno.
\newblock Monodromy preserving deformation of linear ordinary differential
  equations with rational coefficients {I}.
\newblock {\em Physica D}, 2:306--352, 1981.

\bibitem{duma03}
B.~A. Dubrovin and M.~Mazzocco.
\newblock 2003.
\newblock Canonical structure of the Schlesinger systems, in preparation.

\bibitem{fo29}
L.~Ford.
\newblock {\em Automorphic {F}unctions}.
\newblock McGraw--Hill, New York, 1929.

\bibitem{hm00}
J.~Harnad and J.~McKay.
\newblock Modular solutions to equations of generalized {H}alphen type.
\newblock {\em R.  Soc.  Lond.  Proc.  Ser. A.  Math.  Phys.}, 456(1994):261--294, 2000.



\bibitem{hut02}
J.~I. Hutchinson.
\newblock On a class of automorphic functions.
\newblock {\em Trans. Amer. Math. Soc.}, 3:1--11, 1902.

\bibitem{dh01}
B.~Deconinck and M.~van Hoeij.
\newblock Computing {R}iemann matrices of algebraic curves.
\newblock {\em Physica D}, 152-153:28--46, 2001.

\bibitem{muskhel72}
I.~N. Muskhelishvili.
\newblock {\em Singular integral equations}.
\newblock Wolters-Noordhoff, Groningen, 1972.

\bibitem{ma83}
B.~Malgrange.
\newblock Sur les deformation isomonodromiques.
\newblock In A.~Douady A.~Beauville and J.~L. Verdier, editors, {\em
  S\'eminaire E.N.S. 1978-1979, Progress in Mathematics vol. 37}, pages
  401--426, Boston, 1983. Birkh{\"a}user.

\bibitem{th69}
J.~Thomae.
\newblock Beitrag zur {B}estimmung von $\vartheta(0,0,\ldots,0)$ durch die
  {K}lassenmoduln algebraischer {F}unctionen.
\newblock {\em Journ. reine angew. Math.}, 71:201--222, 1870.

\bibitem{ra59}
H.~E. Rauch.
\newblock Weierstrass points, branch points and moduli of {R}iemann surfaces.
\newblock {\em Comm. Pure Appl. Math.}, 12(3):543--560, 1959.

\bibitem{fa92}
J.~D. Fay.
\newblock Kernel functions, analytical torsion and moduli spaces.
\newblock In {\em Memoirs of the {A}merican {M}athematical {S}ociety},
  volume~96, Providence, Rhode Island, 1992. American Mathematical society.


\bibitem{fa73}
J.~D. Fay.
\newblock Theta functions on {R}iemann surfaces.
\newblock In {\em Lectures Notes in Mathematics}, volume 352, Berlin, 1973.
  Springer.

\bibitem{ti78}
A.~N. Tiurin.
\newblock On periods of quadratic differentials.
\newblock {\em Rus. Math. Surv.}, 33(6):149--195, 1978.

\bibitem{fk80}
H.~M. Farkas and I.~Kra.
\newblock {\em Riemann {S}urfaces}.
\newblock Springer, New York, 1980.

\bibitem{diez91}
G.~G. D{i}ez.
\newblock Loci of curves which are prime {G}alois coverings of $\mathbb{P}^1$.
\newblock {\em Proc. London Math. Soc.}, 62:469--489, 1991.

\bibitem{narXX}
M. Narasimhan.
\newblock {\em Lectures on Theta-Functions},  Lectures delivered
at the University of Kaiserslautem,   1987.

\bibitem{br88}
M.~Bershadsky and A.~Radul.
\newblock Fermionic fields on ${Z}_{N}$ curves.
\newblock {\em Commun. Math. Phys.}, 116:689--700, 1988.

\bibitem{kn89}
V.G., Knizhnik, {\em Multi-loop amplitudes in the theory of qunatum string and comples
geometry}. {\em Sov. Phys. Uspekhi}, 57:945-971, 1989.

\bibitem{na97}
A.~Nakayashiki.
\newblock On the {T}homae formula for ${Z}_{N}$ curves.
\newblock {\em Publ. Res. Inst. Math. Sci.}, 33(6):987--1015, 1997.

\bibitem{bo88a}
O.~Bolza.
\newblock On binary sextics with linear transformations onto themselves.
\newblock {\em Amer. Journ. Math.}, 10:47--70, 1888.

\bibitem{earleXX}
C.~Earle.
\newblock Some {J}acobians varieties which split.
\newblock In {\em Lect. Notes Math.}, volume 747, pages 101--107. Springer,
  1979.

\bibitem{farkas74}
H.~Farkas.
\newblock Remarks on automorphisms of compact Riemann surfaces.
\newblock In {\em Discontinuous {G}roups and {R}iemann {S}urfaces, Proceedings
  of the 1973 {C}onference at the {U}niversity of {M}aryland}, Princeton, New
  Jersey, 1974. Princeton University Press and Toyo University Press.

\bibitem{shiga88}
H.~Shiga.
\newblock On the representation of the {P}icard modular function by $\theta$
  constants {I}-{II}.
\newblock {\em Publ. RIMS, Kyoto Univ.}, 24:311--360, 1988.

\bibitem{gonrod99}
V.~Gonza{\'a}lez-Agulera and R.~E. Rodriguez.
\newblock Families of irreducible principally polarized abelian varieties
  isomorphic to a product of elliptic curves.
\newblock {\em Proc. AMS}, 128(3):629--636, 1999.

\bibitem{ig60}
J.~Igusa.
\newblock Arithmetic variety of moduli for curves of genus two.
\newblock {\em Ann. of Math.}, 72:612--649, 1960.


\bibitem{ig62}
J.~Igusa.
\newblock On {S}iegel modular forms of genus two.
\newblock {\em Amer. J. Math.}, 84:175--200, 1962.



\bibitem{lange76}
H.~Lange.
\newblock Uber die {M}odulvariet{\"a}t der {K}urven vom {G}eschlecht $2$.
\newblock {\em Journ. reine angew Math.}, 281:80--96, 1976.

\bibitem{rr88}
G.~Riera and R.~Rodriguez.
\newblock Uniformization of {S}urfaces of {G}enus {T}wo with {A}utomorphisms.
\newblock {\em Math. Ann.}, 282:51--67, 1988.

\bibitem{lawd89}
D.~F. Lawden.
\newblock {\em Elliptic {F}unctions and Applications}.
\newblock Applied Mathematical Sciences, vol. 80. Springer, New York, 1989.


\bibitem{goursat81}
E.~Goursat.
\newblock Sur {L}'\'equation diff\'erentielle lin\'eaire qui admet pour
  int\'egrale la s\'erie hypereg\'eom\'etrique.
\newblock {\em Ann. Sci. \^Ecole Norm. Sup.}, 2(10):3--142, 1881.

\bibitem{Neh52}
Z.~Nehari.
\newblock {\em Conformal Mapping}.
\newblock Dover Publications, New York, 1952.

\bibitem{in56}
E.~L. Ince.
\newblock {\em Ordinary {D}ifferential {E}quations}.
\newblock Dover, New York, 1956.

\bibitem{AbCh96}
S.Chakravarty M.J.Ablowitz.
\newblock Integrability, monodromy evolving deformations,
     and self-dual Bianchi IX systems.
\newblock {\em Phys. Rev. Lett.}, 76:857--860, 1996.

\bibitem{Hi98}
N.~Hitchin. Hypercomplex manifolds and the space of framing,
in \newblock {\em The geometric universe}.
\newblock Oxford University Press, Oxford, England, 1998.

\bibitem{tak92}
L.~A. Takhtadjan.
\newblock A simple example of modular forms as 
$tau$-functions for integrable
  equations.
\newblock {\em Teor. Mat. Fiz.}, 93:330--341, 1992.

\bibitem{chha03}
S.~Chakravarty and R.G. Halburd.
\newblock First integrals of a generalized {D}arboux-{H}alphen system.
\newblock {\em Journal of Mathemaical Physics}, 44:1751--1762, 2003.

\bibitem{Hi95}
N. Hitchin, {\em Twistor spaces, Einstein metrics and 
isomonodromic deformations}, J. Diff. Geom. 41:31-112, 1995.

\bibitem{itno86}
A.~R. Its and V.~Yu. Novokshenov.
\newblock {\em The isomonodromic deformation method in the theory of Painlev\`e
  equations}.
\newblock Lect. Notes. Math. vol 1191. Springer, 1986.

\bibitem{flnew80}
H.~Flaschka and A.~C. Newell.
\newblock Monodromy and spectrum preserving deformations.
\newblock {\em Comm. Math. Phys.}, 76:65--116, 1980.


\bibitem{duma00}
B.~A. Dubrovin and M.~Mazzocco.
\newblock Monodromy of certain {P}ainleve-{VI} transcendents and reflection
  groups.
\newblock {\em Invent. math.}, 141:55--147, 2000.

\bibitem{mazz01}
M.~Mazzocco.
\newblock Picard and {C}hazy solutions to the {P}ainlev\'e VI equation.
\newblock {\em Math. Ann.}, 321:157--195, 2001.

\bibitem{hs66}
N.~S. Hawley and M.~Schiffer.
\newblock Half-order differentials on {R}iemann surfaces.
\newblock {\em Acta Mathematica}, 115:199--236, 1966.
\end{thebibliography}
\end{document}